\documentclass{aa} 
\usepackage{graphicx}
\usepackage[varg]{txfonts}
\usepackage{booktabs}
\usepackage{longtable}
\usepackage{natbib}
\usepackage{pdflscape}
\usepackage{hyperref}
\usepackage{xcolor}
\usepackage{caption}
\usepackage{placeins}
\usepackage{float}

\bibpunct{(}{)}{;}{a}{}{,}

\newcommand{\Ha}{\ensuremath{\mathrm{H}_{\alpha}}}
\newcommand{\Hb}{\ensuremath{\mathrm{H}_{\beta}}}
\newcommand{\Hg}{\ensuremath{\mathrm{H}_{\gamma}}}
\newcommand{\Hd}{\ensuremath{\mathrm{H}_{\delta}}}
\newcommand{\CaTone}{\ensuremath{\mathrm{CaT1}}}
\newcommand{\CaTtwo}{\ensuremath{\mathrm{CaT2}}}

\makeatletter
\renewcommand*\aa@pageof{, page \thepage{} of \pageref*{LastPage}}
\makeatother

\begin{document}

\titlerunning{Radial velocities}
\authorrunning{Baeza-Villagra et al.}
\title{On the use of field RR Lyrae as Galactic probes:}
\subtitle{IX. Radial velocities}

\author{K. Baeza-Villagra \inst{\ref{inst1},\ref{inst2},\ref{inst3}}\orcid{0000-0003-2804-1261}
\and
V. F.  Braga\inst{\ref{inst3}}
\and
M. Fabrizio \inst{\ref{inst4},\ref{inst3}}
\and
G. Bono \inst{\ref{inst1},\ref{inst3},\ref{inst05}}\orcid{0000-0002-4896-8841}
\and
V. D'Orazi
\inst{\ref{inst1},\ref{inst3}}
\and
Z. Prudil
\inst{\ref{inst5}}
\and
F. Berrilli 
\inst{\ref{inst1}}
\and
G. B{\"o}cek Topcu
\inst{\ref{inst1},\ref{inst3}}
\and
G. Ceci
\inst{\ref{inst1},\ref{inst2},\ref{inst3}}
\and
B. Chaboyer
\inst{\ref{inst12}}
\and
M. Dall'Ora\inst{\ref{inst100}}
\and
D. Del Moro
\inst{\ref{inst1}}
\and
M. Di Criscienzo
\inst{\ref{inst3}}
\and
G. Fiorentino
\inst{\ref{inst3}}
\and
M. Gholami \inst{\ref{inst100}}
\and
R.-P. Kudritzki\inst{\ref{inst110},\ref{inst111}}
\and
M. Marengo
\inst{\ref{inst6}}
\and
N. Matsunaga
\inst{\ref{inst9}}
\and
M. Monelli \inst{\ref{inst20},\ref{inst21},\ref{inst22}}
\and
J. P. Mullen
\inst{\ref{inst10}}
\and
A. Nunnari
\inst{\ref{inst3},\ref{inst1},\ref{inst2}}
\and
G. W. Preston
\inst{\ref{inst11}}
\and
C. Sneden
\inst{\ref{inst7}}
\and
M. Tantalo
\inst{\ref{inst3}}
\and F. Thévénin \inst{\ref{inst101}}
\and
I. B. Thompson
\inst{\ref{inst11}}
\and
E. Valenti
\inst{\ref{inst5},\ref{inst14}}
\and
M. Zoccali 
\inst{\ref{inst8}}}
\institute{Dipartimento di Fisica, Università di Roma Tor Vergata, Via della Ricerca Scientifica, 1, Roma 00133, Italy\label{inst1} 
\and Dipartimento di Fisica, Sapienza Università di Roma, P.le A. Moro 5, Roma 00185, Italy\label{inst2} 
\and INAF—Osservatorio Astronomico di Roma, via Frascati 33, I-00078 Monte Porzio Catone, Italy \label{inst3}
\and ASI - Space Science Data Center, via del Politecnico snc, I-00133 Roma, Italy \label{inst4}
\and INFN Sezione di Roma2, Università di Roma Tor Vergata, Via della Ricerca
Scientifica 1, Roma 00133, Italy \label{inst05}
\and European Southern Observatory, Karl-Schwarzschild-Strasse 2, 85748 Garching bei M\"{u}nchen, Germany \label{inst5}
\and
Department of Physics and Astronomy, Dartmouth College, 6127 Wilder Laboratory, Hanover, NH 03755, USA \label{inst12}
\and INAF - Osservatorio Astronomico di Capodimonte, Salita Moiariello 16, I-80131 Napoli, Italy\label{inst100}
\and Institute for Astronomy, University of Hawai’i at Manoa, Honolulu, HI 96822, USA\label{inst110}
\and LMU München, Universitätssternwarte, Scheinerstr. 1, 81679 München, Germany \label{inst111}
\and Florida State University, Department of Physics, 77 Chieftain Way, Tallahassee, FL 32306, USA\label{inst6}
\and Department of Astronomy, School of Science, The University of Tokyo, 7-3-1 Hongo, Bunkyo-ku, Tokyo 113-0033, Japan \label{inst9}
\and INAF-Osservatorio Astronomico d'Abruzzo, via Mentore Maggini s.n.c., 64100 Teramo, Italy \label{inst20}
\and IAC- Instituto de Astrof\'isica de Canarias, Calle V\'ia Lactea s/n, E-38205 La Laguna, Tenerife, Spain \label{inst21}
\and Departmento de Astrof\'isica, Universidad de La Laguna, E-38206 La Laguna, Tenerife, Spain \label{inst22}
\and Department of Physics and Astronomy, Vanderbilt University, Nashville, TN 37240, USA\label{inst10}
\and Carnegie Observatories, 813 Santa Barbara Street, Pasadena, CA 91101-1292, USA\label{inst11}
\and Department of Astronomy and McDonald Observatory, The University of Texas, Austin, TX 78712, USA \label{inst7}
\and Université Côte d’Azur, Observatoire de la Côte d’Azur, CNRS, Laboratoire Lagrange, France \label{inst101} 
\and Excellence Cluster ORIGINS, Boltzmann–Straße 2, D–85748 Garching bei München, Germany \label{inst14}
\and Instituto de Astrof{\'{\i}}sica, Pontificia Universidad Cat{\'o}lica de Chile, Av. Vicu{\~n}a Mackenna 4860, Santiago, Chile \label{inst8}  
} 

\date{Received / Accepted}
\abstract{We present the largest and most homogeneous catalog of radial velocity (RV) measurements for field RR Lyrae (RRL) variables, based on both proprietary and publicly available spectroscopic data. The sample includes 17,563 RRLs pulsating in the fundamental mode (12,353 RRab), in the first overtone (5,011 RRc), and in double-mode (199 RRd). The RV curve (RVC) templates for metallic and Balmer lines were used to derive RV amplitudes and $V_{\gamma}$ velocities, defined as the RV of the stellar barycenter with respect to the Sun. The typical accuracy across the catalog is on average 3.8 km s$^{-1}$ for well-sampled RVCs, 6.5 km s$^{-1}$ for RVCs with 3-7 phase points and 11.3 km s$^{-1}$ for RVCs with fewer than three phase points. The use of different spectroscopic diagnostics and RVC templates provides, within the errors, very similar $V_{\gamma}$ velocities. We found that the metallicity dependence of RV amplitudes is vanishing for metallic lines, but becomes increasingly significant for H$\gamma$ and H$\delta$. Moreover, the scaling relations between photometric (V, $G_{BP}$, G, $G_{RP}$) and RV amplitudes are linear for RRc and nonlinear for RRab variables, independently of the adopted diagnostic. This circumstantial evidence indicates that convection affects more luminosity than RV amplitudes when moving from the blue (hot) to the red (cool) edge of the instability strip. The spectroscopic Bailey diagram (RV amplitude versus period) shows a smooth transition and a reduced spread at a fixed period, when moving from metal-poor to metal-rich RRLs. Finally, we also found evidence that the metallicity distribution function of Blazhko RRLs is skewed toward the metal-intermediate and metal-rich regimes.}

\keywords{Galaxy: kinematics and dynamics - Stars: variables: RR Lyrae - Techniques: radial velocities} 
\maketitle 
\nolinenumbers
\section{Introduction}
RR Lyrae (RRL, hereafter) variables are core helium-burning stars commonly used as tracers of ancient (t $\sim$ 10 Gyr) stellar populations  \citep{Walker1989,Smith1995,Catelan2009,Savino2020}. Recent studies have suggested that RRL stars could also trace intermediate-age populations \citep[e.g.,][]{Bobrick2024,Cuevas-Otahola2025}, but the current empirical evidence indicates that they constitute a minority fraction \citep{Bono2025}. The RRLs provide key insights into the early formation and evolutionary history of the Milky Way \citep[e.g.,][]{Fiorentino2015,Iorio2018,Kunder2022,Crestani2021b,Dorazi2024,Feng2025}. These stars are intrinsic radial pulsators with periods ranging from $\sim$ 0.2 to 1.25 days \citep{CatelanSmith2015,braga2020} and low masses \citep[0.55 to 0.8 $M_\odot$; e.g.,][]{Marconi2015}. Based on their pulsation mode, RRL stars are classified into three main subtypes \citep{Smith1995}: RRab (fundamental mode), RRc (first overtone), and RRd (double mode: they pulsate simultaneously in the fundamental and the first overtone). 
The Bailey diagram (luminosity amplitude versus logarithmic period) can be used to identify the RRL pulsation mode \citep{schwarzschild1940}. The distribution of RRL variables in the Bailey diagram is determined by their physical properties, such as stellar mass, effective temperature, and chemical composition \citep{Bono2020}. The RRab variables occupy the long-period region and exhibit a well-defined decrease in luminosity amplitude as the period increases. In contrast, RRc stars are found in the short-period, low-amplitude region, while RRd stars are located in the transition between RRc and RRab \citep{bragaglia2001}.

The use of RRLs as kinematic stellar tracers requires three fundamental ingredients: photometric time series, spectroscopic time series, and proper motions. Radial velocity (RV) measurements, in particular the barycentric RV with respect to the Sun ($V_{\gamma}$), together with precise individual distances derived from photometric time series and proper motions, are essential for investigating the kinematic properties of these objects. However, the measurement of $V_{\gamma}$ for RRLs requires multiple observations to accurately track its variation throughout the pulsation cycle. For this reason, several authors have built radial velocity curve (RVC) templates to estimate $V_{\gamma}$ using a limited number of RV measurements \citep[e.g.,][]{Oke1966, Liu1991, Sesar2012, Braga2021, Prudil2024}.

The main goal of this work is to provide homogeneous and accurate measurements of $V_{\gamma}$ values and RV amplitudes for a large sample of field RRLs. These new measurements are fundamental for investigating the pulsation and kinematic properties of these ancient stellar tracers. To accomplish this goal, we used RV measurements from a combination of proprietary and publicly available spectroscopic catalogs. The RV amplitudes were derived using two different diagnostics: Balmer lines and metallic lines. The key advantage of using RV amplitudes to investigate the RRL pulsation properties is that they are mainly driven by radius variations, whereas photometric amplitudes are mainly driven by temperature variations \citep[][]{Bono2020}. Moreover, accurate determinations of $V_{\gamma}$ are essential for investigating Galactic dynamics based on RRL variables, since they overcome the errors associated with the random sampling of the RV curves and provide reliable tracers of the underlying stellar kinematics \citep[e.g.,][]{Dambis2013,Kunder2017,kunder2020,Prudil2020}.

The paper is organized as follows. Sections~\ref{sec:2} and~\ref{sec:3} describe the data and the methods used. The results and their discussion are presented in Section~\ref{sec:4}, and Section~\ref{sec:5} summarizes the main results and outlines future developments of this project.

\section{Datasets}\label{sec:2}
To obtain the individual RV measurements, we analyzed a large sample of high-, medium-, and low-resolution spectra (HR, MR, and LR, respectively). We used a combination of our own observations and data from public sources (see Section~\ref{datasets}), together with RVCs of RRLs, to derive the $V_{\gamma}$ and RV amplitude values (see Section~\ref{rvmeasurements}). In addition, the current sample was supplemented with pulsation periods from the Photometric Rome RR Lyrae Catalog (PR3C; \citealt{Braga2021,fabrizio2021,Bono2025}), which is currently being extended by Braga et al. (in prep.), as well as with coordinates (RA, DEC), astrometric and photometric data from the Gaia RR Lyrae catalog \citep[GRRL;][]{Clementini2023,Katz2023}. The latter provides luminosity amplitudes in three photometric bands (Amp(G), Amp($G_{RP}$), Amp($G_{BP}$)) and the RV amplitude (Amp(RV)).

\begin{figure*}
\center
\includegraphics[width=0.9\textwidth]{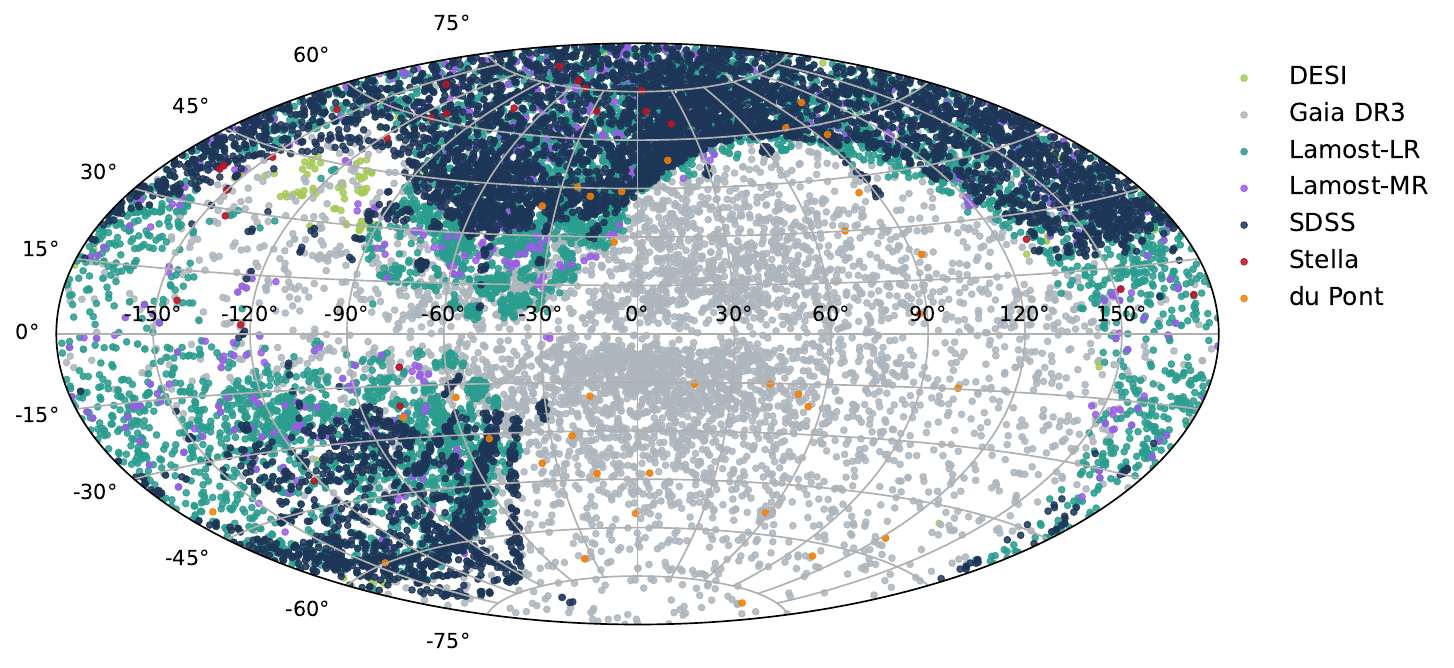}
\caption{Distribution in Galactic coordinates of the entire spectroscopic catalog. The RRLs coming from different spectroscopic datasets are marked with different colors (see labels).}\label{fig:distribucion}
\end{figure*}

Our dataset includes a total of 17,563 RRLs (12,353 RRab, 5,011 RRc, and 199 RRd). The current sample also includes RRLs in nearby globular clusters and in dwarf galaxies. In particular, we identified 246 RRLs associated with either globular clusters or dwarf galaxies according to the classification provided by \citet{CruzReyes2024}. Fig.~\ref{fig:distribucion} shows the sky distribution of the current spectroscopic catalog, which covers the Galactic bulge, halo, thick disk, and thin disk \citep{Maintz2005,Gozha2018,Prudil2020}. 

\begin{figure}
\center
\includegraphics[width=0.45\textwidth]{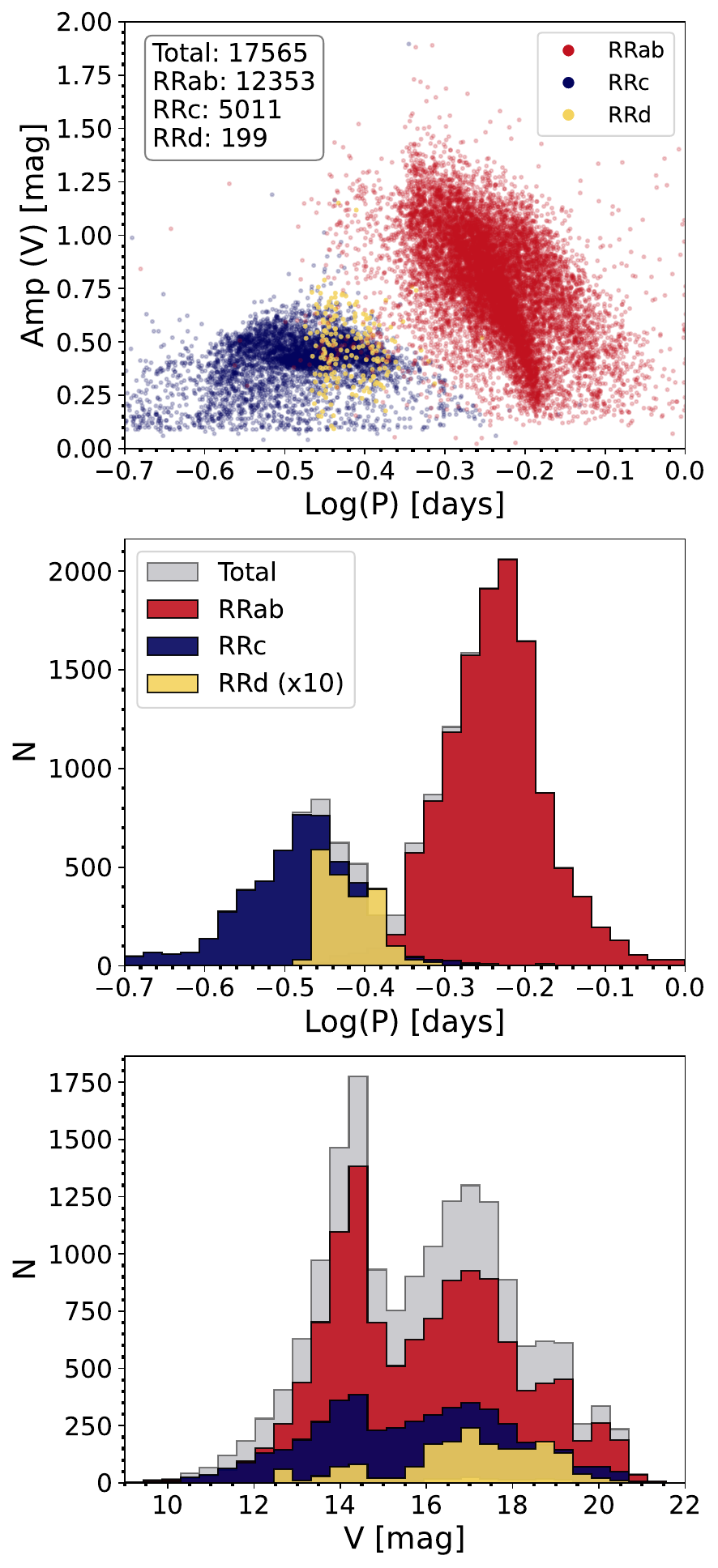}
\caption{Top: Bailey diagram of the spectroscopic sample, showing visual amplitude versus logarithmic period for RRab (red), RRc (blue), and RRd (yellow) variables.
Middle: Logarithmic period distribution for the entire spectroscopic sample (gray), alongside the period distributions for RRab (red), RRc (blue), and RRd (yellow). To facilitate the visualization the RRd counts were multiplied by a factor of ten. 
Bottom: \textit{V}-band magnitude distributions for the entire sample, following the same color code as the top panel. }\label{fig:bailey}
\end{figure}

The period-amplitude (or Bailey) diagram and the period distributions for these RRLs are shown in the top and middle panels of Fig.~\ref{fig:bailey}. Notably, the current sample spans the full period range of RRLs. The vast majority of RRab stars have periods ranging from approximately 0.4 to nearly one day, while RRc stars typically show periods between $\sim$0.2 and $\sim$0.5 days. On the other hand, the RRd variables predominantly have periods between $\sim$0.34 and $\sim$0.45 days. Note that the global fraction of RRc variables accounts for roughly $29\%$ of the entire sample. However, the inclusion of RRc variables poses a challenge due to the presence of short-period eclipsing binaries, which can mimic their luminosity variations \citep{Botan2021}. To mitigate this contamination, we adopted an approach based on the amplitude ratios of optical and mid-infrared data, combined with the visual inspection of the shape of the light curve and the use of Fourier parameters. A more detailed description of this method can be found in \citet{fabrizio2021} and \citet{Mullen2022}. The final sample includes 5,011 RRc variables after applying these selection criteria. Finally, the bottom panel of Fig.~\ref{fig:bailey} shows the distribution of the entire catalog in \textit{V}-band magnitude. The data indicate that the current spectroscopic sample of RRLs covers a wide range of Galactocentric distances, from the solar neighborhood to the outskirts of the Galactic halo ($\sim$100 kpc). Note that main and secondary peaks in the magnitude distribution reflect the limiting magnitudes of the different spectroscopic surveys.

\subsection{Spectroscopy datasets}\label{datasets}
Preliminary versions of this spectroscopic catalog, called the Spectroscopic Rome RRL Catalog (SR3C), have already been used in previous investigations by our group, focused either on RVs and chemical abundances of field RRLs \citep{Fabrizio2019,Bono2020,fabrizio2021,Crestani2021b,Crestani2021a} or on Galactic structure \citep{Bono2025}.
\\
The current version includes LR spectra from the LAMOST LR (7,284 variables), SDSS-DR18 (5,984 variables), and DESI-EDR (594 variables) surveys; MR spectra from LAMOST MR (639 variables) and Gaia DR3 (1,091 variables); and HR spectra collected with the STELLA (14 stars) and du Pont (34 stars) telescopes. This sample was complemented with 4,449 variables for which $V_{\gamma}$ measurements were available in the literature.
\\
The current spectroscopic catalog (SR3C-2), when compared with the previous one (SR3C), differs in the following aspects:\\
i) the total number of objects (17,563 vs. 16,000);\\
ii) the adoption of a new validation test, which led to the removal of 117 variables identified as spurious;\\
iii) the new $V_{\gamma}$ measurements are mainly based on template fitting (8,664 variables), whereas the previous ones were primarily based on the mean of the measurements (see Section~\ref{sec:3} for further details).\\
The key features of each dataset are summarized in Table~\ref{tab:datasets}, while the stars in common between the different datasets are listed in Table~\ref{tab:starscommon}.

Fig.~\ref{fig:lines} shows representative spectra for RRab (black) and RRc (violet) stars from the LR (top panel) and HR (bottom panels) datasets. Spectra are shown for four RRL stars, covering the Balmer lines, the Mg I b triplet, the Na I doublet, and two lines of the near-infrared Ca II triplet.

\subsection{RV measurements}\label{rvmeasurements}
The observed RV variation in pulsating stars reflects the combination of the systemic velocity $V_{\gamma}$ and the radial motion of the atmospheric layers where the absorption lines are formed. By comparing the observed spectrum with reference spectra, we can identify shifts in the wavelengths of specific lines and calculate the RV. The RRLs are radial variables and different spectral lines trace the radial displacements of the atmosphere at different optical depths, even when observed at the same pulsation phase. This is because different lines form at different radial distances and in different physical conditions along the pulsation cycle \citep{Chadid2017,Sneden2017,Bono2020,Sneden2024}.

Following the approach adopted by \citet{fabrizio2011} and \citet{Braga2021}, we performed RV measurements separately for each spectroscopic diagnostic, using the same code and procedure described in those works. The analysis included four Balmer lines (\Ha, \Hb, \Hg, \Hd), two lines of the near-infrared Ca II triplet (\CaTone{} and \CaTtwo), a set of Fe and Sr lines (three lines from the Fe I multiplet 43 and the Sr II resonance line), the two lines of the Na I doublet (D$_1$ and D$_2$), and the Mg I $b$ triplet (Mg $b_1$, Mg $b_2$, Mg $b_3$). The laboratory air wavelengths of these absorption features are listed in Table~\ref{tab:wave}.
The RVs were determined by fitting a Lorentzian profile to each absorption line using an automated procedure written in IDL. The wavelength range adopted by the fitting algorithm was fixed according to the spectral resolution of each spectrograph. Specifically, we selected a wavelength interval extending ten times the full width at half maximum (FWHM) on either side of the line center. The FWHM was estimated as $FWHM=2.355 \times \frac{\lambda_{obs}}{R}$, where $\lambda_{obs}$ is the wavelength of the diagnostic and $R$ is the spectral resolution. 

Our analysis also includes RV measurements provided by Gaia’s Radial Velocity Spectrometer, which covers a wavelength range from 8450 $\text{\AA}$ to 8720 $\text{\AA}$, encompassing the near-infrared Ca II triplet region as well as other metallic lines \citep{Cropper2018, Katz2023}. The Gaia DR3 catalog provides individual RV measurements for 1,091 RRLs. Among them, 1,086 stars have at least seven individual RV measurements obtained across different epochs \citep{Clementini2023}.

The RV estimates were flagged as invalid when they exceeded the range |RV| $\geq$ 1000 km s$^{-1}$, or when the corresponding spectral line was deemed unreliable. In the latter case, a visual inspection of the spectra was performed to identify measurements affected by low signal-to-noise ratios, spurious features, or poorly defined line profiles. Measurements classified as invalid according to these criteria were excluded from the final analysis.

\section{Radial velocity curve templates}\label{sec:3}
In this study, we adopted the RVC templates for RRL stars provided by \citet{Braga2021}. They were developed for three groups of metallic lines (Fe, Mg, and Na) and four Balmer lines (\Ha, \Hb, \Hg, \Hd). In principle, these templates require prior knowledge of the pulsation period, pulsation mode, photometric amplitude in the $V$ band, and the reference epoch. Depending on the number of available RV measurements, we adopted different implementations of the template-fitting procedure. When four or more RV measurements are available, we adopted the free-amplitude method introduced by \citet{Braga2024}, which allows the RV amplitude to be directly fit from the data without assuming a prior luminosity amplitude. When fewer measurements are available, but at least three RV points are present, we adopted the fixed-amplitude method described by \citet{Braga2021}, which does not require prior knowledge of the reference epoch. 
The reader interested in a more quantitative analysis of the different approaches adopted to estimate the systemic velocity is referred to Table~\ref{tab:npoints}.

The RVC templates can, in principle, be applied even with a single RV measurement. However, the precision of the inferred mean RV strongly depends on both the number of measurements and their phase coverage. RV measurements spanning a limited phase interval are expected to provide less accurate mean velocities than the same number of measurements distributed over the full pulsation cycle. A more quantitative analysis of this effect is presented in Appendix~\ref{sec:Test}. For stars that did not meet the requirements for template fitting, we estimated the $V_{\gamma}$ by adopting the weighted mean of the available RV measurements. 

\citet{Prudil2024} showed that RVC templates constructed from Fe I lines are suitable for RV measurements derived from the Ca II triplet region, such as those provided by Gaia DR3. Based on this result, we adopted Fe-based RVC templates for both the Gaia DR3 radial velocities and the RVs derived from the Ca II triplet lines analyzed in this study. Throughout this paper, we refer to these measurements as RVs derived from metallic lines.

To investigate the presence of possible systematics in both the $V_{\gamma}$ measurements based on template fitting and the $V_{\gamma}$ estimates based on the weighted mean of the RV measurements, we selected two calibration samples. The former sample is based on HR spectra collected with du Pont and includes 19 RRLs (16 RRab, 3 RRc); the latter sample is based on Gaia DR3 and includes 56 RRLs (39 RRab, 17 RRc). The calibration sample was composed exclusively of RRLs with RV measurements covering the entire pulsation cycle, and the final selection was confirmed by visual inspection. We found that, on average, the difference between $V_{\gamma}$ values obtained from template fitting and those reported in the literature is smaller than 2 km s$^{-1}$ for both RRab and RRc stars. In contrast, the simple mean method exhibits systematic offsets and a larger scatter of approximately 3 km s$^{-1}$. The difference in RV amplitudes between the free-amplitude template method and the literature is, on average, smaller than 4 km s$^{-1}$ for RRab and 3 km s$^{-1}$ for RRc stars.
The reader interested in a more quantitative discussion concerning the selection criteria and the approach adopted to validate the current measurements and the estimate of the uncertainties is referred to Appendix~\ref{sec:Test}.

\begin{figure}
\center
\includegraphics[width=0.45\textwidth]{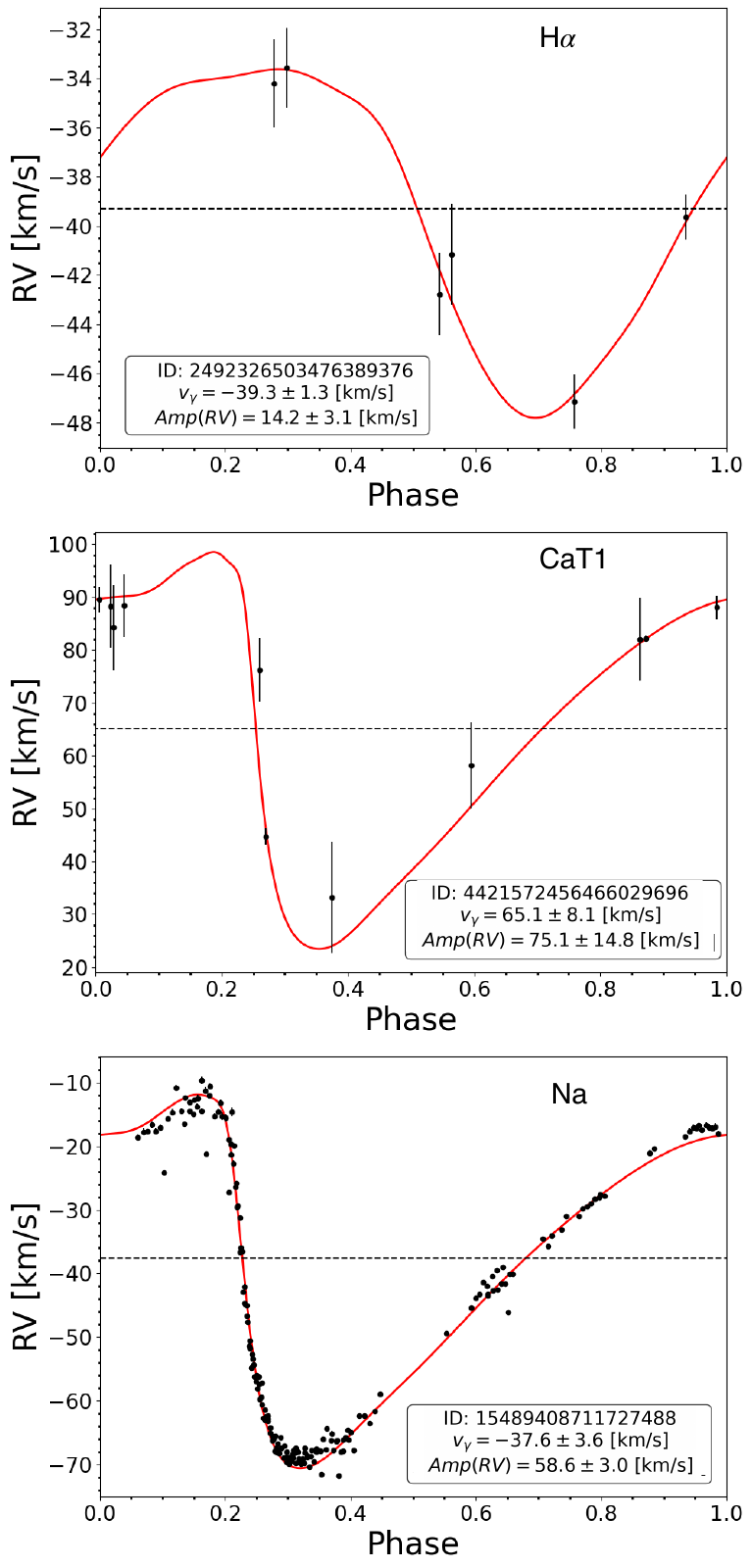}
\caption{Selected RV curves as a function of phase for three RRLs. Individual measurements are shown as black dots, and the solid red line displays the RVC template based on the \Ha{} (top), \CaTone{} (middle), and Na lines (bottom), obtained using the free-amplitude method. The dashed black line indicates the mean $V_{\gamma}$ value.}\label{fig:RVC}
\end{figure}

Fig.~\ref{fig:RVC} shows selected RV curves as a function of phase for three RRLs with different numbers of RV measurements derived from the \Ha, \CaTone, and Na lines, respectively. The RV curves in all panels were obtained using the free-amplitude method.

\subsection{Cleaning of the RV curves}

To ensure a robust modeling of the RVC templates, we applied a sigma clipping procedure designed to identify and remove outlier measurements\footnote{In this work, we define an outlier as a measurement that significantly deviates from the mean locus traced by the bulk of the data for a given variable star in its RVC; quantitatively, outliers are those lying more than $3\sigma$ from the fit template.}. The approach relies on two steps. First, we removed measurements with unusually large uncertainties by calculating the mean and the standard deviation ($\sigma_{\mathrm{err}}$) of the RV measurement errors. Only data points with RV uncertainties within a $2\sigma_{\mathrm{err}}$ threshold were retained. This step minimizes the impact of highly uncertain data on the fitting procedure. Second, we fit a preliminary RVC using a fixed-amplitude template and computed the vertical distance between each observed RV point and the fit template at the corresponding phase. The standard deviation ($\sigma_{\mathrm{res}}$) of these vertical residuals was then used to apply a second round of sigma clipping, in which we rejected all measurements deviating by more than $3\sigma_{\mathrm{res}}$ from the template fit. This conservative approach ensures that only reliable RV measurements across the pulsation cycle are used to estimate the $V_{\gamma}$ velocity.

\subsection{Estimates of barycentric radial velocity and RV amplitudes}
In the free-amplitude method, the RV amplitude is obtained directly from the best-fitting template by computing the difference between the maximum and minimum values of the fit RVC. To estimate the uncertainty on the derived amplitude ($\sigma_{\mathrm{Amp(RV)}}$), we adopted an empirical method introduced by \citet{Braga2016}. This method combines both observational errors and the scatter of the data points around the fit template in the regions that most strongly constrain the amplitude.

We identified the RV measurements lying within a range of $\pm20\%$ of Amp(RV) from the minimum and maximum of the fit template. We then computed the median observational error of these points, $\tilde{\sigma}_{\mathrm{RV}}$, and their squared residuals with respect to the fit template, given by

\begin{equation} \Delta^2 = \sum_{i=1}^{n} \left( \mathrm{RV}_i - \mathrm{RV}_{\text{template}}(\phi_i) \right)^2.
\end{equation}

where $n = n_{\text{min}} + n_{\text{max}}$ is the total number of selected points around the extrema and $\phi_i$ is the phase of the individual measurement $RV_i$. Finally, the amplitude uncertainty was estimated as

\begin{equation} \sigma_{\mathrm{Amp(RV)}} = \sqrt{ \frac{\Delta^2}{n} + \tilde{\sigma}_{\mathrm{RV}}^2 + \left( \frac{0.2\cdot \mathrm{Amp(RV)}}{\sqrt{n_{\text{min}} + 1}} \right)^2 + \left( \frac{0.2\cdot\mathrm{Amp(RV)}}{\sqrt{n_{\text{max}} + 1}} \right)^2 },
\label{eq:err_amp}
\end{equation}

This procedure yields a robust estimate of the amplitude uncertainty, even when the RVC is either sparsely sampled or the phase coverage is irregular. If no points are found within $0.2\mathrm{Amp(RV)}$ of at least one of the extrema, we adopted the standard deviation of the measurements as a representative error. In Table~\ref{table:properties}, we provide an indicator to specify the method used to derive the uncertainty: a value of 0 indicates that equation~\ref{eq:err_amp} was applied, while a value of 1 denotes that the standard deviation was adopted due to insufficient data near the minimum or maximum. Moreover, we list the basic pulsation properties and the RV amplitudes for the variables adopted in the amplitude scaling relation.

\section{Results}\label{sec:4}

\subsection{Barycentric radial velocity}

\begin{figure*}
\center
\includegraphics[width=1.\textwidth]{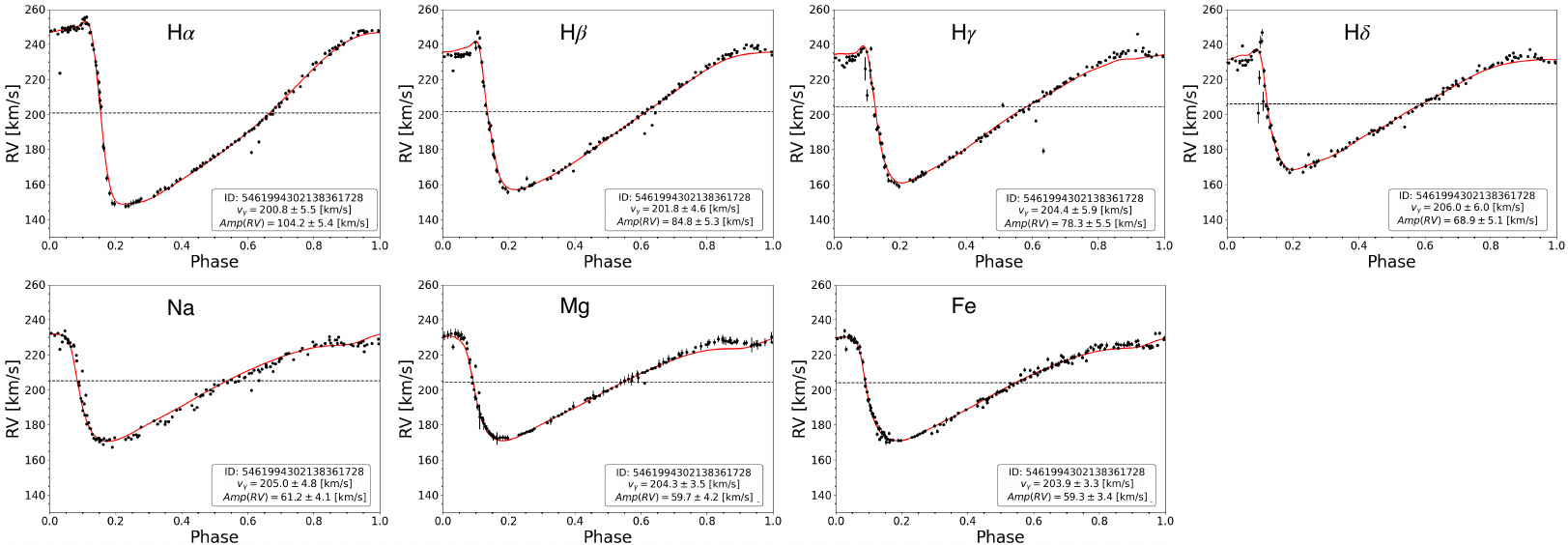}
\caption{Radial velocity curves as a function of phase for the RRab star, Gaia DR3: 5461994302138361728, based on Balmer lines (\Ha, \Hb, \Hd, \Hg) and metallic lines (Na, Mg, Fe). Measurements from the du Pont spectrographs are shown as black dots, while the solid red line shows the RVC template obtained using the free-amplitude method. The RV amplitude (Amp(RV)) and $V_{\gamma}$ are also labeled.}\label{fig:WY}
\end{figure*}

An example of the RV curves is shown in Fig.~\ref{fig:WY} for the RRab star WY Ant (Gaia DR3 source ID: 5461994302138361728). The RV measurements plotted in this figure display the expected difference between Balmer and metallic lines. The variation arises from the distinct physical conditions under which each spectral line is formed; lines formed at smaller optical depths exhibit larger RV amplitudes \citep{Preston1959,Bono2020,Braga2021}. In particular, the \Ha{} line shows the largest amplitude (Amp(RV) = 104.2 km s$^{-1}$), followed by \Hb{} (84.8 km s$^{-1}$), \Hg{} (78.3 km s$^{-1}$), and \Hd{} (68.9 km s$^{-1}$), reflecting a progressive decrease in amplitude. In contrast, the metallic lines (Na, Mg, and Fe), which form deeper in the atmosphere, exhibit significantly lower amplitudes around 59-61 km s$^{-1}$. The amplitude difference for WY Ant between the \Ha{} and Fe lines is approximately 43$\%$. Regarding the $V_{\gamma}$, a small variation is observed across the different lines. While \Ha{} yields a $V_{\gamma}$ of 200.8 $\pm$ 5.5 km s$^{-1}$, the value increases gradually to 206.0 $\pm$ 4.8 km s$^{-1}$ for \Hd{}. However, the difference is always within 1$\sigma$, indicating that the $V_{\gamma}$ is essentially the same across all lines. Moreover, the difference across the entire sample is randomly distributed between the different diagnostics.

\begin{figure*}
\center
\includegraphics[width=0.9\textwidth]{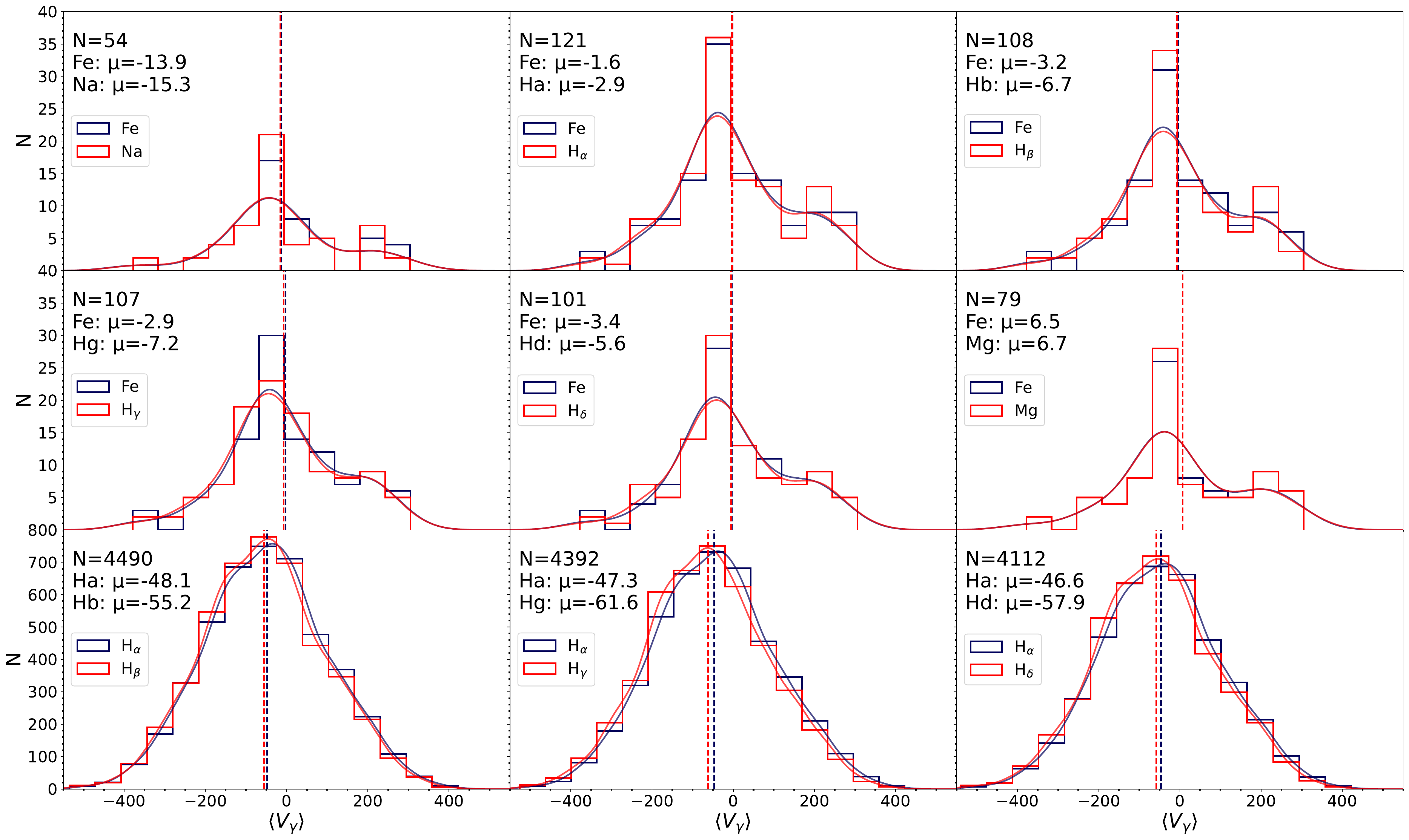}
\caption{Comparison of the $V_{\gamma}$ distributions obtained from different spectral lines using the fixed-amplitude method. Each panel shows the histogram of the $V_{\gamma}$ values for two spectral lines, with vertical dashed lines marking their respective mean values. The solid lines display the smoothed distributions. Each histogram includes only stars with $V_{\gamma}$ measurements available for both spectral lines being compared. The corresponding number of stars (N) and mean $V_{\gamma}$ are indicated in each panel.}\label{fig:GV}
\end{figure*}

Fig.~\ref{fig:GV} shows the comparison of the $V_{\gamma}$ distributions derived from two different spectral lines using the fixed-amplitude method. 
In each panel, the histograms are constructed using only the subset of stars that have reliable $V_{\gamma}$ measurements in both spectral lines under comparison; therefore, the number of stars varies between panels. The resulting distributions and mean values are very similar across the diagnostics. The differences observed arise from the distinct shapes of the RV curves, smoother for metallic lines compared to the Balmer lines.
Additionally, a secondary peak can be observed in some of the distributions. To assess the statistical relevance of this feature, we calculated the excess counts of the secondary peak relative to the local background level in the histogram and compared it to the expected Poisson fluctuations in each bin. The resulting significance values for the secondary peak range approximately between 1.0$\sigma$ and 1.8$\sigma$, depending on the specific pair of spectral lines analyzed. This evidence indicates that the secondary peak cannot be considered statistically significant.

In addition, the spectral lines were divided into three groups: Balmer lines (\Ha, \Hb, \Hd, \Hg), metallic lines (Fe, Na, Mg), and calcium lines (\CaTone{}, \CaTtwo{}). In this step, we computed both the simple mean and the weighted average for each star with two or more lines available within a given group, while the original value was retained for stars with only a single measurement. In particular, the decision to treat \CaTone{} and \CaTtwo{} as a separate group stems from the limited number of RV measurements available for these diagnostics. As a result, we were able to apply the fixed-amplitude method to 695 stars and the free-amplitude method to 314 stars, whereas for 2,077 stars we computed a weighted average. Owing to these differences, we analyzed this group separately to allow for a more careful comparison.

Finally, the adopted value of $V_{\gamma}$ for each variable was determined as the global mean of the available measurements. Note that we applied a filtering procedure based on the individual $V_{\gamma}$ uncertainties. 
In particular, we neglected measurements with uncertainties larger than 50 km s$^{-1}$ and recursively removed outliers that deviate by more than $2\sigma$ from the mean. This approach ensures a robust estimate of $V_{\gamma}$ and minimizes the impact of spurious measurements on the mean. After applying the quality criteria to our measurements and combining them with values from the literature, the final catalog comprises 15,955 RRL stars, including 11,270 RRab, 4,489 RRc, and 196 RRd variables. The set of $V_{\gamma}$ values and their standard deviations are listed in Table~\ref{table:Vgammafinal}.

Values exceeding an absolute velocity of 500 km s$^{-1}$ are considered indicative of high-velocity RRL candidates. Such objects have been previously detected in the Galactic bulge and appear to be gravitationally bound and kinematically associated with the Milky Way halo, despite currently passing through the bulge. These stars therefore provide valuable insights into the high-velocity component of the ancient stellar population in the Galactic center \citep{kunder2015, kunder2020, prudil2022,Fernandez2025}. In total, we identified 43 high-velocity RRL candidates, which will be discussed in detail in a forthcoming paper (Baeza-Villagra et al., in prep.).

\begin{table}[ht!]
\caption{Systemic velocities ($V_{\gamma}$) of the RRL  stars in our sample.}
\label{table:Vgammafinal}
\centering
\footnotesize
\begin{tabular}{r l r@{$\,\pm\,$}r c}
\hline\hline
ID\tablefootmark{a} &
Type &
\multicolumn{2}{c}{$V_{\gamma}$} &
flag\_work\tablefootmark{b} \\
& &
\multicolumn{2}{c}{(km s$^{-1}$)} &
\\
\hline
4006995996571691008 & RRab &  89.4  & 0.2  & 0 \\
828629693980693248  & RRab & -25.1  & 10.8 & 0 \\
2678622049970867200 & RRc  & -114.0 & 3.6  & 0 \\
1432938460184080640 & RRab & -288.2 & 4.0  & 0 \\
368963499186050432  & RRab & -179.2 & 7.2  & 0 \\
794010127272856192  & RRc  & -141.1 & 2.6  & 0 \\
2745376772444159488 & RRab & -237.2 & 2.4  & 0 \\
2738382847698520960 & RRab & -114.6 & 1.4  & 0 \\
3724035644076002432 & RRc  & -13.8  & 2.4  & 0 \\
3725550049544011648 & RRc  & 43.0   & 1.7  & 0 \\
\hline
\end{tabular}
\tablefoot{This table is available entirely at the CDS. A portion is provided here as an example to illustrate its structure and content.\\
\tablefoottext{a}{Gaia DR3 identification number.}\\
\tablefoottext{b}{A value of
0 indicates a measurement obtained in this work, whereas a value of 1 indicates a value adopted from the literature.}}
\end{table}

\subsection{RV amplitudes}\label{sec:rv}
New empirical relations between RV and photometric amplitudes are quite relevant in the study of field RRLs. However, light curves not only vastly outnumber RVCs, but also contain orders of magnitude more phase points. Such relations provide an estimate of the RV amplitude by using photometric data. This approach not only enhances the scientific exploitation of extensive photometric surveys such as Gaia, but also mitigates the limitations associated with the high observational cost of spectroscopic measurements.

Based on the photometric amplitudes from the Gaia catalog (Amp($G_{BP}$), Amp($G$), Amp($G_{RP}$)) and Amp($V$), we established empirical scaling relations linking these quantities to the RV amplitudes derived in this study using the free-amplitude method. Previous studies, such as \citet{Sesar2012} and \citet{Braga2021}, used a linear scaling relation between the photometric and RV amplitudes. However, \citet{Prudil2024} found a quadratic relation for RRab stars and a linear one for RRc, consistent with \citet{Jurcsik2017}, who noted that the RRab relation is not linear. To enable a direct comparison with previous studies, we selected the RV amplitudes derived from free-amplitude templates based on Fe line, considering only stars with amplitude values less than or equal to 120 km s$^{-1}$ for RRab-type stars and 50 km s$^{-1}$ for RRc-type stars. Based on this selection, we obtained a subsample of 902 RRab and 164 RRc stars. We explored different analytical fitting functions while explicitly accounting for uncertainties on both axes. We then employed a Bayesian inference framework to estimate the fit parameters and their associated uncertainties. For each case, we computed the reduced chi-squared, the Akaike information criterion (AIC), the Bayesian information criterion (BIC), and the root-mean-square error (RMSE) to quantitatively evaluate the quality of the fits. According to these diagnostics, a linear fit is preferred for RRc stars, whereas for RRab stars the logarithmic fit provides the most adequate representation of the data, consistent with the trend reported by \citet{Prudil2024}. 

\begin{figure*}
\center
\includegraphics[width=0.95\textwidth]{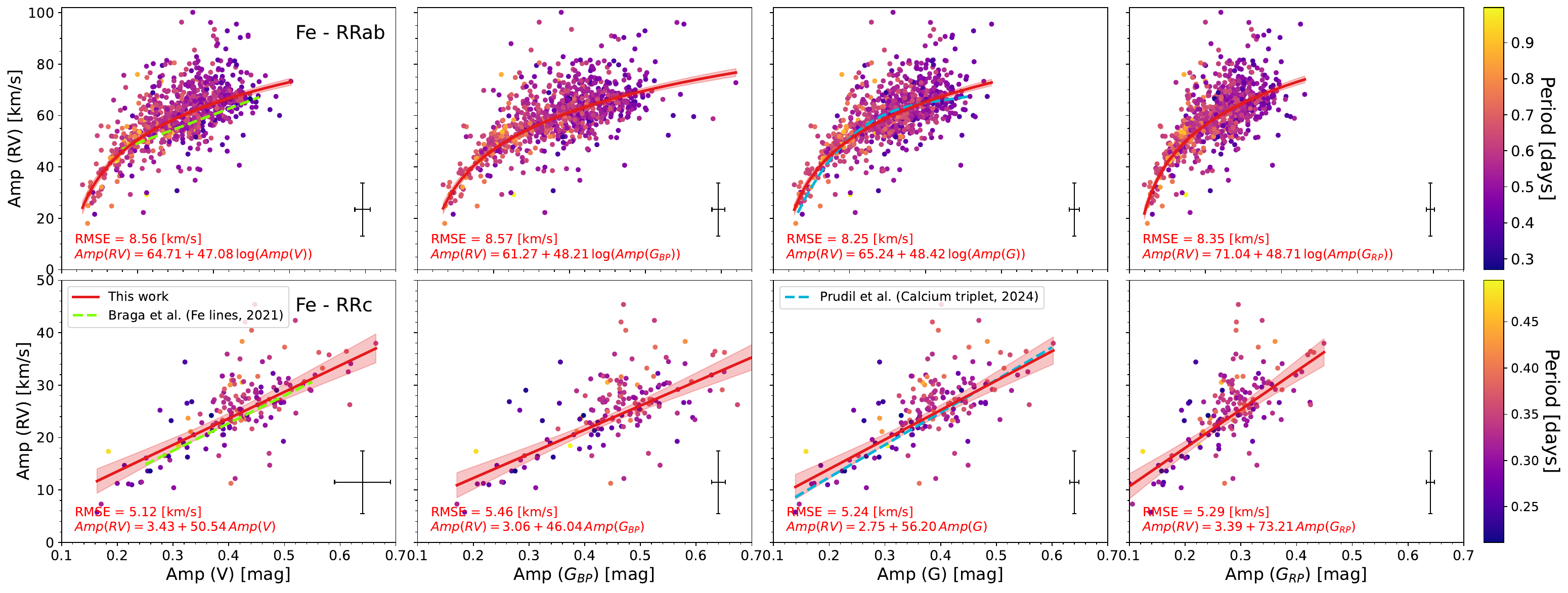}
\caption{Relations between the photometric amplitudes, Amp($V$), Amp($G_{BP}$), Amp($G$), and Amp($G_{RP}$), and the RV amplitudes derived using Fe templates. The upper panels show RRab stars, while the lower panels show RRc variables. Both RRL subclasses are color-coded by pulsation period. The best-fit relations are shown as solid red lines, and the shaded red regions represent the 95$\%$ confidence intervals of the fits, reflecting the uncertainties in the fit parameters. The dashed green and cyan lines correspond to the relations from \citet{Braga2021} and \citet{Prudil2024}, respectively. The analytical form of each fit relation and the corresponding RMSE value are displayed at the bottom of each panel, while the mean uncertainties are indicated in the bottom-right corner.}\label{fig:Amp_gaia}
\end{figure*}

Fig.~\ref{fig:Amp_gaia} shows the scaling relations for both RRL subclasses, which can be expressed by the following equations 

\begin{align}
\text{RRab :}\quad \mathrm{Amp(RV)} &= a + b\,\log(\mathrm{Amp(x)}), \\
\text{RRc :}\quad \mathrm{Amp(RV)} &= a + b\,\mathrm{Amp(x)},
\end{align}
where $a$ and $b$ are the fit coefficients (see Table~\ref{table:coeff2}), and $x$ corresponds to the photometric amplitudes in the $V$, $G_{BP}$, $G$, and $G_{RP}$ bands. 

We note a break in the relation for first-overtone pulsators occurring at approximately $\sim$0.4 mag (corresponding to $\sim$46 km s$^{-1}$ in the $V$ band, $\sim$42 km s$^{-1}$ in the $G_{BP}$ band, and $\sim$47 km s$^{-1}$ in the $G$ band). For the $G_{RP}$ band, the break occurs at a lower amplitude, around $\sim$0.3 mag (corresponding to $\sim$46 km s$^{-1}$).
The nonlinearity of the scaling relations was previously discussed by \citet{Prudil2024}, who concluded that mass, luminosity, and metallicity alone cannot explain the observed trend of photometric and RV amplitudes. In particular, they suggested that variation in the efficiency of convective transport could explain the observed bending. To investigate on a more quantitative basis this working hypothesis, Fig.~\ref{fig:Amp_gaia} shows the amplitude ratios color-coded by pulsation period. The data plotted in this figure show that the bending in the amplitude ratios, is mainly driven by long-period RRab stars. This circumstantial evidence indicates that the bending is mainly caused by the different trends of luminosity and RV amplitudes as a function of the pulsation period. Optical luminosity amplitudes decrease steadily with increasing pulsation period, declining by a factor of five to seven from the blue (hotter) to the red (cooler) edge of the fundamental instability region. The RV amplitudes display a similar trend, but they only decrease by a factor of four when moving from the blue to the red edge of the instability strip. This finding supports the suggestion by \citet{Prudil2024} that the increased efficiency in the convective transport plays a key role in the bending of the amplitude ratios. However, evolutionary effects can also affect the trend and could contribute to the observed spread at a fixed range in periods. The bending is not present in RRc variables, because they cover a narrower range in period and they are also systematically hotter when compared with RRab variables (convective transport is not very efficient).

After deriving the relations, we examined the RV amplitudes derived from the Balmer lines. In this case, in order to select a sub-sample of stars with RV measurements covering the full pulsation cycle and for which the template fitting provides reliable amplitude estimates, we restricted the analysis to RRab stars with more than nine RV measurements and RRc stars with more than four. Note that a higher threshold was adopted for RRab stars due to their more complex RV light curve morphology, whereas RRc stars typically exhibit near-sinusoidal shape. Finally, we neglected 1195 stars from the SDSS catalog to derive the relations, since they show a larger scatter. After applying these selection criteria and complementing them with a visual inspection of each setting, the resulting subsample includes the following number of stars per spectral line: 90 RRab and 61 RRc for \Ha, 80 RRab and 53 RRc for \Hb, 61 RRab and 44 RRc for \Hg, and 34 RRab and 14 RRc for \Hd. 

\begin{figure*}
\center
\includegraphics[width=0.92\textwidth]{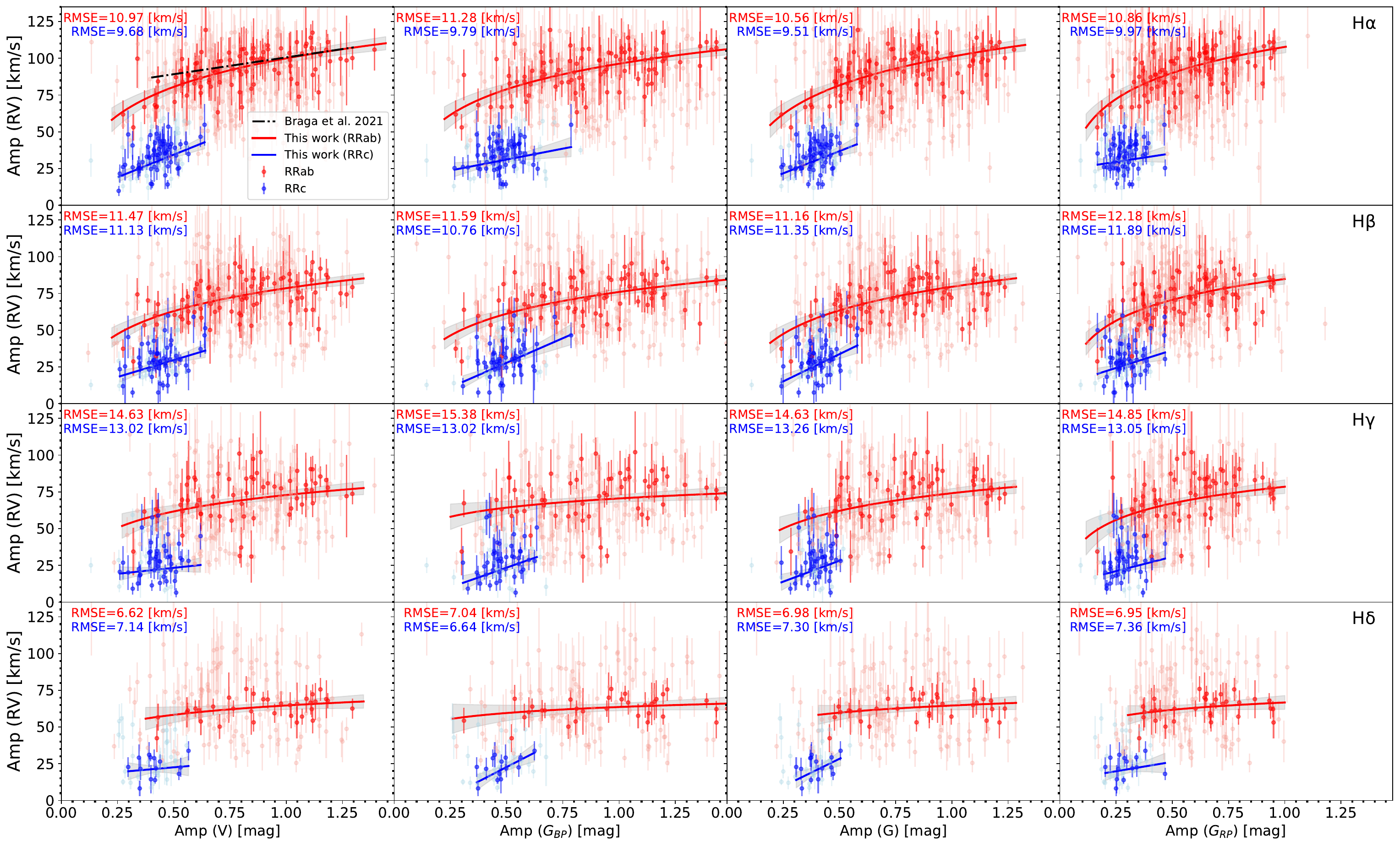}
\caption{Relations between the photometric amplitudes (V, $G_{BP}$, G, and $G_{RP}$) and the RV amplitudes derived from the Balmer lines. From top to bottom, the panels show the relations based on the \Ha{}, \Hb{}, \Hg{}, and \Hd{} lines, respectively. The selected RRab and RRc variables are marked with red and blue symbols, respectively, with lighter tones representing the full sample. The solid red and blue lines indicate the best-fit relations obtained in this work, while the shaded gray region represents the 95$\%$ confidence interval of the fit. Additionally, the dashed black line corresponds to the linear relations from \citet{Braga2021}. }\label{fig:Amp_balmer}
\end{figure*}

Fig.~\ref{fig:Amp_balmer} shows the final results obtained from these sub-samples and the coefficients of the fit are listed in Table~\ref{table:coeff2}. This figure shows that, in contrast to previous studies, Balmer lines follow a logarithmic rather than a linear relation for RRab stars, whereas RRc stars follow a linear trend. However, it is more difficult to identify a clear and well-defined break in the relations for RRab variables. This is likely due to the smaller sample size and the larger dispersion of the Balmer-line data compared to the metallic lines. 
Although approximate break locations can be suggested, these should be interpreted with caution. For \Ha, the transition appears to occur around $\sim$0.68 mag (90 km s$^{-1}$ for the V band, 89 km s$^{-1}$ for the $G_{BP}$ band, and 91 km s$^{-1}$ for the G band), while for the $G_{RP}$ band it occurs at approximately $\sim$0.4 mag (87 km s$^{-1}$). Similarly, for \Hb, a transition can be tentatively identified at $\sim$0.68 mag (70 km s$^{-1}$ for the V band, 69 km s$^{-1}$ for the $G_{BP}$ band, and 71 km s$^{-1}$ for the G band), and at $\sim$0.4 mag (67 km s$^{-1}$) for the $G_{RP}$ band.

\begin{figure*}
\center
\includegraphics[width=0.92\textwidth]{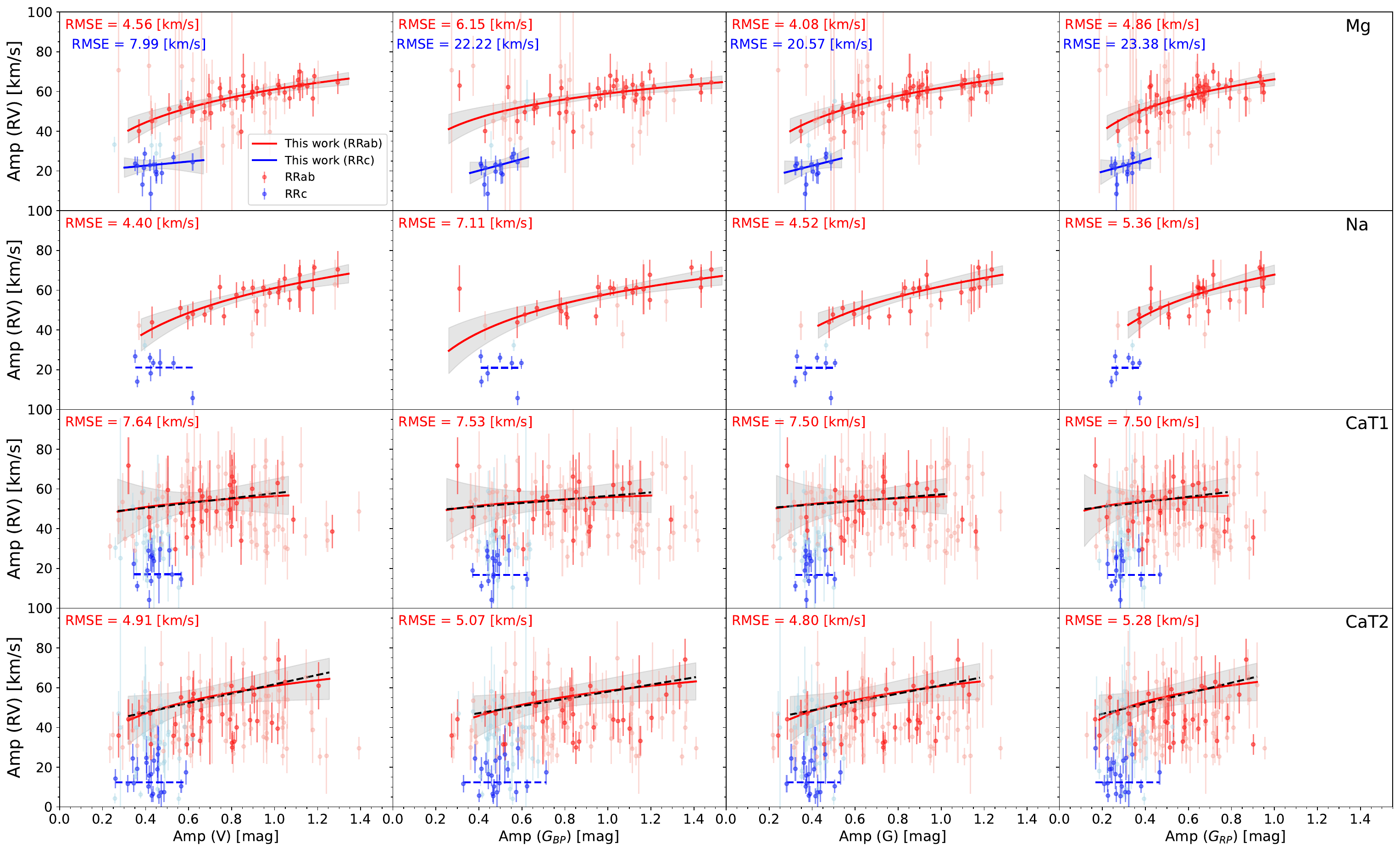}
\caption{Same as Fig.~\ref{fig:Amp_balmer}, but showing the relation based on RV amplitudes derived from the Mg, Na, \CaTone, and \CaTtwo{} lines. Additionally, the dashed black lines correspond to the linear relations for RRab and the mean for RRc.}\label{fig:Amp_metallic}
\end{figure*}

Finally, we checked the behavior of the RV amplitudes from Mg, Na, \CaTone{}, and \CaTtwo{} lines. Following the same selection criteria adopted for the Balmer lines, we selected a sub-sample consisting of 39 RRab and 13 RRc stars for the Mg line, 24 RRab and 8 RRc for Na, 27 RRab and 15 RRc for \CaTone, and 34 RRab and 21 RRc for \CaTtwo. The RV amplitudes were fit using a logarithmic relation for RRab stars and a linear relation for RRc stars (see Fig.~\ref{fig:Amp_metallic}). In particular, for the Na, \CaTone{}, and \CaTtwo{} lines, we used the mean for RRc stars, motivated by the large scatter in the data. Additionally, we performed a linear fit for the amplitude ratios of RRab variables based on \CaTone{} and \CaTtwo{} lines. The coefficients of the fit are listed in Table~\ref{table:coeff2}. In passing, we note that the main difference between the analytical fits based on the Balmer or metallic lines and those based on Fe templates lies in the sample size. In the former data sets, the number of very low-amplitude RRLs is quite limited.

\begin{table}
\caption{Coefficients of the Amp(RV)–Amp(V, $G_{BP}$, $G$, $G_{RP}$) relations for RRab and RRc stars. }
\label{table:coeff2}
\centering
\footnotesize
\resizebox{\columnwidth}{!}{
\begin{tabular}{ccccc|ccc}
\toprule
\multicolumn{5}{c|}{RRab [$y = a + b\log(x)$]} &
\multicolumn{3}{c}{RRc [$y = a + bx$]} \\
\cmidrule(r){1-5} \cmidrule(l){6-8}
Line & Band & $a \pm \epsilon_a$ & $b \pm \epsilon_b$ & RMSE  
& $a \pm \epsilon_a$ & $b \pm \epsilon_b$& RMSE  \\
& & & & (km s$^{-1}$) & & &  (km s$^{-1}$)\\
\midrule
Fe &$V$ & 64.71$\pm$ 0.45 & 47.08 $\pm$ 1.72 &8.56& 3.43 $\pm$ 1.91 & 50.54 $\pm$ 4.75 &5.12\\
&$G_{BP}$ & 61.27 $\pm$ 0.38 & 48.21 $\pm$ 1.57& 8.57& 3.06 $\pm$ 1.92 & 46.04 $\pm$ 4.32&5.46 \\
&$G$ & 65.24 $\pm$ 0.45 & 48.42 $\pm$ 1.54 &8.25& 2.75 $\pm$ 1.88 & 56.20 $\pm$ 5.11&5.24 \\
&$G_{RP}$ & 71.04 $\pm$ 0.60 & 48.71 $\pm$ 1.56 & 8.35&3.39 $\pm$ 1.87 & 73.21 $\pm$ 6.82&5.29 \\
\midrule
\Ha & $V$        & $99.97 \pm 1.37$  & $64.48 \pm 7.45$ &10.97 & $3.70 \pm 4.20$   & $61.48 \pm 9.84$&9.68 \\
    & $G_{BP}$   & $96.33 \pm 1.18$  & $57.85 \pm 7.01$  &11.28 &$16.26 \pm 5.22$  & $29.69 \pm 10.73$&9.79 \\
    & $G$        & $101.16 \pm 1.45$ & $64.93 \pm 7.26$  &10.56& $6.64 \pm 5.65$   & $60.37 \pm 14.12$&9.51 \\
    & $G_{RP}$   & $107.75 \pm 2.05$ & $58.80 \pm 6.78$  &10.86& $23.71 \pm 4.49$  & $23.31 \pm 15.06$&9.97 \\
\midrule
\Hb & $V$        & $78.59 \pm 1.20$  & $51.81 \pm 6.13$  &11.47& $6.63 \pm 4.77$   & $46.09 \pm 10.91$&11.13\\
    & $G_{BP}$   & $76.03 \pm 1.06$  & $49.23 \pm 5.88$ & 11.59& $-5.61 \pm 5.45$  & $66.89 \pm 11.22$&10.76 \\
    & $G$        & $79.55 \pm 1.27$  & $53.25 \pm 6.06$ & 11.16& $-3.37 \pm 5.87$  & $74.18 \pm 14.42$ &11.35\\
    & $G_{RP}$   & $84.86 \pm 1.80$  & $47.16 \pm 5.73$ &12.18 & $12.26 \pm 4.23$  & $48.30 \pm 14.05$ &11.89\\
\midrule
\Hg & $V$        & $72.86 \pm 1.44$  & $37.03 \pm 8.79$  &14.63&$15.19 \pm 5.57$  & $16.24 \pm 12.69$&13.02\\
    & $G_{BP}$   & $70.54 \pm 1.28$  & $20.75 \pm 7.94$  &15.38& $-3.29 \pm 6.32$  & $53.48 \pm 12.81$&13.02  \\
    & $G$        & $74.12 \pm 1.58$  & $39.81 \pm 8.64$  &14.63 &$-0.34 \pm 6.63$  & $57.03 \pm 16.17$&13.26  \\
    & $G_{RP}$   & $78.51 \pm 2.37$  & $37.56 \pm 8.35$  &14.85& $11.32 \pm 4.69$  & $39.02 \pm 15.62$&13.05  \\
\midrule
\Hd & $V$        & $64.67 \pm 1.41$  & $21.07 \pm 10.30$ &6.62 &$15.94 \pm 9.19$  & $13.28 \pm 21.60$&7.14 \\
    & $G_{BP}$   & $63.63 \pm 1.28$  & $13.60 \pm 8.82$  &7.04& $-16.62 \pm 10.02$& $78.78 \pm 20.52$&6.64 \\
    & $G$        & $64.64 \pm 1.52$  & $16.18 \pm 10.00$ &6.98& $-9.09 \pm 9.89$  & $74.95 \pm 24.00$&7.30 \\
    & $G_{RP}$   & $66.70 \pm 2.40$  & $16.51 \pm 9.77$  &6.95 &$13.52 \pm 6.57$  & $25.56 \pm 20.53$&7.36 \\
\midrule
Mg & $V$        & $61.13 \pm 0.91$  & $42.93 \pm 6.86$  &4.56& $18.34 \pm 7.07$  & $10.86 \pm 15.76$&7.99 \\
   & $G_{BP}$    & $58.98 \pm 0.85$  & $30.03 \pm 6.34$  &6.15& $3.12 \pm 10.00$  & $40.22 \pm 19.95$&22.22 \\
   & $G$         & $61.97 \pm 1.02$  & $42.91 \pm 6.71$  &4.08& $7.42 \pm 9.18$   & $37.95 \pm 21.97$&20.57 \\
   & $G_{RP}$    & $66.43 \pm 1.63$  & $39.02 \pm 6.68$  &4.86&$7.85 \pm 9.52$   & $49.09 \pm 30.34$&23.38 \\
\midrule
Na & $V$        & $61.11 \pm 1.26$  & $56.51 \pm 10.35$ &4.40& $\dotsc$          & $\dotsc$ & $\dotsc$\\
   & $G_{BP}$    & $58.11 \pm 1.16$  & $48.41 \pm 9.52$  &7.11& $\dotsc$          & $\dotsc$& $\dotsc$ \\
   & $G$         & $62.04 \pm 1.43$  & $53.79 \pm 9.64$  &4.52& $\dotsc$          & $\dotsc$ & $\dotsc$\\
   & $G_{RP}$    & $68.06 \pm 2.33$  & $51.75 \pm 9.70$  &5.36& $\dotsc$          & $\dotsc$& $\dotsc$ \\
\midrule
\CaTone & $V$     & $56.48 \pm 4.50$  & $14.46 \pm 21.30$ &7.64 &$\dotsc$          & $\dotsc$ & $\dotsc$\\
        & $G_{BP}$& $55.87 \pm 3.39$  & $10.92 \pm 16.98$ &7.53& $\dotsc$          & $\dotsc$& $\dotsc$ \\
        & $G$     & $56.26 \pm 4.49$  & $9.65 \pm 19.08$  &7.50& $\dotsc$          & $\dotsc$& $\dotsc$ \\
        & $G_{RP}$& $57.71 \pm 6.11$  & $9.47 \pm 15.81$  &7.50&$\dotsc$          & $\dotsc$& $\dotsc$ \\
\midrule
\CaTtwo & $V$     & $61.26 \pm 3.92$  & $36.74 \pm 17.09$ &4.91& $\dotsc$          & $\dotsc$& $\dotsc$ \\
        & $G_{BP}$& $58.54 \pm 3.11$  & $32.57 \pm 15.52$ &5.07 &$\dotsc$          & $\dotsc$ & $\dotsc$\\
        & $G$     & $61.07 \pm 3.83$  & $32.91 \pm 15.13$ &4.80 &$\dotsc$          & $\dotsc$& $\dotsc$ \\
        & $G_{RP}$& $64.45 \pm 5.37$  & $29.09 \pm 14.35$ &5.28 &$\dotsc$          & $\dotsc$ & $\dotsc$\\
\bottomrule
\end{tabular}}
\tablefoot{A logarithmic fit was adopted for RRab stars, whereas a linear fit was used for RRc stars. The columns display the coefficients (a, b) and their uncertainties for the RRab (left) and RRc (right) variables, respectively. The mean values do not have  coefficients, since they are the average of the RV amplitude of the data.}
\end{table}

\subsection{The spectroscopic Bailey diagram}\label{sec:BD}
The Bailey diagram is a solid diagnostic to investigate the pulsation properties of radial variables because it is distance- and reddening-independent. This evidence is even more compelling for RRLs, because the Bailey diagram can be safely adopted not only for mode identification, but also for constraining the impact that temperature and radius variations have on the morphology of light and RVC \citep{Bono2020}. 

\begin{figure}
\center
\includegraphics[width=0.45\textwidth]{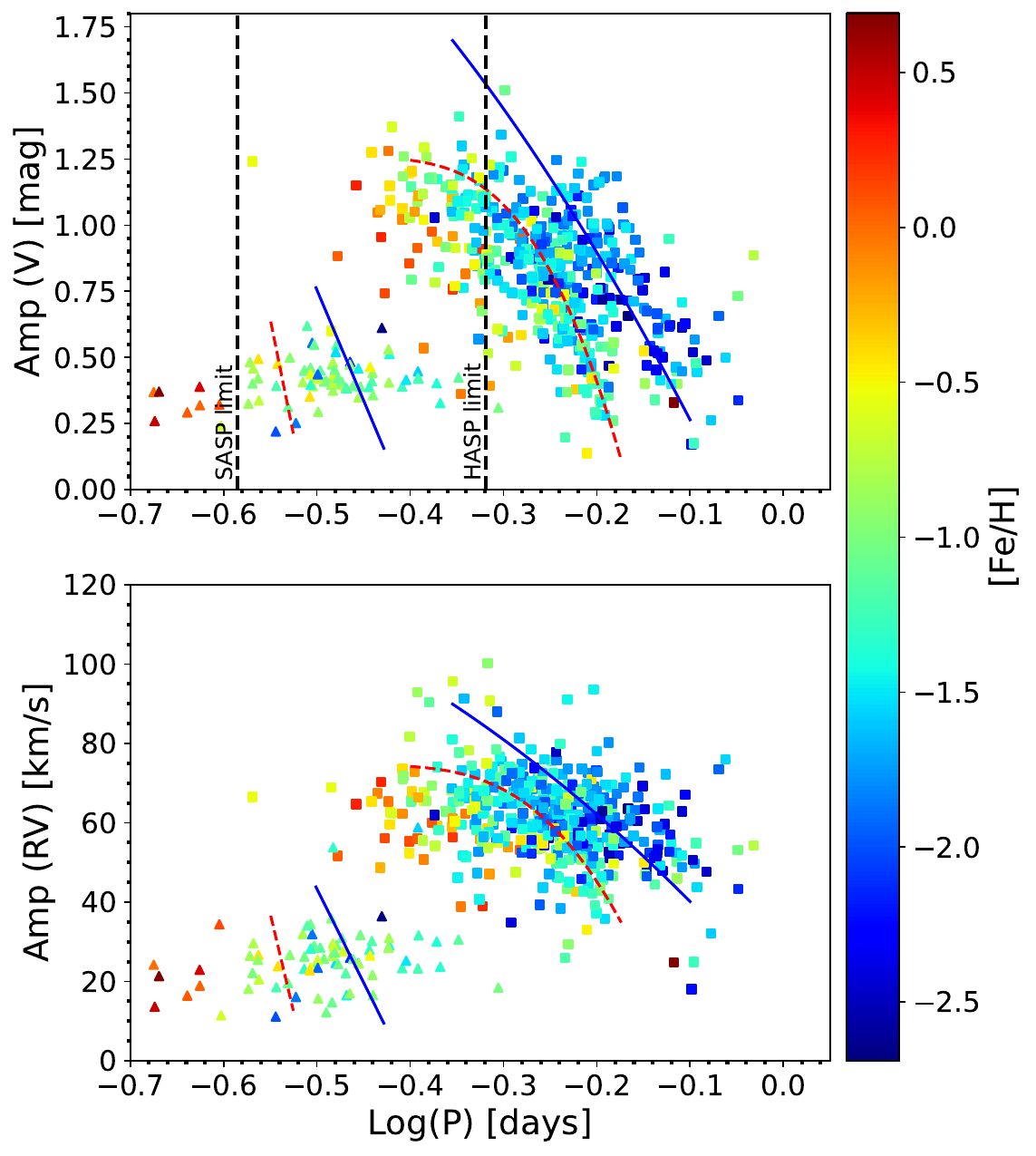}
\caption{Top: Bailey diagram based on the photometric amplitudes in the \textit{V}-band as a function of the logarithmic period. The RRab and RRc variables are marked with squares and triangles, respectively. Metallicity is color-coded, and the color bar is shown on the right. The lines show the analytical relations from \citet{fabrizio2021}, tracing more metal-rich (dashed red lines) and more metal-poor (solid blue lines) overdensities for both RRab and RRc variables. The vertical dashed lines indicate the lower limits in period for HASP and SASP variables. Bottom: Same as the top panel, but based on the RV amplitudes derived from Fe templates (free-amplitude method).}\label{fig:Amp_fe}
\end{figure}

Fig.~\ref{fig:Amp_fe} shows the Bailey diagram of photometric and RV amplitudes as a function of logarithmic period. The RV measurements are based on Fe line, with RV amplitudes estimated using free-amplitude templates. Note that the symbols are color-coded and move from dark blue (very metal-poor) to dark red (very metal-rich) RRLs. The individual metallicities were adopted from SR3C. To properly identify the short-period (SP) and the long-period (LP) sequences between the RRab and RRc variables, we used the analytical relations provided by \citet{fabrizio2021}. 

The separation between short- and long-period RRLs closely resembles the classical Oosterhoff dichotomy, with Oosterhoff I (Oo I) variables being, on average, more metal-rich and characterized by shorter periods compared to Oosterhoff II (Oo II) variables. Empirical evidence indicates that both periods and luminosity amplitudes decrease steadily from very metal-poor to very metal-rich RRLs. \citet{Fabrizio2019,fabrizio2021} offered an explanation for the origin of the Oosterhoff dichotomy, linking it to an anticorrelation between the location of RRLs in the Bailey diagram and their metallicity, identified for both RRab and RRc types. They found that higher metallicities correspond to progressively shorter pulsation periods and lower visual amplitudes. According to their interpretation, the scarcity of Galactic globular clusters with intermediate metallicities that contain substantial numbers of RRL stars could account for this dichotomy. This interpretation has been recently reinforced by \citet{Medina2025}, who reported similar evidence based on a large spectroscopic dataset collected by the DESI survey. 

To further constrain the real nature of this phenomenon we investigated the variation in the Bailey  diagram in terms of RV amplitudes by transforming the analytical relations provided by \citet{fabrizio2021} into the RV-amplitude domain. The resulting relations for the SP (OoI) and LP (OoII) sequences are therefore not new fits, they are directly obtained from those of \citet{fabrizio2021} after applying the transformation. The resulting relations describing the SP and LP sequences for RRab variables are the following 

\begin{align}
\text{SP:}\quad \mathrm{Amp(RV)} &=
-90.41 - 1027.51\,\log(P) - 1816.71\,\log(P)^2 \notag \\
&\quad + 1730.33\,\log(P)^4, \\
\text{LP:}\quad \mathrm{Amp(RV)} &=
15.05 - 266.76\,\log(P) - 156.79\,\log(P)^2,
\end{align}

where P is the pulsation period. Following the same approach we derived the relations for RRc variables as 

\begin{align}
\text{SP:}\quad \mathrm{Amp(RV)} &=
-506.59 - 988.49\,\log(P), \\
\text{LP:}\quad \mathrm{Amp(RV)} &=
-192.74 - 472.19\,\log(P).
\end{align}
The data plotted in the top panel of Fig.~\ref{fig:Amp_fe} show that the distribution of RRLs is clearly inhomogeneous. Most RRab stars are aligned along the SP sequence, accounting for approximately $70\%$ of the sample, while the LP sequence hosts the remaining fraction. However, we note that a precise separation between the two sequences is challenging in the overlapping region. Although \citet{fabrizio2021} reported that the RRc variables tend to exhibit the opposite trend with respect to RRab stars, favoring the LP sequence over the SP sequence, the present analysis, based on a more limited sample of RRc stars, yields comparable proportions, with approximately half of the objects associated with each sequence.

The two vertical dashed lines show the two lower limits for the high-amplitude and short-period RRab (HASP, $\log$P$\sim$-0.319) and the small-amplitude and short-period RRc (SASP, $\log$P$\sim$-0.585). These limits \citep{Fiorentino2015, fiorentino2022} indicate the regions forbidden to RRL stars in dwarf galaxies with masses smaller than Fornax dwarf spheroidal. The strong correlation between the presence of short period, metal-rich variables is very clear and suggests that classical dwarf galaxies played a minor role in the early formation of the main Galactic components (halo, disks).

The bottom panel of Fig.~\ref{fig:Amp_fe} displays the Bailey diagram based on RV amplitudes of the same RRLs plotted in the top panel. The distribution of the RRLs in this plane is quite different since the fractional variation in RV is roughly a factor of two smaller when compared with the variation in the \textit{V}-band amplitude. The difference is expected, since RV variations mainly trace radius variations along the pulsation cycle while $V$-band luminosity variations trace both temperature and radius variation \citep{Bono2020}. Based on the results obtained, two interesting features are worth discussing in more detail:

\begin{enumerate}
    \item The RV amplitudes display variations as a function of the metal content similar to the luminosity amplitudes. Indeed, they show a similar ranking between SP and LP sequences, the so-called Oosterhoff groups;

    \item RV amplitudes show, together with luminosity amplitudes, overdensities that are associated with the chemical enrichment history of field RRLs \citep{Bono2025}, but both of them display a smooth variation as a function of the metal content \citet{Fabrizio2019,fabrizio2021}. 
\end{enumerate}

\begin{figure*}
\center
\includegraphics[width=0.9\textwidth]{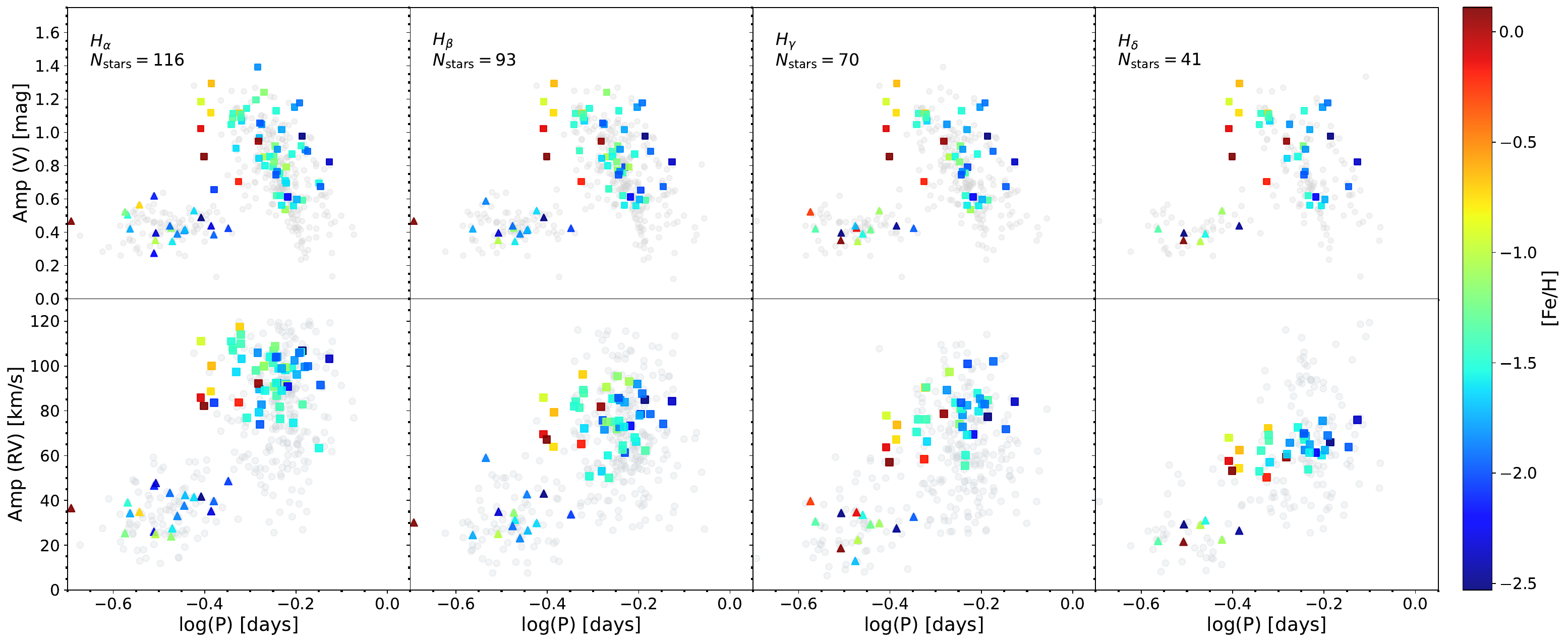}
\caption{Same as Fig.~\ref{fig:Amp_fe}, but using RV amplitudes from the Balmer lines (H$\alpha$–H$\delta$) for the RRab sample. The selected RRab variables are color-coded, while the light gray dots show the full sample.}\label{fig:Amp_balmer_new}
\end{figure*}

We repeated the entire analysis using the Balmer diagnostics. For consistency, we adopted the same RRab subsample used in Fig.~\ref{fig:Amp_balmer}. The resulting spectroscopic Bailey diagrams, shown in Fig.~\ref{fig:Amp_balmer_new}, reveal that all Balmer tracers reproduce the qualitative trends obtained from the Fe templates and that the separation between the short- and long-period sequences is preserved. This confirms that the observed behavior is not line-dependent and supports the robustness of the trend across different spectral diagnostics.

The opportunity to deal with a large and homogeneous spectroscopic dataset allowed us to investigate the distribution of candidate Blazhko stars in the Bailey diagram. Candidate Blazhko RRLs were identified by cross-matching the current spectroscopic RRL sample with recent compilations of Blazhko candidates \citep{Skarka2013, Netzel2018, Skarka2020, Donev2025}. Overall, we found 333 variables in common, including 287 RRab and 46 RRc. The objects in common with our spectroscopic sample are limited to \citet{Skarka2013} and \citet{Donev2025}, with 226 and 108 matches, respectively, while no additional matches were found with the remaining two samples.

We adopted the amplitude of the current fit of the RVC for candidate Blazhko stars, without attempting to model their modulation. We note that no specific templates for Blazhko RRLs are currently available. Therefore, our estimates should be regarded as a first-order approximation of the global amplitude.

\begin{figure}
\center
\includegraphics[width=0.45\textwidth]{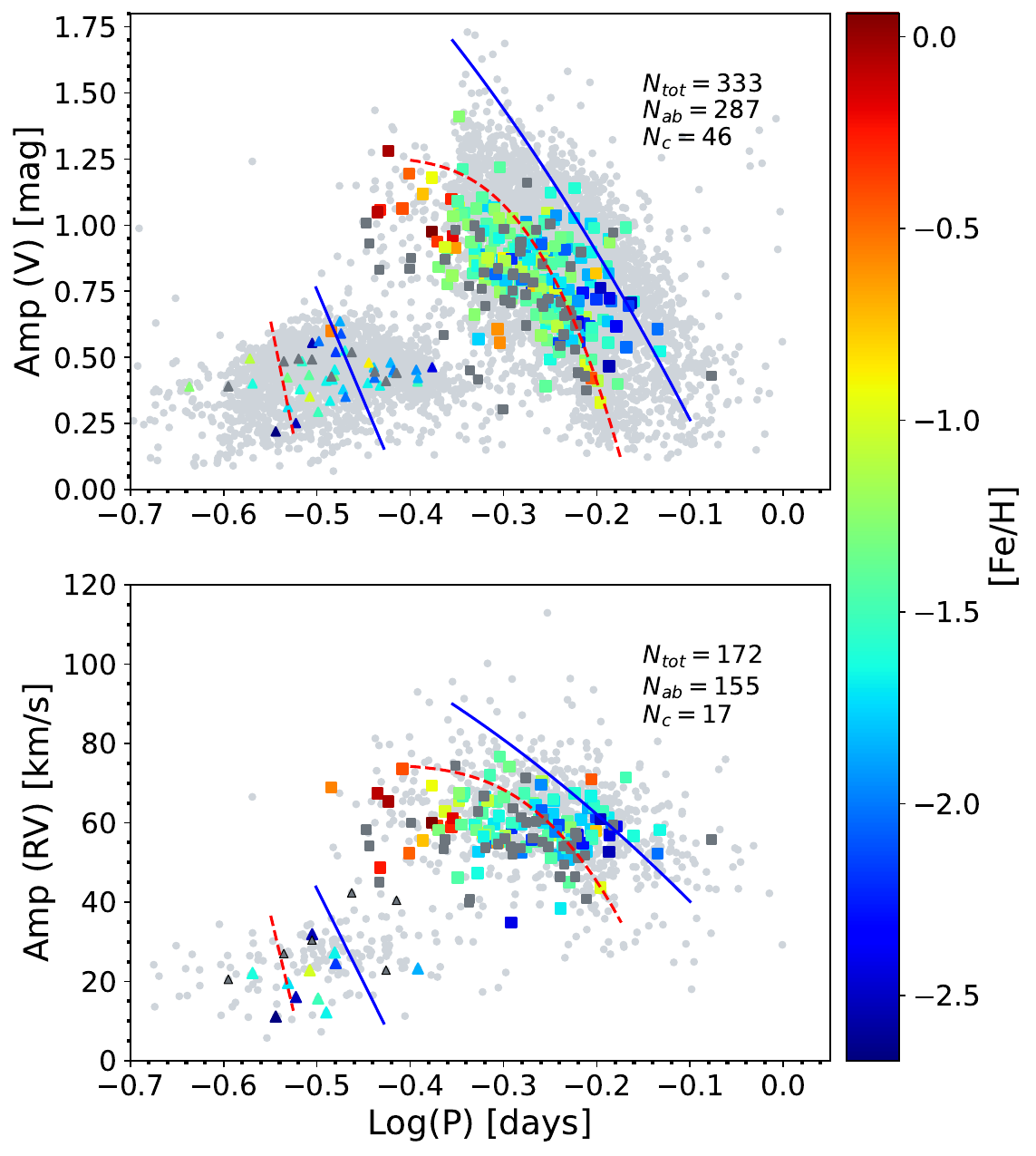}
\caption{Same as Fig.~\ref{fig:Amp_fe}, but showing the candidate Blazhko stars for both RRab (square symbols) and RRc (triangle symbols). The light gray markers represent the full sample of RRL stars. Blazhko RRLs with available iron abundances are plotted with colored symbols, while those without metallicity measurements are shown in dark gray. The number of candidate Blazhko RRab and RRc stars are labeled.}\label{fig:Amp_fe_blazhko}
\end{figure}

The top panel of Fig.~\ref{fig:Amp_fe_blazhko} shows the cross-match of candidate Blazhko stars in the photometric 
Bailey diagram, based on the $V$-band luminosity amplitudes, while the bottom panel shows the cross-match in the spectroscopic Bailey diagram, based on RV amplitudes measured with the free-amplitude method. This is the reason why the latter includes a smaller number of objects.

Data plotted in Fig.~\ref{fig:Amp_fe_blazhko} show quite clearly that candidate Blazhko RRLs are mainly distributed 
in the short period range. In particular, $\approx$80$\%$ of the candidate Blazhko RRab have periods shorter than 0.6 days in both the photometric and the spectroscopic Bailey samples. This confirms the paucity of long-period Blazhko candidates, in agreement with early empirical evidence by \citet{Smith1981}.

The mean spectroscopic metallicities do not indicate a significant offset between the candidate Blazhko stars and the corresponding parent samples. However, a sizeable fraction of the Blazhko candidates is located in the metal-intermediate and metal-rich regime. Indeed, half of the candidate Blazhko stars are more metal-rich than [Fe/H] $\gtrsim$ -1.5, in both photometric and spectroscopic Bailey samples. Furthermore, several candidate Blazhko RRLs are located in the HASP region. These objects are only observed in stellar systems more metal-rich than [{$\rm$ Fe/H}] $\sim$ -1.5 \citep{Fiorentino2015} and imply the occurrence of metal-rich Blazhko RRLs, as originally suggested by \citet{Braga2016}. 

This evidence is further supported by the reduced dispersion, at a fixed period, of RV amplitudes when compared with luminosity amplitudes. Indeed, the spectroscopic Bailey diagram is systematically less affected by nonlinear phenomena, such as shock formation and propagation, and by off-ZAHB evolution \citep{Bono2020}. The increase in sample size and the accuracy of the RV amplitudes make this trend quite clear for both fundamental and first-overtone RRLs. The preliminary findings concerning the metallicity dependence of Blazhko candidates should be treated with caution, since the current spectroscopic catalog and the catalogs of Blazhko RRLs are far from complete.

\subsection{Dependence of RV amplitudes on metallicity}

\begin{figure}
\center
\includegraphics[width=0.49\textwidth]{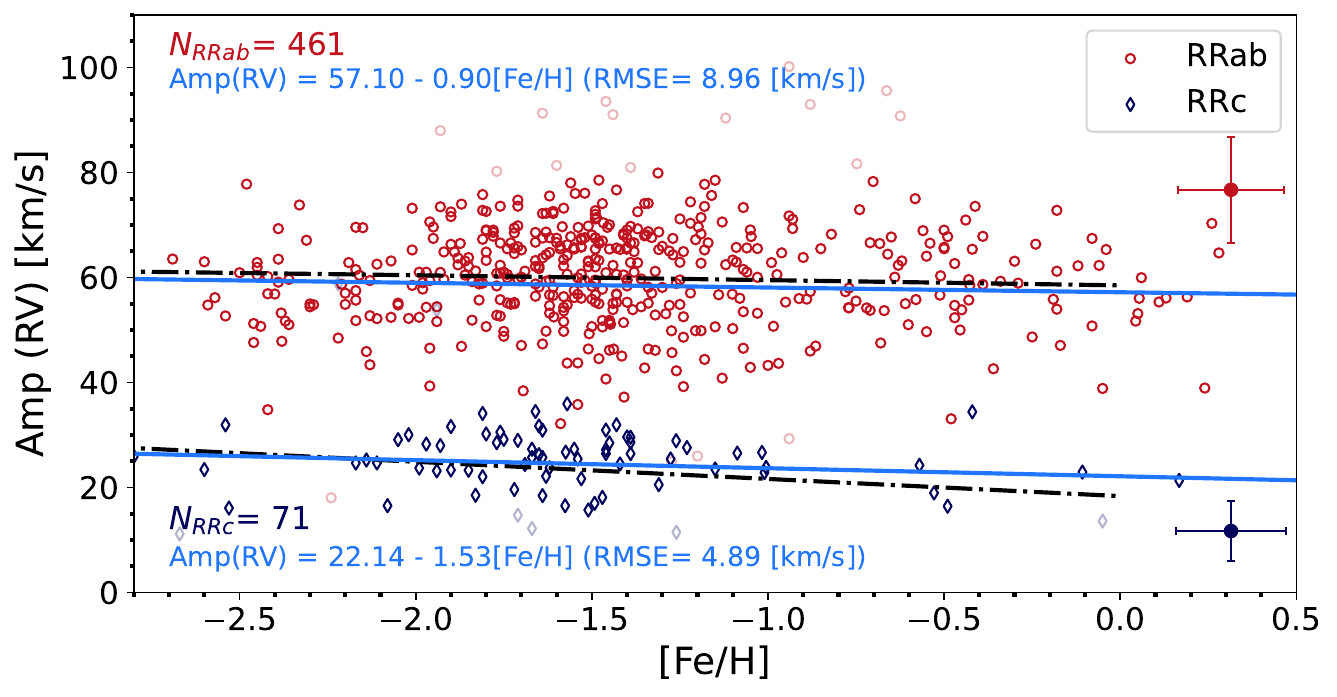}
\caption{RV amplitude as a function of iron abundance for RRab (open red circles) and RRc (open blue diamonds) variables. The dashed black lines show the relations transformed from \citet{fabrizio2021}, while the solid electric blue lines indicate the linear regressions derived in this work; the corresponding coefficients are labeled. The mean uncertainties in [Fe/H] and RV amplitude are indicated on the right-hand side of the panel.}\label{fig:Amprvfe}
\end{figure}

In the above subsection we outlined the key advantages in discussing the pulsation properties of radial variables in the spectroscopic Bailey diagram. Therefore, we investigated on a more quantitative basis the dependence of the RV amplitudes on the metal content. Fig.~\ref{fig:Amprvfe} shows the RV amplitude obtained from the Fe templates as a function of the iron abundance. The solid light blue lines display the linear fit over the data and the analytical relations for RRab and RRc variables (see Table~\ref{table:coefffe}). 

The statistical significance of the slope is limited ($\le2\sigma$) and suggests that RV amplitudes based on metallic lines display a minimal, if any, dependence on the iron abundance. To investigate whether the current finding is affected by possible systematics in the metallicity range and/or on the sample size we also plotted the analytical relations provided by \citet{fabrizio2021} by using the $V$-band luminosity amplitudes. The reason why we adopted these analytical relations is because the $V$-band amplitudes display a steady decrease when moving from metal-poor to metal-rich RRLs. To transform these relations into RV amplitudes, we adopted the ratio between visual and RV amplitudes discussed in Section~\ref{sec:rv}. The two sets of analytical relations, the direct ones (solid lines) and the transformed ones (dash-dotted lines), are in remarkable agreement across the entire metallicity range. This result further supports the marginal dependence of RV amplitude of Fe-lines on the iron abundance. The RV amplitudes associated with first-overtone RRLs show a very similar trend, since they display a minimal variation over the entire metallicity range. Once again, the direct and the transformed analytical relations agree quite well with each other.

\begin{figure*}
\center
\includegraphics[width=0.9\textwidth]{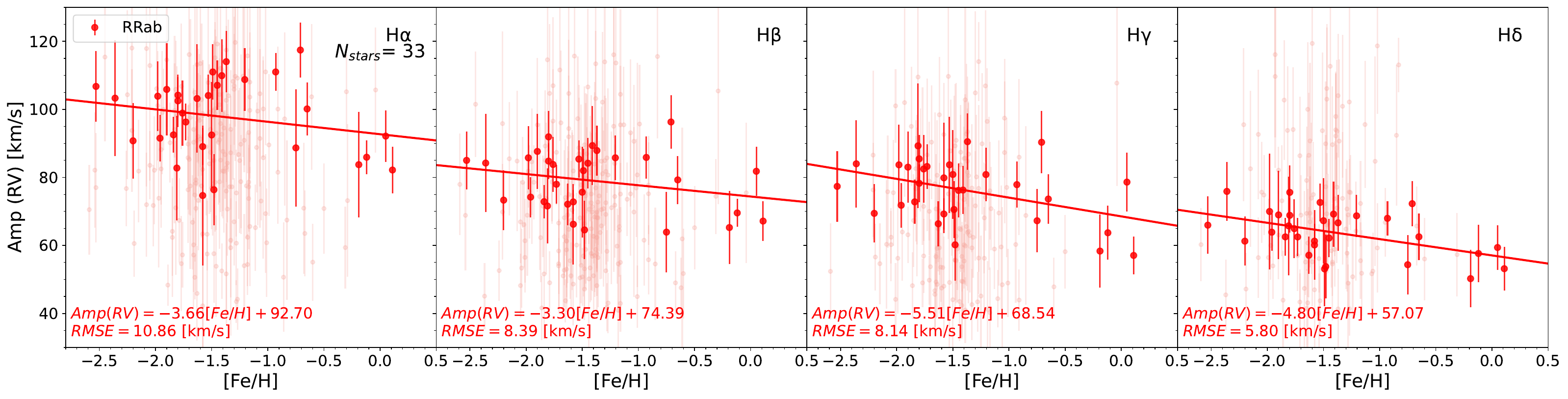}
\caption{Same as Fig.~\ref{fig:Amprvfe}, but using the Balmer lines for RRab stars. 
The red line shows the linear regression derived in this work, and the corresponding coefficients are shown in each panel.}\label{fig:metalicidad_balmer}
\end{figure*}

To check this result, we repeated the same analysis, but using the Balmer lines (see Fig.~\ref{fig:metalicidad_balmer}). We focused our attention on RRab variables, since the number of RRc variables with accurate RV amplitudes based on Balmer lines is significantly smaller. For consistency, we adopted the same subsample used in Fig.~\ref{fig:Amp_balmer}, but selecting the 33 stars in common with the adopted Balmer lines. The coefficients of the fits are listed in Table~\ref{table:coefffe}.

\begin{table}
\caption{Coefficients of the Amp(RV)–[Fe/H] relations for RRab and RRc stars.}
\label{table:coefffe}
\centering
\footnotesize
\resizebox{\columnwidth}{!}{
\begin{tabular}{cccr|ccc}
\toprule
\multicolumn{4}{c|}{RRab [$y = a + bx$]} &
\multicolumn{3}{c}{RRc [$y = a + bx$]} \\
\cmidrule(r){1-4} \cmidrule(l){5-7}
Line & $a \pm \epsilon_a$ & $b \pm \epsilon_b$ & RMSE 
&
$a \pm \epsilon_a$ & $b \pm \epsilon_b$& RMSE  \\
 &  & & (km s$^{-1}$) & & & (km s$^{-1}$)  \\
\midrule
Fe &  57.10 $\pm$ 1.13 & -0.90 $\pm$ 0.74 & 8.96 & 22.14 $\pm$ 1.84 & -1.53 $\pm$ 1.12 & 4.89 \\
\Ha & 92.70 $\pm$ 4.63  & -3.66 $\pm$ 3.00& 10.86 & $\dots$  & $\dots$  &  $\dots$\\
\Hb & 74.39 $\pm$ 3.58  & -3.30 $\pm$ 2.32 &8.38 & $\dots$   & $\dots$ & $\dots$\\
\Hg & 68.54 $\pm$ 3.47 & -5.51 $\pm$ 2.25& 8.14 & $\dots$  & $\dots$ &$\dots$ \\
\Hd & 57.07 $\pm$ 2.47 &-4.80 $\pm$ 1.60 & 5.80 & $\dots$  & $\dots$ &$\dots$ \\
\bottomrule
\end{tabular}}
\tablefoot{The left and right columns display the linear fit coefficients (a, b) and their uncertainties for the RRab and RRc variables, respectively.}
\end{table}

The data plotted in this figure also suggest a gradual decrease in the RV amplitudes from H$\alpha$ to H$\delta$, although the trend is not equally evident in all relations because of the large scatter and the limited number of measurements available for some lines.
The slopes listed in Table~\ref{table:coefffe} indicate that the dependence of the RV amplitudes on metallicity becomes progressively stronger from H$\alpha$/H$\beta$ to H$\gamma$/H$\delta$, in contrast with the almost constant trend observed for the metallic lines. More importantly, the relative uncertainty of the slope decreases steadily from H$\alpha$ to H$\delta$. The H$\alpha$ relation exhibits a relatively large slope, but its large uncertainty makes the trend only marginally significant. In contrast, the H$\delta$ relation combines a steeper slope with a substantially smaller relative uncertainty, making the metallicity dependence considerably more robust. This suggests that the effect becomes increasingly significant for the higher-order Balmer lines, particularly H$\gamma$ and H$\delta$. These results further support the idea that spectroscopic Bailey diagrams and RV amplitudes provide complementary diagnostics for investigating the pulsation properties and evolutionary status of field RRL stars.

\section{Summary and final remarks}\label{sec:5}

We present new and homogeneous RV measurements for 17,563 field RRL stars by using a mix of proprietary and publicly available spectroscopic data. The sample includes stars pulsating in the fundamental mode (12,353 RRab), in the first overtone (5,011 RRc), and in double-mode (199 RRd), which are well distributed across the Galactic spheroid. 

By applying RVC templates for metallic lines (Fe, Mg, Na), two lines of the near-infrared Ca II triplet (\CaTone{}, \CaTtwo{}), and Balmer lines (\Ha, \Hb, \Hg, \Hd) to both RRab and RRc stars, we derived a homogeneous set of $V_{\gamma}$ velocities and RV amplitudes for 5,966 RRab and 2,698 RRc stars. When combined with literature data, the current dataset includes measurements for a total of 15,955 field RRLs. 
The typical accuracy on individual $V_{\gamma}$ velocities across the catalog is 3.8 km s$^{-1}$ for well-sampled RVCs, 6.5 km s$^{-1}$ for RVCs with 3-7 phase points, and 11.3 km s$^{-1}$ for RVCs with fewer than three randomly distributed phase points. The typical accuracy on RV amplitudes ranges from 5.6 km s$^{-1}$ for well-sampled RVCs to 9.7 km s$^{-1}$ for RVCs with 4-7 randomly distributed phase points.
Moreover, and even more importantly, the dispersion in $V_{\gamma}$ velocities obtained with the template-fitting method used in this investigation is approximately 1.5 times smaller than the dispersion based on the mean of the RV measurements typically adopted in the literature, demonstrating improved precision even with limited phase coverage. In the following, we summarize the most relevant results: 

\begin{enumerate}
    \item We investigated the spectroscopic Bailey diagram (RV amplitudes versus period) and showed, for the first time, that RV amplitudes display a smooth transition from the short-period to the long-period regime. The observed trend is very similar to that of the photometric Bailey diagram; however, the RV amplitudes exhibit a smaller dispersion compared with luminosity amplitudes. This difference is mainly due to the fact that RV amplitudes primarily trace radius variations and are only marginally affected by nonlinear phenomena and off-ZAHB evolution when compared with optical luminosity amplitudes \citep{Bono2020}. Moreover, the same behavior is observed when adopting either metallic or Balmer diagnostics, confirming that the trend is not line-dependent and supporting the robustness of the result across different spectral diagnostics.
    
    \item  We extended this analysis to the candidate Blazhko stars (155 RRab and 17 RRc) and their period distribution shows a clear lack of Blazhko RRLs with periods greater than 0.7 days. For the first time, we found clear evidence that their metallicity distribution function is skewed toward the metal-intermediate and metal-rich regime, as suggested by the presence of several Blazhko RRLs in the HASP region. These findings provide robust evidence supporting the existence of metal-rich Blazhko stars, confirming what was previously suggested by \citet{Braga2016}.

    \item We investigated for the first time the variation in RV amplitudes as a function of metallicity. The RV amplitudes based on metallic lines display a marginal dependence on iron abundance, since the statistical significance of the observed trends remains below the $2\sigma$ level. In contrast, the Balmer lines reveal a different behavior. Although the H$\alpha$ and H$\beta$ relations remain only marginally significant because of their large uncertainties, the dependence of RV amplitudes on metallicity becomes progressively stronger in absolute value and statistically more robust from H$\gamma$ to H$\delta$. In particular, the relative uncertainty of the slope steadily decreases toward the higher-order Balmer lines, indicating that the metallicity dependence becomes increasingly steeper and more significant in atmospheric layers probed by these transitions. These findings suggest that the spectroscopic Bailey diagram and RV amplitudes provide complementary diagnostics for investigating the pulsation properties and atmospheric structure of RRL stars.

    \item We derived scaling relations between the photometric ($V$, $G_{BP}$, $G$, $G_{RP}$) and RV amplitudes. As expected, the scaling between photometric and RV amplitudes for RRab variables is nonlinear when using metallic lines. Interestingly, we find for the first time that the same nonlinear behavior also applies to the Balmer and calcium lines in RRab stars. By contrast, the trends for RRc variables are linear within the uncertainties. Plain physical arguments indicate that the nonlinearity, and in particular the bending, in the amplitude ratios, is mainly driven by long-period RRab variables, suggesting that convection affects luminosity amplitudes more strongly than RV amplitudes when moving from the blue (hotter) to the red (cooler) edge of the instability strip.

\end{enumerate} 

Our results are expected to be particularly valuable for studies investigating the structure and kinematics of the Galactic stellar populations. The large, homogeneous dataset of $V_{\gamma}$ and RV amplitudes we provide can serve as a robust training set for near-future spectroscopic surveys (WEAVE\footnote{\href{https://weave-project.atlassian.net/wiki/spaces/WEAVE/overview}{https://weave-project.atlassian.net/wiki/spaces/WEAVE/overview}}\citep{weave}, 4MOST\footnote{\href{https://www.4most.eu/cms/home/}{https://www.4most.eu/cms/home/}}\citep{4most}, MOONS\footnote{\href{https://vltmoons.org/}{https://vltmoons.org/}}\citep{moons}, PFS\footnote{\href{https://pfs.ipmu.jp/}{https://pfs.ipmu.jp/}}\citep{pfs})
to calibrate and validate their RV measurements. Moreover, they are also useful for investigating the stellar kinematics of old stellar populations in different Galactic components and in merger relics.

\section{Data availability}
Tables~\ref{table:Vgammafinal} and~\ref{table:properties} are only available in electronic form at the CDS via anonymous ftp to cdsarc.u-strasbg.fr (130.79.128.5) or via \url{http://cdsweb.u-strasbg.fr/cgi-bin/qcat?J/A+A/}.

\begin{acknowledgements}
We thank the anonymous referee for providing helpful comments and pertinent suggestions on an earlier version of this manuscript, which significantly improved its content and readability.
Support for this project is provided by Italian Research Center on High Performance Computing Big Data and Quantum Computing (ICSC), project funded by European Union - NextGenerationEU - and National
Recovery and Resilience Plan (NRRP) - Mission 4 Component 2 within the activities of Spoke 3 (Astrophysics and Cosmos Observations).
Several of us thank the support from Project PRIN MUR 2022 (code 2022ARWP9C) ‘Early Formation and Evolution of Bulge and HalO (EFEBHO)’ (PI: M. Marconi), funded by the European Union—Next Generation EU, and from the Large grant INAF 2023 MOVIE (PI: M. Marconi). 
This project has been supported by the Munich Institute for Astro-, Particle and BioPhysics (MIAPbP), which is funded by the Deutsche Forschungsgemeinschaft (DFG, German Research Foundation) under Germany’s Excellence Strategy—EXC-2094—390783311.
G.B. acknowledges partial support from the INFN InDARK project.
M.F. acknowledges financial support from the ASI-INAF agreement no. 2022-14-HH.0.
A.N., M.T. and G.F. acknowledge support for this research from the Next Generation EU funds within the National Recovery and Resilience Plan (PNRR), Mission 4 - Education and Research, Component 2 - From Research to Business (M4C2), Investment Line 3.1 - Strengthening and creation of Research Infrastructures, Project IR0000034 – “STILES - Strengthening the Italian Leadership in ELT and SKA”, CUP C33C22000640006.
MZ acknowledges financial support from the ANID BASAL Center for Astrophysics and Associated Technologies (CATA)
through grant FB210003, and by FONDECYT Regular grant No. 1230731.
M. Marengo is supported by the National Science Foundation under Grant No. AST-2605928.
M.Mo. acknowledge support from Spanish Ministry of Science, Innovation and Universities (MICIU) through the Spanish State Research Agency under the grants "RR Lyrae stars, a lighthouse to distant galaxies and early galaxy evolution" and the European Regional Development Fund (ERDF) with reference PID2021-127042OB-I00.
M.Mo. and G.F. acknowledge the INAF projects "Participation in LSST - Large Synoptic Survey Telescope" (LSST inkind contribution ITA-INA-S22, PI: G. Fiorentino), OB.FU. 1.05.03.06 and "MINI-GRANTS (2023) DI RSN2" (PI: G. Fiorentino), OB.FU. 1.05.23.04.02.
This research has also made use of the GaiaPortal catalogs access tool, ASI - Space Science Data Center, Rome, Italy (https://gaiaportal.ssdc.asi.it).
\\

This paper is based on data obtained with the Dark Energy Spectroscopic Instrument (DESI). DESI construction and operations is managed by the Lawrence Berkeley National Laboratory. This material is based upon work supported by the U.S. Department of Energy, Office of Science, Office of High-Energy Physics, under Contract No. DE–AC02–05CH11231, and by the National Energy Research Scientific Computing Center, a DOE Office of Science User Facility under the same contract. Additional support for DESI was provided by the U.S. National Science Foundation (NSF), Division of Astronomical Sciences under Contract No. AST-0950945 to the NSF’s National Optical-Infrared Astronomy Research Laboratory; the Science and Technology Facilities Council of the United Kingdom; the Gordon and Betty Moore Foundation; the Heising-Simons Foundation; the French Alternative Energies and Atomic Energy Commission (CEA); the National Council of Humanities, Science and Technology of Mexico (CONAHCYT); the Ministry of Science and Innovation of Spain (MICINN), and by the DESI Member Institutions: www.desi.lbl.gov/collaborating-institutions. The DESI collaboration is honored to be permitted to conduct scientific research on I’oligam Du’ag (Kitt Peak), a mountain with particular significance to the Tohono O’odham Nation. Any opinions, findings, and conclusions or recommendations expressed in this material are those of the author(s) and do not necessarily reflect the views of the U.S. National Science Foundation, the U.S. Department of Energy, or any of the listed funding agencies.
\\

This work has made use of data from the European Space Agency (ESA) mission {\it Gaia} (\url{https://www.cosmos.esa.int/gaia}), processed by the {\it Gaia} Data Processing and Analysis Consortium (DPAC,
\url{https://www.cosmos.esa.int/web/gaia/dpac/consortium}). Funding for the DPAC has been provided by national institutions, in particular the institutions participating in the {\it Gaia} Multilateral Agreement.
\\

Guoshoujing Telescope (the Large Sky Area Multi-Object Fiber Spectroscopic Telescope, LAMOST) is a National Major Scientific Project built by the Chinese Academy of Sciences. Funding for the project has been provided by the National Development and Reform Commission. LAMOST is operated and managed by the National Astronomical Observatories, Chinese Academy of Sciences.
\\

Funding for the Sloan Digital Sky Survey V has been provided by the Alfred P. Sloan Foundation, the Heising-Simons Foundation, the National Science Foundation, and the Participating Institutions. SDSS acknowledges support and resources from the Center for High-Performance Computing at the University of Utah. SDSS telescopes are located at Apache Point Observatory, funded by the Astrophysical Research Consortium and operated by New Mexico State University, and at Las Campanas Observatory, operated by the Carnegie Institution for Science. The SDSS web site is \url{www.sdss.org}.SDSS is managed by the Astrophysical Research Consortium for the Participating Institutions of the SDSS Collaboration, including the Carnegie Institution for Science, Chilean National Time Allocation Committee (CNTAC) ratified researchers, Caltech, the Gotham Participation Group, Harvard University, Heidelberg University, The Flatiron Institute, The Johns Hopkins University, L'Ecole polytechnique f\'{e}d\'{e}rale de Lausanne (EPFL), Leibniz-Institut f\"{u}r Astrophysik Potsdam (AIP), Max-Planck-Institut f\"{u}r Astronomie (MPIA Heidelberg), Max-Planck-Institut f\"{u}r Extraterrestrische Physik (MPE), Nanjing University, National Astronomical Observatories of China (NAOC), New Mexico State University, The Ohio State University, Pennsylvania State University, Smithsonian Astrophysical Observatory, Space Telescope Science Institute (STScI), the Stellar Astrophysics Participation Group, Universidad Nacional Aut\'{o}noma de M\'{e}xico, University of Arizona, University of Colorado Boulder, University of Illinois at Urbana-Champaign, University of Toronto, University of Utah, University of Virginia, Yale University, and Yunnan University. \\

This research has also made use of the GaiaPortal catalogs access tool, ASI - Space Science Data Center, Rome, Italy
(\url{https://gaiaportal.ssdc.asi.it}) and the TOPCAT catalog handling tool (Taylor 2005, 2017). 

\end{acknowledgements}

\bibliographystyle{aa} 
\bibliography{main}

@BOOK{Smith1995,
       author = {{Smith}, H.~A.},
        title = "{RR Lyrae Stars (Cambridge University Press, Cambridge)}",
      journal = {Cambridge Astrophysics Series},
         year = 1995,
       adsurl = {https://ui.adsabs.harvard.edu/abs/1995CAS....27.....S}
}

@ARTICLE{Clementini2023,
       author = {{Clementini}, G. and {Ripepi}, V. and {Garofalo}, A. and {Molinaro}, R. and {Muraveva}, T. and {Leccia}, S. and {Rimoldini}, L. and {Holl}, B. and {Jevardat de Fombelle}, G. and {Sartoretti}, P. and {Marchal}, O. and {Audard}, M. and {Nienartowicz}, K. and {Andrae}, R. and {Marconi}, M. and {Szabados}, L. and {Evans}, D.~W. and {Lecoeur-Taibi}, I. and {Mowlavi}, N. and {Musella}, I. and {Eyer}, L.},
        title = "{Gaia Data Release 3. Specific processing and validation of all-sky RR Lyrae and Cepheid stars: The RR Lyrae sample}",
      journal = {\aap},
         year = 2023,
        month = jun,
       volume = {674},
          eid = {A18},
        pages = {A18},
          doi = {10.1051/0004-6361/202243964},
archivePrefix = {arXiv},
       eprint = {2206.06278},
 primaryClass = {astro-ph.SR},
       adsurl = {https://ui.adsabs.harvard.edu/abs/2023A&A...674A..18C}
}

@ARTICLE{Cropper2018,
       author = {{Cropper}, M. and {Katz}, D. and {Sartoretti}, P. and {Prusti}, T. and {de Bruijne}, J.~H.~J. and {Chassat}, F. and {Charvet}, P. and {Boyadjian}, J. and {Perryman}, M. and {Sarri}, G. and {Gare}, P. and {Erdmann}, M. and {Munari}, U. and {Zwitter}, T. and {Wilkinson}, M. and {Arenou}, F. and {Vallenari}, A. and {G{\'o}mez}, A. and {Panuzzo}, P. and {Seabroke}, G. and {Allende Prieto}, C. and {Benson}, K. and {Marchal}, O. and {Huckle}, H. and {Smith}, M. and {Dolding}, C. and {Jan{\ss}en}, K. and {Viala}, Y. and {Blomme}, R. and {Baker}, S. and {Boudreault}, S. and {Crifo}, F. and {Soubiran}, C. and {Fr{\'e}mat}, Y. and {Jasniewicz}, G. and {Guerrier}, A. and {Guy}, L.~P. and {Turon}, C. and {Jean-Antoine-Piccolo}, A. and {Th{\'e}venin}, F. and {David}, M. and {Gosset}, E. and {Damerdji}, Y.},
        title = "{Gaia Data Release 2. Gaia Radial Velocity Spectrometer}",
      journal = {\aap},
         year = 2018,
        month = aug,
       volume = {616},
          eid = {A5},
        pages = {A5},
          doi = {10.1051/0004-6361/201832763},
archivePrefix = {arXiv},
       eprint = {1804.09369},
 primaryClass = {astro-ph.IM},
       adsurl = {https://ui.adsabs.harvard.edu/abs/2018A&A...616A...5C}
}

@ARTICLE{Katz2023,
       author = {{Katz}, D. and {Sartoretti}, P. and {Guerrier}, A. and {Panuzzo}, P. and {Seabroke}, G.~M. and {Th{\'e}venin}, F. and {Cropper}, M. and {Benson}, K. and {Blomme}, R. and {Haigron}, R. and {Marchal}, O. and {Smith}, M. and {Baker}, S. and {Chemin}, L. and {Damerdji}, Y. and {David}, M. and {Dolding}, C. and {Fr{\'e}mat}, Y. and {Gosset}, E. and {Jan{\ss}en}, K. and {Jasniewicz}, G. and {Lobel}, A. and {Plum}, G. and {Samaras}, N. and {Snaith}, O. and {Soubiran}, C. and {Vanel}, O. and {Zwitter}, T. and {Antoja}, T. and {Arenou}, F. and {Babusiaux}, C. and {Brouillet}, N. and {Caffau}, E. and {Di Matteo}, P. and {Fabre}, C. and {Fabricius}, C. and {Fragkoudi}, F. and {Haywood}, M. and {Huckle}, H.~E. and {Hottier}, C. and {Lasne}, Y. and {Leclerc}, N. and {Mastrobuono-Battisti}, A. and {Royer}, F. and {Teyssier}, D. and {Zorec}, J. and {Crifo}, F. and {Jean-Antoine Piccolo}, A. and {Turon}, C. and {Viala}, Y.},
        title = "{Gaia Data Release 3. Properties and validation of the radial velocities}",
      journal = {\aap},
         year = 2023,
        month = jun,
       volume = {674},
          eid = {A5},
        pages = {A5},
          doi = {10.1051/0004-6361/202244220},
archivePrefix = {arXiv},
       eprint = {2206.05902},
 primaryClass = {astro-ph.GA},
       adsurl = {https://ui.adsabs.harvard.edu/abs/2023A&A...674A...5K}
}

@ARTICLE{Sneden2017,
       author = {{Sneden}, Christopher and {Preston}, George W. and {Chadid}, Merieme and {Adam{\'o}w}, Monika},
        title = "{The RRc Stars: Chemical Abundances and Envelope Kinematics}",
      journal = {\apj},
         year = 2017,
        month = oct,
       volume = {848},
       number = {1},
          eid = {68},
        pages = {68},
          doi = {10.3847/1538-4357/aa8b10},
archivePrefix = {arXiv},
       eprint = {1709.00494},
 primaryClass = {astro-ph.SR},
       adsurl = {https://ui.adsabs.harvard.edu/abs/2017ApJ...848...68S}
}

@ARTICLE{Chadid2017,
       author = {{Chadid}, Merieme and {Sneden}, Christopher and {Preston}, George W.},
        title = "{Spectroscopic Comparison of Metal-rich RRab Stars of the Galactic Field with their Metal-poor Counterparts}",
      journal = {\apj},
         year = 2017,
        month = feb,
       volume = {835},
       number = {2},
          eid = {187},
        pages = {187},
          doi = {10.3847/1538-4357/835/2/187},
archivePrefix = {arXiv},
       eprint = {1611.02368},
 primaryClass = {astro-ph.SR},
       adsurl = {https://ui.adsabs.harvard.edu/abs/2017ApJ...835..187C}
}

@ARTICLE{Sneden2024,
       author = {{Sneden}, Christopher and {Preston}, George W.},
        title = "{Variations in Line Profiles of Atomic Transitions in RR Lyrae Stars}",
      journal = {\aj},
         year = 2024,
        month = jun,
       volume = {167},
       number = {6},
          eid = {268},
        pages = {268},
          doi = {10.3847/1538-3881/ad3c34},
archivePrefix = {arXiv},
       eprint = {2404.01420},
 primaryClass = {astro-ph.SR},
       adsurl = {https://ui.adsabs.harvard.edu/abs/2024AJ....167..268S}
}

@ARTICLE{Medina2025,
       author = {{Medina}, Gustavo E. and {Li}, Ting S. and {Allende Prieto}, C. and {Beraldo e Silva}, L. and {Bystrom}, A. and {Carlberg}, R.~G. and {Koposov}, S.~E. and {Lambert}, M. and {Najita}, J.~R. and {Rockosi}, C.~M. and {Kizhuprakkat}, N. and {Riley}, A. and {Aguilar}, J. and {Ahlen}, S. and {Bianchi}, D. and {Brooks}, D. and {Claybaugh}, T. and {Cooper}, A.~P. and {de la Macorra}, A. and {Dey}, A. and {Doel}, P. and {Forero-Romero}, J. and {Gazta{\~n}aga}, E. and {Gontcho}, S. Gontcho A and {Gutierrez}, G. and {Ishak}, M. and {Kehoe}, R. and {Kisner}, T. and {Landriau}, M. and {Le Guillou}, L. and {Meisner}, A. and {Miquel}, R. and {Prada}, F. and {Perez-Rafols}, I. and {Rossi}, G. and {Sanchez}, E. and {Schlegel}, D.~J. and {Silber}, J.~H. and {Sprayberry}, D. and {Tarle}, G. and {Weaver}, B.~A. and {Zhou}, R.},
        title = "{The DESI Y1 RR Lyrae catalog II: The metallicity dependency of pulsational properties, the shape of the RR Lyrae instability strip, and metal rich RR Lyrae}",
      journal = {\apj},
      note = {in press},
         year = 2025,
        month = may,
          eid = {arXiv:2505.10614},
        pages = {arXiv:2505.10614},
          doi = {10.48550/arXiv.2505.10614},
archivePrefix = {arXiv},
       eprint = {2505.10614},
 primaryClass = {astro-ph.GA},
       adsurl = {https://ui.adsabs.harvard.edu/abs/2025arXiv250510614M}
}

@ARTICLE{Braga2021,
       author = {{Braga}, V.~F. and {Crestani}, J. and {Fabrizio}, M. and {Bono}, G. and {Sneden}, C. and {Preston}, G.~W. and {Storm}, J. and {Kamann}, S. and {Latour}, M. and {Lala}, H. and {Lemasle}, B. and {Prudil}, Z. and {Altavilla}, G. and {Chaboyer}, B. and {Dall'Ora}, M. and {Ferraro}, I. and {Gilligan}, C.~K. and {Fiorentino}, G. and {Iannicola}, G. and {Inno}, L. and {Kwak}, S. and {Marengo}, M. and {Marinoni}, S. and {Marrese}, P.~M. and {Mart{\'\i}nez-V{\'a}zquez}, C.~E. and {Monelli}, M. and {Mullen}, J.~P. and {Matsunaga}, N. and {Neeley}, J. and {Stetson}, P.~B. and {Valenti}, E. and {Zoccali}, M.},
        title = "{On the Use of Field RR Lyrae as Galactic Probes. V. Optical and Radial Velocity Curve Templates}",
      journal = {\apj},
         year = 2021,
        month = oct,
       volume = {919},
       number = {2},
          eid = {85},
        pages = {85},
          doi = {10.3847/1538-4357/ac1074},
archivePrefix = {arXiv},
       eprint = {2107.00923},
 primaryClass = {astro-ph.SR},
       adsurl = {https://ui.adsabs.harvard.edu/abs/2021ApJ...919...85B}
}

@ARTICLE{fabrizio2021,
       author = {{Fabrizio}, M. and {Braga}, V.~F. and {Crestani}, J. and {Bono}, G. and {Ferraro}, I. and {Fiorentino}, G. and {Iannicola}, G. and {Preston}, G.~W. and {Sneden}, C. and {Th{\'e}venin}, F. and {Altavilla}, G. and {Chaboyer}, B. and {Dall'Ora}, M. and {da Silva}, R. and {Grebel}, E.~K. and {Gilligan}, C.~K. and {Lala}, H. and {Lemasle}, B. and {Magurno}, D. and {Marengo}, M. and {Marinoni}, S. and {Marrese}, P.~M. and {Mart{\'\i}nez-V{\'a}zquez}, C.~E. and {Matsunaga}, N. and {Monelli}, M. and {Mullen}, J.~P. and {Neeley}, J. and {Nonino}, M. and {Prudil}, Z. and {Salaris}, M. and {Stetson}, P.~B. and {Valenti}, E. and {Zoccali}, M.},
        title = "{On the Use of Field RR Lyrae As Galactic Probes: IV. New Insights Into and Around the Oosterhoff Dichotomy}",
      journal = {\apj},
         year = 2021,
        month = oct,
       volume = {919},
       number = {2},
          eid = {118},
        pages = {118},
          doi = {10.3847/1538-4357/ac1115},
archivePrefix = {arXiv},
       eprint = {2107.00919},
 primaryClass = {astro-ph.SR},
       adsurl = {https://ui.adsabs.harvard.edu/abs/2021ApJ...919..118F}
}

@ARTICLE{Savino2020,
       author = {{Savino}, A. and {Koch}, A. and {Prudil}, Z. and {Kunder}, A. and {Smolec}, R.},
        title = "{The age of the Milky Way inner stellar spheroid from RR Lyrae population synthesis}",
      journal = {\aap},
         year = 2020,
        month = sep,
       volume = {641},
          eid = {A96},
        pages = {A96},
          doi = {10.1051/0004-6361/202038305},
archivePrefix = {arXiv},
       eprint = {2006.12507},
 primaryClass = {astro-ph.GA},
       adsurl = {https://ui.adsabs.harvard.edu/abs/2020A&A...641A..96S}
}

@ARTICLE{Catelan2009,
       author = {{Catelan}, M.},
        title = "{Horizontal branch stars: the interplay between observations and theory, and insights into the formation of the Galaxy}",
      journal = {\apss},
         year = 2009,
        month = apr,
       volume = {320},
        pages = {261-309},
          doi = {10.1007/s10509-009-9987-8},
archivePrefix = {arXiv},
       eprint = {astro-ph/0507464},
 primaryClass = {astro-ph},
       adsurl = {https://ui.adsabs.harvard.edu/abs/2009Ap&SS.320..261C}
}

@ARTICLE{Walker1989,
       author = {{Walker}, Alistair R.},
        title = "{A Survey for RR Lyrae Variables in Five Small Magellanic Cloud Clusters}",
      journal = {\pasp},
         year = 1989,
        month = jun,
       volume = {101},
        pages = {570},
          doi = {10.1086/132470},
       adsurl = {https://ui.adsabs.harvard.edu/abs/1989PASP..101..570W}
}

@ARTICLE{Kunder2022,
       author = {{Kunder}, Andrea M.},
        title = "{RR Lyrae Variables as Tracers of the Galactic Bulge Kinematic Structure}",
      journal = {Universe},
         year = 2022,
        month = mar,
       volume = {8},
       number = {4},
          eid = {206},
        pages = {206},
          doi = {10.3390/universe8040206},
       adsurl = {https://ui.adsabs.harvard.edu/abs/2022Univ....8..206K}
}

@ARTICLE{Gozha2018,
       author = {{Gozha}, M.~L. and {Marsakov}, V.~A. and {Koval'}, V.~V.},
        title = "{Abundance of Chemical Elements in RR Lyrae Variables and their Kinematic Parameters}",
      journal = {Astrophysics},
         year = 2018,
        month = mar,
       volume = {61},
       number = {1},
        pages = {41-49},
          doi = {10.1007/s10511-018-9514-0},
       adsurl = {https://ui.adsabs.harvard.edu/abs/2018Ap.....61...41G}
}

@ARTICLE{Dambis2013,
       author = {{Dambis}, A.~K. and {Berdnikov}, L.~N. and {Kniazev}, A.~Y. and {Kravtsov}, V.~V. and {Rastorguev}, A.~S. and {Sefako}, R. and {Vozyakova}, O.~V.},
        title = "{RR Lyrae variables: visual and infrared luminosities, intrinsic colours and kinematics}",
      journal = {\mnras},
         year = 2013,
        month = nov,
       volume = {435},
       number = {4},
        pages = {3206-3220},
          doi = {10.1093/mnras/stt1514},
archivePrefix = {arXiv},
       eprint = {1308.4727},
 primaryClass = {astro-ph.GA},
       adsurl = {https://ui.adsabs.harvard.edu/abs/2013MNRAS.435.3206D}
}

@ARTICLE{Kunder2017,
       author = {{Kunder}, Andrea and {Kordopatis}, Georges and {Steinmetz}, Matthias and {Zwitter}, Toma{\v{z}} and {McMillan}, Paul J. and {Casagrande}, Luca and {Enke}, Harry and {Wojno}, Jennifer and {Valentini}, Marica and {Chiappini}, Cristina and {Matijevi{\v{c}}}, Gal and {Siviero}, Alessandro and {de Laverny}, Patrick and {Recio-Blanco}, Alejandra and {Bijaoui}, Albert and {Wyse}, Rosemary F.~G. and {Binney}, James and {Grebel}, E.~K. and {Helmi}, Amina and {Jofre}, Paula and {Antoja}, Teresa and {Gilmore}, Gerard and {Siebert}, Arnaud and {Famaey}, Benoit and {Bienaym{\'e}}, Olivier and {Gibson}, Brad K. and {Freeman}, Kenneth C. and {Navarro}, Julio F. and {Munari}, Ulisse and {Seabroke}, George and {Anguiano}, Borja and {{\v{Z}}erjal}, Maru{\v{s}}a and {Minchev}, Ivan and {Reid}, Warren and {Bland-Hawthorn}, Joss and {Kos}, Janez and {Sharma}, Sanjib and {Watson}, Fred and {Parker}, Quentin A. and {Scholz}, Ralf-Dieter and {Burton}, Donna and {Cass}, Paul and {Hartley}, Malcolm and {Fiegert}, Kristin and {Stupar}, Milorad and {Ritter}, Andreas and {Hawkins}, Keith and {Gerhard}, Ortwin and {Chaplin}, W.~J. and {Davies}, G.~R. and {Elsworth}, Y.~P. and {Lund}, M.~N. and {Miglio}, A. and {Mosser}, B.},
        title = "{The Radial Velocity Experiment (RAVE): Fifth Data Release}",
      journal = {\aj},
         year = 2017,
        month = feb,
       volume = {153},
       number = {2},
          eid = {75},
        pages = {75},
          doi = {10.3847/1538-3881/153/2/75},
archivePrefix = {arXiv},
       eprint = {1609.03210},
 primaryClass = {astro-ph.SR},
       adsurl = {https://ui.adsabs.harvard.edu/abs/2017AJ....153...75K}
}

@ARTICLE{Maintz2005,
       author = {{Maintz}, G. and {de Boer}, K.~S.},
        title = "{RR Lyrae stars: kinematics, orbits and z-distribution}",
      journal = {\aap},
         year = 2005,
        month = oct,
       volume = {442},
       number = {1},
        pages = {229-237},
          doi = {10.1051/0004-6361:20053231},
archivePrefix = {arXiv},
       eprint = {astro-ph/0507604},
 primaryClass = {astro-ph},
       adsurl = {https://ui.adsabs.harvard.edu/abs/2005A&A...442..229M}
}

@ARTICLE{Prudil2020,
       author = {{Prudil}, Z. and {D{\'e}k{\'a}ny}, I. and {Grebel}, E.~K. and {Kunder}, A.},
        title = "{Evidence for Galactic disc RR Lyrae stars in the solar neighbourhood}",
      journal = {\mnras},
         year = 2020,
        month = mar,
       volume = {492},
       number = {3},
        pages = {3408-3419},
          doi = {10.1093/mnras/staa046},
archivePrefix = {arXiv},
       eprint = {2001.02486},
 primaryClass = {astro-ph.SR},
       adsurl = {https://ui.adsabs.harvard.edu/abs/2020MNRAS.492.3408P}
}

@ARTICLE{dorazi2024,
       author = {{D'Orazi}, Valentina and {Storm}, Nicholas and {Casey}, Andrew R. and {Braga}, Vittorio F. and {Zocchi}, Alice and {Bono}, Giuseppe and {Fabrizio}, Michele and {Sneden}, Christopher and {Massari}, Davide and {Giribaldi}, Riano E. and {Bergemann}, Maria and {Campbell}, Simon W. and {Casagrande}, Luca and {de Grijs}, Richard and {De Silva}, Gayandhi and {Lugaro}, Maria and {Zucker}, Daniel B. and {Bragaglia}, Angela and {Feuillet}, Diane and {Fiorentino}, Giuliana and {Chaboyer}, Brian and {Dall'Ora}, Massimo and {Marengo}, Massimo and {Mart{\'\i}nez-V{\'a}zquez}, Clara E. and {Matsunaga}, Noriyuki and {Monelli}, Matteo and {Mullen}, Joseph P. and {Nataf}, David and {Tantalo}, Maria and {Thevenin}, Frederic and {Vitello}, Fabio R. and {Kudritzki}, Rolf-Peter and {Bland-Hawthorn}, Joss and {Buder}, Sven and {Freeman}, Ken and {Kos}, Janez and {Lewis}, Geraint F. and {Lind}, Karin and {Martell}, Sarah and {Sharma}, Sanjib and {Stello}, Dennis and {Zwitter}, Toma{\v{z}}},
        title = "{The GALAH survey: tracing the Milky Way's formation and evolution through RR Lyrae stars}",
      journal = {\mnras},
         year = 2024,
        month = jun,
       volume = {531},
       number = {1},
        pages = {137-162},
          doi = {10.1093/mnras/stae1149},
archivePrefix = {arXiv},
       eprint = {2405.04580},
 primaryClass = {astro-ph.GA},
       adsurl = {https://ui.adsabs.harvard.edu/abs/2024MNRAS.531..137D}
}

@ARTICLE{crestani2021a,
       author = {{Crestani}, J. and {Fabrizio}, M. and {Braga}, V.~F. and {Sneden}, C. and {Preston}, G. and {Ferraro}, I. and {Iannicola}, G. and {Bono}, G. and {Alves-Brito}, A. and {Nonino}, M. and {D'Orazi}, V. and {Inno}, L. and {Monelli}, M. and {Storm}, J. and {Altavilla}, G. and {Chaboyer}, B. and {Dall'Ora}, M. and {Fiorentino}, G. and {Gilligan}, C. and {Grebel}, E.~K. and {Lala}, H. and {Lemasle}, B. and {Marengo}, M. and {Marinoni}, S. and {Marrese}, P.~M. and {Mart{\'\i}nez-V{\'a}zquez}, C.~E. and {Matsunaga}, N. and {Mullen}, J.~P. and {Neeley}, J. and {Prudil}, Z. and {da Silva}, R. and {Stetson}, P.~B. and {Th{\'e}venin}, F. and {Valenti}, E. and {Walker}, A. and {Zoccali}, M.},
        title = "{On the Use of Field RR Lyrae as Galactic Probes. II. A New {\ensuremath{\Delta}}S Calibration to Estimate Their Metallicity}",
      journal = {\apj},
         year = 2021,
        month = feb,
       volume = {908},
       number = {1},
          eid = {20},
        pages = {20},
          doi = {10.3847/1538-4357/abd183},
archivePrefix = {arXiv},
       eprint = {2012.02284},
 primaryClass = {astro-ph.SR},
       adsurl = {https://ui.adsabs.harvard.edu/abs/2021ApJ...908...20C}
}

@ARTICLE{crestani2021b,
       author = {{Crestani}, J. and {Braga}, V.~F. and {Fabrizio}, M. and {Bono}, G. and {Sneden}, C. and {Preston}, G. and {Ferraro}, I. and {Iannicola}, G. and {Nonino}, M. and {Fiorentino}, G. and {Th{\'e}venin}, F. and {Lemasle}, B. and {Prudil}, Z. and {Alves-Brito}, A. and {Altavilla}, G. and {Chaboyer}, B. and {Dall'Ora}, M. and {D'Orazi}, V. and {Gilligan}, C. and {Grebel}, E.~K. and {Koch-Hansen}, A.~J. and {Lala}, H. and {Marengo}, M. and {Marinoni}, S. and {Marrese}, P.~M. and {Mart{\'\i}nez-V{\'a}zquez}, C. and {Matsunaga}, N. and {Monelli}, M. and {Mullen}, J.~P. and {Neeley}, J. and {da Silva}, R. and {Stetson}, P.~B. and {Salaris}, M. and {Storm}, J. and {Valenti}, E. and {Zoccali}, M.},
        title = "{On the Use of Field RR Lyrae as Galactic Probes. III. The {\ensuremath{\alpha}}-element Abundances}",
      journal = {\apj},
         year = 2021,
        month = jun,
       volume = {914},
       number = {1},
          eid = {10},
        pages = {10},
          doi = {10.3847/1538-4357/abfa23},
archivePrefix = {arXiv},
       eprint = {2104.08113},
 primaryClass = {astro-ph.GA},
       adsurl = {https://ui.adsabs.harvard.edu/abs/2021ApJ...914...10C}
}

@ARTICLE{Mullen2022,
       author = {{Mullen}, Joseph P. and {Marengo}, Massimo and {Mart{\'\i}nez-V{\'a}zquez}, Clara E. and {Bono}, Giuseppe and {Braga}, Vittorio F. and {Chaboyer}, Brian and {Crestani}, Juliana and {Dall'Ora}, Massimo and {Fabrizio}, Michele and {Fiorentino}, Giuliana and {Monelli}, Matteo and {Neeley}, Jillian R. and {Stetson}, Peter B. and {Th{\'e}venin}, Fr{\'e}d{\'e}ric},
        title = "{Metallicity of Galactic RR Lyrae from Optical and Infrared Light Curves. II. Period-Fourier-Metallicity Relations for First Overtone RR Lyrae}",
      journal = {\apj},
         year = 2022,
        month = jun,
       volume = {931},
       number = {2},
          eid = {131},
        pages = {131},
          doi = {10.3847/1538-4357/ac67ee},
archivePrefix = {arXiv},
       eprint = {2204.07627},
 primaryClass = {astro-ph.SR},
       adsurl = {https://ui.adsabs.harvard.edu/abs/2022ApJ...931..131M}
}

@ARTICLE{Jurcsik2017,
       author = {{Jurcsik}, J. and {Smitola}, P. and {Hajdu}, G. and {S{\'o}dor}, {\'A}. and {Nuspl}, J. and {Kolenberg}, K. and {F{\H{u}}r{\'e}sz}, G. and {Bal{\'a}zs}, L.~G. and {Pilachowski}, C. and {Saha}, A. and {Mo{\'o}r}, A. and {Kun}, E. and {P{\'a}l}, A. and {Bakos}, J. and {Kelemen}, J. and {Kov{\'a}cs}, T. and {Kriskovics}, L. and {S{\'a}rneczky}, K. and {Szalai}, T. and {Szing}, A. and {Vida}, K.},
        title = "{Photometric and radial-velocity time series of RR Lyrae stars in M3: analysis of single-mode variables}",
      journal = {\mnras},
         year = 2017,
        month = jun,
       volume = {468},
       number = {2},
        pages = {1317-1337},
          doi = {10.1093/mnras/stx382},
archivePrefix = {arXiv},
       eprint = {1702.03264},
 primaryClass = {astro-ph.SR},
       adsurl = {https://ui.adsabs.harvard.edu/abs/2017MNRAS.468.1317J}
}

@ARTICLE{Feng2025,
       author = {{Feng}, Yuting and {Guhathakurta}, Puragra and {Peng}, Eric W. and {Cunningham}, Emily C. and {C{\^o}t{\'e}}, Patrick and {Ferrarese}, Laura and {Gwyn}, Stephen D.~J.},
        title = "{Kinematics of Distant Milky Way Halo RR Lyrae Stars out to 160 kpc}",
      journal = {\apj},
         year = 2026,
        month = feb,
       volume = {998},
       number = {1},
          eid = {157},
        pages = {157},
          doi = {10.3847/1538-4357/ae313b},
archivePrefix = {arXiv},
       eprint = {2512.09795},
 primaryClass = {astro-ph.GA},
       adsurl = {https://ui.adsabs.harvard.edu/abs/2026ApJ...998..157F}
}

@ARTICLE{Iorio2018,
       author = {{Iorio}, G. and {Belokurov}, V. and {Erkal}, D. and {Koposov}, S.~E. and {Nipoti}, C. and {Fraternali}, F.},
        title = "{The first all-sky view of the Milky Way stellar halo with Gaia+2MASS RR Lyrae}",
      journal = {\mnras},
         year = 2018,
        month = feb,
       volume = {474},
       number = {2},
        pages = {2142-2166},
          doi = {10.1093/mnras/stx2819},
archivePrefix = {arXiv},
       eprint = {1707.03833},
 primaryClass = {astro-ph.GA},
       adsurl = {https://ui.adsabs.harvard.edu/abs/2018MNRAS.474.2142I}
}

@ARTICLE{Bobrick2024,
       author = {{Bobrick}, Alexey and {Iorio}, Giuliano and {Belokurov}, Vasily and {Vos}, Joris and {Vu{\v{c}}kovi{\'c}}, Maja and {Giacobbo}, Nicola},
        title = "{RR Lyrae from binary evolution: abundant, young, and metal-rich}",
      journal = {\mnras},
         year = 2024,
        month = feb,
       volume = {527},
       number = {4},
        pages = {12196-12218},
          doi = {10.1093/mnras/stad3996},
archivePrefix = {arXiv},
       eprint = {2208.04332},
 primaryClass = {astro-ph.SR},
       adsurl = {https://ui.adsabs.harvard.edu/abs/2024MNRAS.52712196B}
}

@ARTICLE{Cuevas-Otahola2025,
       author = {{Cuevas-Otahola}, Bolivia and {Mateu}, Cecilia and {Cabrera-Ziri}, Ivan and {Bruzual}, Gustavo and {Hern{\'a}ndez-P{\'e}rez}, Fabiola and {Magris}, Gladis and {Baumgardt}, Holger},
        title = "{RR Lyrae stars in intermediate-age Magellanic clusters: membership probabilities and delay time distribution}",
      journal = {\mnras},
         year = 2025,
        month = aug,
       volume = {541},
       number = {2},
        pages = {1434-1448},
          doi = {10.1093/mnras/staf1095},
archivePrefix = {arXiv},
       eprint = {2411.12741},
 primaryClass = {astro-ph.GA},
       adsurl = {https://ui.adsabs.harvard.edu/abs/2025MNRAS.541.1434C}
}

@ARTICLE{kunder2020,
       author = {{Kunder}, Andrea and {P{\'e}rez-Villegas}, Angeles and {Rich}, R. Michael and {Ogata}, Jonathan and {Murari}, Emma and {Boren}, Emilie and {Johnson}, Christian I. and {Nataf}, David and {Walker}, Alistair and {Bono}, Giuseppe and {Koch}, Andreas and {Propris}, Roberto De and {Storm}, Jesper and {Wojno}, Jennifer},
        title = "{The Bulge Radial Velocity Assay for RR Lyrae Stars (BRAVA-RR) DR2: A Bimodal Bulge?}",
      journal = {\aj},
         year = 2020,
        month = jun,
       volume = {159},
       number = {6},
          eid = {270},
        pages = {270},
          doi = {10.3847/1538-3881/ab8d35},
archivePrefix = {arXiv},
       eprint = {2004.11382},
 primaryClass = {astro-ph.SR},
       adsurl = {https://ui.adsabs.harvard.edu/abs/2020AJ....159..270K}
}

@ARTICLE{kunder2015,
       author = {{Kunder}, Andrea and {Rich}, R.~M. and {Hawkins}, K. and {Poleski}, R. and {Storm}, J. and {Johnson}, C.~I. and {Shen}, J. and {Li}, Z. -Y. and {Cordero}, M.~J. and {Nataf}, D.~M. and {Bono}, G. and {Walker}, A.~R. and {Koch}, A. and {De Propris}, R. and {Udalski}, A. and {Szyma{\'n}ski}, M.~K. and {Soszy{\'n}ski}, I. and {Pietrzy{\'n}ski}, G. and {Ulaczyk}, K. and {Wyrzykowski}, {\L}. and {Pietrukowicz}, P. and {Skowron}, J. and {Koz{\l}owski}, S. and {Mr{\'o}z}, P.},
        title = "{A High-velocity Bulge RR Lyrae Variable on a Halo-like Orbit}",
      journal = {\apjl},
         year = 2015,
        month = jul,
       volume = {808},
       number = {1},
          eid = {L12},
        pages = {L12},
          doi = {10.1088/2041-8205/808/1/L12},
archivePrefix = {arXiv},
       eprint = {1506.02664},
 primaryClass = {astro-ph.SR},
       adsurl = {https://ui.adsabs.harvard.edu/abs/2015ApJ...808L..12K}
}

@ARTICLE{Netzel2018,
       author = {{Netzel}, H. and {Smolec}, R. and {Soszy{\'n}ski}, I. and {Udalski}, A.},
        title = "{Blazhko effect in the first overtone RR Lyrae stars of the OGLE Galactic bulge collection}",
      journal = {\mnras},
         year = 2018,
        month = oct,
       volume = {480},
       number = {1},
        pages = {1229-1246},
          doi = {10.1093/mnras/sty1883},
archivePrefix = {arXiv},
       eprint = {1812.05409},
 primaryClass = {astro-ph.SR},
       adsurl = {https://ui.adsabs.harvard.edu/abs/2018MNRAS.480.1229N}
}

@ARTICLE{Donev2025,
       author = {{Donev}, Ema and {Ivezi{\'c}}, {\v{Z}}eljko},
        title = "{Search for the Blazhko Effect in Field RR Lyrae Stars Using LINEAR and ZTF Light Curves}",
      journal = {\aj},
         year = 2025,
        month = jun,
       volume = {169},
       number = {6},
          eid = {310},
        pages = {310},
          doi = {10.3847/1538-3881/adccc2},
archivePrefix = {arXiv},
       eprint = {2504.05434},
 primaryClass = {astro-ph.SR},
       adsurl = {https://ui.adsabs.harvard.edu/abs/2025AJ....169..310D}
}

@ARTICLE{Skarka2020,
       author = {{Skarka}, M. and {Prudil}, Z. and {Jurcsik}, J.},
        title = "{Blazhko effect in the Galactic bulge fundamental mode RR Lyrae stars - II. Modulation shapes, amplitudes, and periods}",
      journal = {\mnras},
         year = 2020,
        month = may,
       volume = {494},
       number = {1},
        pages = {1237-1249},
          doi = {10.1093/mnras/staa673},
archivePrefix = {arXiv},
       eprint = {2001.00754},
 primaryClass = {astro-ph.SR},
       adsurl = {https://ui.adsabs.harvard.edu/abs/2020MNRAS.494.1237S}
}

@ARTICLE{fabrizio2011,
       author = {{Fabrizio}, M. and {Nonino}, M. and {Bono}, G. and {Ferraro}, I. and {Fran{\c{c}}ois}, P. and {Iannicola}, G. and {Monelli}, M. and {Th{\'e}venin}, F. and {Stetson}, P.~B. and {Walker}, A.~R. and {Buonanno}, R. and {Caputo}, F. and {Corsi}, C.~E. and {Dall'Ora}, M. and {Gilmozzi}, R. and {James}, C.~R. and {Merle}, T. and {Pulone}, L. and {Romaniello}, M.},
        title = "{The Carina Project. IV. Radial Velocity Distribution}",
      journal = {\pasp},
         year = 2011,
        month = apr,
       volume = {123},
       number = {902},
        pages = {384},
          doi = {10.1086/659743},
archivePrefix = {arXiv},
       eprint = {1102.3038},
 primaryClass = {astro-ph.GA},
       adsurl = {https://ui.adsabs.harvard.edu/abs/2011PASP..123..384F}
}

@ARTICLE{Marconi2015,
       author = {{Marconi}, M. and {Coppola}, G. and {Bono}, G. and {Braga}, V. and {Pietrinferni}, A. and {Buonanno}, R. and {Castellani}, M. and {Musella}, I. and {Ripepi}, V. and {Stellingwerf}, R.~F.},
        title = "{On a New Theoretical Framework for RR Lyrae Stars. I. The Metallicity Dependence}",
      journal = {\apj},
         year = 2015,
        month = jul,
       volume = {808},
       number = {1},
          eid = {50},
        pages = {50},
          doi = {10.1088/0004-637X/808/1/50},
archivePrefix = {arXiv},
       eprint = {1505.02531},
 primaryClass = {astro-ph.SR},
       adsurl = {https://ui.adsabs.harvard.edu/abs/2015ApJ...808...50M}
}

@BOOK{CatelanSmith2015,
       author = {{Catelan}, M. and {Smith}, H.~A.},
        title = "{Pulsating Stars (Wiley-VCH, Weinheim)}",
         year = 2015,
       adsurl = {https://ui.adsabs.harvard.edu/abs/2015pust.book.....C}
}

@ARTICLE{Fiorentino2015,
       author = {{Fiorentino}, Giuliana and {Bono}, Giuseppe and {Monelli}, Matteo and {Stetson}, Peter B. and {Tolstoy}, Eline and {Gallart}, Carme and {Salaris}, Maurizio and {Mart{\'\i}nez-V{\'a}zquez}, Clara E. and {Bernard}, Edouard J.},
        title = "{Weak Galactic Halo-Dwarf Spheroidal Connection from RR Lyrae Stars}",
      journal = {\apjl},
         year = 2015,
        month = jan,
       volume = {798},
       number = {1},
          eid = {L12},
        pages = {L12},
          doi = {10.1088/2041-8205/798/1/L12},
archivePrefix = {arXiv},
       eprint = {1411.7300},
 primaryClass = {astro-ph.SR},
       adsurl = {https://ui.adsabs.harvard.edu/abs/2015ApJ...798L..12F}
}

@INPROCEEDINGS{fiorentino2022,
       author = {{Fiorentino}, G. and {Bono}, G. and {Braga}, V.~F. and {Monelli}, M. and {Fabrizio}, M. and {Crestani}, J. and {Fern{\'a}ndez Alvar}, E. and {Salaris}, M. and {Stetson}, P.~B. and {Martinez-Vazsquez}, C.~E. and {Dall'Ora}, M. and {Di Criscienzo}, M. and {D'Orazi}, V. and {Marengo}, M. and {Mullen}, J.~P. and {Kwak}, S. and {Tantalo}, M.},
        title = "{On the early formation of the Galactic halo traced by RR Lyrae stars}",
    booktitle = {Memorie della Societa Astronomica Italiana},
         year = 2022,
       volume = {93},
        month = dec,
        pages = {47},
          doi = {10.36116/MEMSAIT_93N4.2022.47},
       adsurl = {https://ui.adsabs.harvard.edu/abs/2022MmSAI..93d..47F}
}

@ARTICLE{Preston1959,
       author = {{Preston}, George W.},
        title = "{A Spectroscopic Study of the RR Lyrae Stars.}",
      journal = {\apj},
         year = 1959,
        month = sep,
       volume = {130},
        pages = {507},
          doi = {10.1086/146743},
       adsurl = {https://ui.adsabs.harvard.edu/abs/1959ApJ...130..507P}}

@ARTICLE{Botan2021,
       author = {{Botan}, E. and {Saito}, R.~K. and {Minniti}, D. and {Kanaan}, A. and {Contreras Ramos}, R. and {Ferreira}, T.~S. and {Gramajo}, L.~V. and {Navarro}, M.~G.},
        title = "{Unveiling short-period binaries in the inner VVV bulge}",
      journal = {\mnras},
         year = 2021,
        month = jun,
       volume = {504},
       number = {1},
        pages = {654-666},
          doi = {10.1093/mnras/stab888},
archivePrefix = {arXiv},
       eprint = {2103.16023},
 primaryClass = {astro-ph.SR},
       adsurl = {https://ui.adsabs.harvard.edu/abs/2021MNRAS.504..654B}
}

@ARTICLE{Fabrizio2019,
       author = {{Fabrizio}, M. and {Bono}, G. and {Braga}, V.~F. and {Magurno}, D. and {Marinoni}, S. and {Marrese}, P.~M. and {Ferraro}, I. and {Fiorentino}, G. and {Giuffrida}, G. and {Iannicola}, G. and {Monelli}, M. and {Altavilla}, G. and {Chaboyer}, B. and {Dall'Ora}, M. and {Gilligan}, C.~K. and {Layden}, A. and {Marengo}, M. and {Nonino}, M. and {Preston}, G.~W. and {Sesar}, B. and {Sneden}, C. and {Valenti}, E. and {Th{\'e}venin}, F. and {Zoccali}, E.},
        title = "{On the Use of Field RR Lyrae as Galactic Probes. I. The Oosterhoff Dichotomy Based on Fundamental Variables}",
      journal = {\apj},
         year = 2019,
        month = sep,
       volume = {882},
       number = {2},
          eid = {169},
        pages = {169},
          doi = {10.3847/1538-4357/ab3977},
archivePrefix = {arXiv},
       eprint = {1908.02064},
 primaryClass = {astro-ph.SR},
       adsurl = {https://ui.adsabs.harvard.edu/abs/2019ApJ...882..169F}
}

@ARTICLE{Bono2020,
       author = {{Bono}, G. and {Braga}, V.~F. and {Crestani}, J. and {Fabrizio}, M. and {Sneden}, C. and {Marconi}, M. and {Preston}, G.~W. and {Mullen}, J.~P. and {Gilligan}, C.~K. and {Fiorentino}, G. and {Pietrinferni}, A. and {Altavilla}, G. and {Buonanno}, R. and {Chaboyer}, B. and {da Silva}, R. and {Dall'Ora}, M. and {Degl'Innocenti}, S. and {Di Carlo}, E. and {Ferraro}, I. and {Grebel}, E.~K. and {Iannicola}, G. and {Inno}, L. and {Kovtyukh}, V. and {Kunder}, A. and {Lemasle}, B. and {Marengo}, M. and {Marinoni}, S. and {Marrese}, P.~M. and {Mart{\'\i}nez-V{\'a}zquez}, C.~E. and {Matsunaga}, N. and {Monelli}, M. and {Neeley}, J. and {Nonino}, M. and {Moroni}, P.~G. Prada and {Prudil}, Z. and {Stetson}, P.~B. and {Th{\'e}venin}, F. and {Tognelli}, E. and {Valenti}, E. and {Walker}, A.~R.},
        title = "{On the Metamorphosis of the Bailey Diagram for RR Lyrae Stars}",
      journal = {\apjl},
         year = 2020,
        month = jun,
       volume = {896},
       number = {1},
          eid = {L15},
        pages = {L15},
          doi = {10.3847/2041-8213/ab9538},
archivePrefix = {arXiv},
       eprint = {2005.11566},
 primaryClass = {astro-ph.SR},
       adsurl = {https://ui.adsabs.harvard.edu/abs/2020ApJ...896L..15B}
}

@ARTICLE{prudil2022,
       author = {{Prudil}, Z. and {Koch-Hansen}, A.~J. and {Lemasle}, B. and {Grebel}, E.~K. and {Marchetti}, T. and {Hansen}, C.~J. and {Crestani}, J. and {Braga}, V.~F. and {Bono}, G. and {Chaboyer}, B. and {Fabrizio}, M. and {Dall'Ora}, M. and {Mart{\'\i}nez-V{\'a}zquez}, C.~E.},
        title = "{Milky Way archaeology using RR Lyrae and type II Cepheids. II. High-velocity RR Lyrae stars and Milky Way mass}",
      journal = {\aap},
         year = 2022,
        month = aug,
       volume = {664},
          eid = {A148},
        pages = {A148},
          doi = {10.1051/0004-6361/202142251},
archivePrefix = {arXiv},
       eprint = {2206.00417},
 primaryClass = {astro-ph.GA},
       adsurl = {https://ui.adsabs.harvard.edu/abs/2022A&A...664A.148P}
}

@ARTICLE{CruzReyes2024,
       author = {{Cruz Reyes}, Mauricio and {Anderson}, Richard I. and {Johansson}, Lucas and {Netzel}, Henryka and {Medaric}, Zo{\'e}},
        title = "{Variable stars in galactic globular clusters. I. The population of RR Lyrae stars}",
      journal = {\aap},
         year = 2024,
        month = apr,
       volume = {684},
          eid = {A173},
        pages = {A173},
          doi = {10.1051/0004-6361/202348961},
archivePrefix = {arXiv},
       eprint = {2402.08843},
 primaryClass = {astro-ph.SR},
       adsurl = {https://ui.adsabs.harvard.edu/abs/2024A&A...684A.173C}
}

@ARTICLE{Skarka2013,
       author = {{Skarka}, M.},
        title = "{Known Galactic field Blazhko stars}",
      journal = {\aap},
         year = 2013,
        month = jan,
       volume = {549},
          eid = {A101},
        pages = {A101},
          doi = {10.1051/0004-6361/201220398},
archivePrefix = {arXiv},
       eprint = {1210.7120},
 primaryClass = {astro-ph.SR},
       adsurl = {https://ui.adsabs.harvard.edu/abs/2013A&A...549A.101S}
}

@ARTICLE{Bono2025,
       author = {{Bono}, G. and {Braga}, V.~F. and {Fabrizio}, M. and {Tantalo}, M. and {Baeza-Villagra}, K. and {Crestani}, J. and {D'Orazi}, V. and {Dall'Ora}, M. and {Di Criscienzo}, M. and {Fiorentino}, G. and {Gholami}, M. and {Marengo}, M. and {Mart{\'\i}nez-V{\'a}zquez}, C.~E. and {Monelli}, M. and {Mullen}, J.~P. and {Nunnari}, A. and {Pipwala}, V.~D. and {Prudil}, Z. and {Sneden}, C. and {Altavilla}, G. and {Bergemann}, M. and {B{\"o}cek Topcu}, G. and {Buonanno}, R. and {Calamida}, A. and {Carretta}, E. and {Ceci}, G. and {Chaboyer}, B. and {Correnti}, M. and {da Silva}, R. and {Ferraro}, I. and {G{\'o}mez}, F.~A. and {Iannicola}, G. and {Kudritzki}, R.-P. and {Kunder}, A. and {Kwak}, S. and {Marconi}, M. and {Marinoni}, S. and {Matsunaga}, N. and {Matteucci}, F. and {Monachesi}, A. and {Musella}, I. and {Navarro Ovando}, M.~G. and {Preston}, G.~W. and {Ripepi}, V. and {Salaris}, M. and {S{\'a}nchez-Benavente}, M. and {Spitoni}, E. and {Stetson}, P.~B. and {Th{\'e}venin}, F. and {Thompson}, I.~B. and {Tissera}, P.~B. and {Tsujimoto}, T. and {Valenti}, E. and {Vivas}, A.~K. and {Walker}, A.~R. and {Zoccali}, M. and {Zocchi}, A.},
        title = "{On the Use of Field RR Lyrae as Galactic Probes. VIII. Early Formation of the Galactic Spheroid}",
      journal = {\apj},
         year = 2026,
        month = feb,
       volume = {998},
       number = {1},
          eid = {86},
        pages = {86},
          doi = {10.3847/1538-4357/ae2c60},
archivePrefix = {arXiv},
       eprint = {2601.16523},
 primaryClass = {astro-ph.GA},
       adsurl = {https://ui.adsabs.harvard.edu/abs/2026ApJ...998...86B}
}

@ARTICLE{Bragaglia2001,
       author = {{Bragaglia}, A. and {Gratton}, R.~G. and {Carretta}, E. and {Clementini}, G. and {Di Fabrizio}, L. and {Marconi}, M.},
        title = "{Metallicities for Double-Mode RR Lyrae Stars in the Large Magellanic Cloud}",
      journal = {\aj},
         year = 2001,
        month = jul,
       volume = {122},
       number = {1},
        pages = {207-219},
          doi = {10.1086/321116},
archivePrefix = {arXiv},
       eprint = {astro-ph/0103515},
 primaryClass = {astro-ph},
       adsurl = {https://ui.adsabs.harvard.edu/abs/2001AJ....122..207B}
}

@ARTICLE{schwarzschild1940,
       author = {{Schwarzschild}, Martin},
        title = "{On the variables in Messier 3}",
      journal = {Harvard College Observatory Circular},
         year = 1940,
        month = mar,
       volume = {437},
        pages = {1-12},
       adsurl = {https://ui.adsabs.harvard.edu/abs/1940HarCi.437....1S}
}

@ARTICLE{braga2020,
       author = {{Braga}, V.~F. and {Bono}, G. and {Fiorentino}, G. and {Stetson}, P.~B. and {Dall'Ora}, M. and {Salaris}, M. and {da Silva}, R. and {Fabrizio}, M. and {Marinoni}, S. and {Marrese}, M.~P. and {Mateo}, M. and {Matsunaga}, N. and {Monelli}, M. and {Wallerstein}, G.},
        title = "{Separation between RR Lyrae and type II Cepheids and their importance for a distance determination: the case of omega Cen}",
      journal = {\aap},
         year = 2020,
        month = dec,
       volume = {644},
          eid = {A95},
        pages = {A95},
          doi = {10.1051/0004-6361/202039145},
archivePrefix = {arXiv},
       eprint = {2010.06368},
 primaryClass = {astro-ph.SR},
       adsurl = {https://ui.adsabs.harvard.edu/abs/2020A&A...644A..95B}
}

@ARTICLE{Braga2024,
       author = {{Braga}, V.~F. and {Monelli}, M. and {Dall'Ora}, M. and {Mullen}, J.~P. and {Molinaro}, R. and {Marconi}, M. and {Szab{\'o}}, R. and {Gallart}, C.},
        title = "{On the use of field RR Lyrae as Galactic probes: VII. Light curve templates in the LSST photometric system}",
      journal = {\aap},
         year = 2024,
        month = sep,
       volume = {689},
          eid = {A349},
        pages = {A349},
          doi = {10.1051/0004-6361/202450971},
archivePrefix = {arXiv},
       eprint = {2407.20813},
 primaryClass = {astro-ph.SR},
       adsurl = {https://ui.adsabs.harvard.edu/abs/2024A&A...689A.349B}
}

@ARTICLE{Braga2016,
       author = {{Braga}, V.~F. and {Stetson}, P.~B. and {Bono}, G. and {Dall'Ora}, M. and {Ferraro}, I. and {Fiorentino}, G. and {Freyhammer}, L.~M. and {Iannicola}, G. and {Marengo}, M. and {Neeley}, J. and {Valenti}, E. and {Buonanno}, R. and {Calamida}, A. and {Castellani}, M. and {da Silva}, R. and {Degl'Innocenti}, S. and {Di Cecco}, A. and {Fabrizio}, M. and {Freedman}, W.~L. and {Giuffrida}, G. and {Lub}, J. and {Madore}, B.~F. and {Marconi}, M. and {Marinoni}, S. and {Matsunaga}, N. and {Monelli}, M. and {Persson}, S.~E. and {Piersimoni}, A.~M. and {Pietrinferni}, A. and {Prada-Moroni}, P. and {Pulone}, L. and {Stellingwerf}, R. and {Tognelli}, E. and {Walker}, A.~R.},
        title = "{On the RR Lyrae Stars in Globulars. IV. {\ensuremath{\omega}} Centauri Optical UBVRI Photometry}",
      journal = {\aj},
         year = 2016,
        month = dec,
       volume = {152},
       number = {6},
          eid = {170},
        pages = {170},
          doi = {10.3847/0004-6256/152/6/170},
archivePrefix = {arXiv},
       eprint = {1609.04916},
 primaryClass = {astro-ph.SR},
       adsurl = {https://ui.adsabs.harvard.edu/abs/2016AJ....152..170B}
}

@ARTICLE{Smith1981,
       author = {{Smith}, H.~A.},
        title = "{Period distributions of irregularity variable RR LYR stars.}",
      journal = {\pasp},
         year = 1981,
        month = dec,
       volume = {93},
        pages = {721-727},
          doi = {10.1086/130915},
       adsurl = {https://ui.adsabs.harvard.edu/abs/1981PASP...93..721S}
}

@ARTICLE{Sesar2012,
       author = {{Sesar}, Branimir},
        title = "{Template RR Lyrae H{\ensuremath{\alpha}}, H{\ensuremath{\beta}}, and H{\ensuremath{\gamma}} Velocity Curves}",
      journal = {\aj},
         year = 2012,
        month = oct,
       volume = {144},
       number = {4},
          eid = {114},
        pages = {114},
          doi = {10.1088/0004-6256/144/4/114},
archivePrefix = {arXiv},
       eprint = {1208.1997},
 primaryClass = {astro-ph.SR},
       adsurl = {https://ui.adsabs.harvard.edu/abs/2012AJ....144..114S}
}

@ARTICLE{Liu1991,
       author = {{Liu}, Tianxing},
        title = "{Synthetic RR Lyrae Velocity Curves}",
      journal = {\pasp},
         year = 1991,
        month = feb,
       volume = {103},
        pages = {205},
          doi = {10.1086/132809},
       adsurl = {https://ui.adsabs.harvard.edu/abs/1991PASP..103..205L}
}

@ARTICLE{Oke1966,
       author = {{Oke}, J.~B.},
        title = "{A Spectrophotometric Study of X ARIETIS}",
      journal = {\apj},
         year = 1966,
        month = aug,
       volume = {145},
        pages = {468},
          doi = {10.1086/148787},
       adsurl = {https://ui.adsabs.harvard.edu/abs/1966ApJ...145..468O}
}

@ARTICLE{Prudil2024,
       author = {{Prudil}, Z. and {Smolec}, R. and {Kunder}, A. and {Koch-Hansen}, A.~J. and {D{\'e}k{\'a}ny}, I.},
        title = "{The Galactic bulge exploration. II. Line-of-sight velocity templates for single-mode RR Lyrae stars}",
      journal = {\aap},
         year = 2024,
        month = may,
       volume = {685},
          eid = {A153},
        pages = {A153},
          doi = {10.1051/0004-6361/202347340},
archivePrefix = {arXiv},
       eprint = {2404.05279},
 primaryClass = {astro-ph.SR},
       adsurl = {https://ui.adsabs.harvard.edu/abs/2024A&A...685A.153P}
}

@inproceedings{4most,
       author = {{de Jong}, Roelof S. and {Bellido-Tirado}, Olga and {Chiappini}, Cristina and {Depagne}, {\'E}ric and {Haynes}, Roger and {Johl}, Diana and {Schnurr}, Olivier and {Schwope}, Axel and {Walcher}, Jakob and {Dionies}, Frank and {Haynes}, Dionne and {Kelz}, Andreas and {Kitaura}, Francisco S. and {Lamer}, Georg and {Minchev}, Ivan and {M{\"u}ller}, Volker and {Nuza}, Sebasti{\'a}n. E. and {Olaya}, Jean-Christophe and {Piffl}, Tilmann and {Popow}, Emil and {Steinmetz}, Matthias and {Ural}, Ugur and {Williams}, Mary and {Winkler}, Roland and {Wisotzki}, Lutz and {Ansorge}, Wolfgang R. and {Banerji}, Manda and {Gonzalez Solares}, Eduardo and {Irwin}, Mike and {Kennicutt}, Robert C. and {King}, Dave and {McMahon}, Richard G. and {Koposov}, Sergey and {Parry}, Ian R. and {Sun}, David and {Walton}, Nicholas A. and {Finger}, Gert and {Iwert}, Olaf and {Krumpe}, Mirko and {Lizon}, Jean-Louis and {Vincenzo}, Mainieri and {Amans}, Jean-Philippe and {Bonifacio}, Piercarlo and {Cohen}, Mathieu and {Francois}, Patrick and {Jagourel}, Pascal and {Mignot}, Shan B. and {Royer}, Fr{\'e}d{\'e}ric and {Sartoretti}, Paola and {Bender}, Ralf and {Grupp}, Frank and {Hess}, Hans-Joachim and {Lang-Bardl}, Florian and {Muschielok}, Bernard and {B{\"o}hringer}, Hans and {Boller}, Thomas and {Bongiorno}, Angela and {Brusa}, Marcella and {Dwelly}, Tom and {Merloni}, Andrea and {Nandra}, Kirpal and {Salvato}, Mara and {Pragt}, Johannes H. and {Navarro}, Ram{\'o}n and {Gerlofsma}, Gerrit and {Roelfsema}, Ronald and {Dalton}, Gavin B. and {Middleton}, Kevin F. and {Tosh}, Ian A. and {Boeche}, Corrado and {Caffau}, Elisabetta and {Christlieb}, Norbert and {Grebel}, Eva K. and {Hansen}, Camilla and {Koch}, Andreas and {Ludwig}, Hans-G. and {Quirrenbach}, Andreas and {Sbordone}, Luca and {Seifert}, Walter and {Thimm}, Guido and {Trifonov}, Trifon and {Helmi}, Amina and {Trager}, Scott C. and {Feltzing}, Sofia and {Korn}, Andreas and {Boland}, Wilfried},
        title = "{4MOST: 4-metre multi-object spectroscopic telescope}",
    booktitle = {Ground-based and Airborne Instrumentation for Astronomy IV},
         year = 2012,
       editor = {{McLean}, Ian S. and {Ramsay}, Suzanne K. and {Takami}, Hideki},
       series = {Society of Photo-Optical Instrumentation Engineers (SPIE) Conference Series},
       volume = {8446},
        month = sep,
          eid = {84460T},
        pages = {84460T},
          doi = {10.1117/12.926239},
archivePrefix = {arXiv},
       eprint = {1206.6885},
 primaryClass = {astro-ph.IM},
       adsurl = {https://ui.adsabs.harvard.edu/abs/2012SPIE.8446E..0TD}
}

@ARTICLE{Fernandez2025,
       author = {{Fern{\'a}ndez-Alvar}, Emma and {Ruiz-Lara}, Tom{\'a}s and {Gallart}, Carme and {Cassisi}, Santi and {Surot}, Francisco and {Gonz{\'a}lez-Koda}, Yllari K. and {Callingham}, Thomas M. and {Queiroz}, Anna B. and {Battaglia}, Giuseppina and {Thomas}, Guillaume and {Chiappini}, Cristina and {Hill}, Vanessa and {Dodd}, Emma and {Helmi}, Amina and {Aznar-Menargues}, Guillem and {de la Cueva}, Alejandro and {Mirabal}, David and {Quintana-Ansaldo}, M{\'o}nica and {Rivero}, Alicia},
        title = "{Chronology of our Galaxy from Gaia colour─magnitude diagram fitting (ChronoGal): II. Unveiling the formation and evolution of the kinematically selected thick and thin discs}",
      journal = {\aap},
         year = 2025,
        month = dec,
       volume = {704},
          eid = {A258},
        pages = {A258},
          doi = {10.1051/0004-6361/202553814},
archivePrefix = {arXiv},
       eprint = {2503.19536},
 primaryClass = {astro-ph.GA},
       adsurl = {https://ui.adsabs.harvard.edu/abs/2025A&A...704A.258F}
}

@INPROCEEDINGS{moons,
       author = {{Taylor}, William and {Cirasuolo}, Michele and {Afonso}, Jose and {Carollo}, Marcella and {Evans}, Chris and {Flores}, Hector and {Maiolino}, Roberto and {Paltani}, Stephane and {Vanzi}, Leonardo and {Abreu}, Manuel and {Amans}, Jean-Phillipe and {Atkinson}, David and {Barrett}, Joe and {Beard}, Steven and {B{\'e}chet}, Celementine and {Black}, Martin and {Boettger}, David and {Brierley}, Saskia and {Buscher}, David and {Cabral}, Alexandre and {Cochrane}, William and {Coelho}, Jo{\~a}o. and {Colling}, Miriam and {Conzelmann}, Ralf and {Dalessio}, Francesco and {Dauvin}, Louise and {Davidson}, George and {Drass}, Holger and {D{\"u}nner}, Rolando and {Fairley}, Alasdair and {Fasola}, Giles and {Ferruzzi}, Debora and {Fisher}, Martin and {Flores}, Mauricio and {Garilli}, Bianca and {Gaudemard}, Julien and {Gonzalez}, Oscar and {Guinouard}, Isabelle and {Gutierrez}, Pablo and {Hammersley}, Peter and {Haigron}, R{\'e}gis and {Haniff}, Chris and {Hayati}, Mahmoud and {Ives}, Derek and {Iwert}, Olaf and {Laporte}, Philippe and {Lee}, David and {Li Causi}, Gianluca and {Luco}, Yerko and {Macleod}, Alastair and {Mainieri}, Vincenzo and {Maire}, Charles and {Melse}, Basile-Thierry and {Nix}, Johannes and {Oliva}, Ernesto and {Oliveira}, Ant{\'o}nio and {Origlia}, Livia and {Parry}, Ian and {Pedichini}, Fernando and {Piazzesi}, Roberto and {Rees}, Phil and {Reix}, Florent and {Rodrigues}, Myriam and {Rojas}, Felipe and {Rota}, Stefano and {Royer}, Fr{\'e}d{\'e}ric and {Santos}, Pedro and {Schnell}, Robin and {Shen}, Tzu-Chiang and {Sordet}, Michael and {Strachan}, Jonathan and {Sun}, Xiaowei and {Tait}, Graham and {Torres}, Miguel and {Tozzi}, Andre and {Tulloch}, Simon and {Navarro}, {\'A}lvaro Valenzuela and {Von Dran}, Lauren and {Waring}, Chris and {Watson}, Stephen and {Woodward}, Brian and {Yang}, Yanbin},
        title = "{Rising MOONS: an update on the VLT's next multi-object spectrograph as it begins to grow}",
    booktitle = {Ground-based and Airborne Instrumentation for Astronomy VII},
         year = 2018,
       editor = {{Evans}, Christopher J. and {Simard}, Luc and {Takami}, Hideki},
       series = {Society of Photo-Optical Instrumentation Engineers (SPIE) Conference Series},
       volume = {10702},
        month = jul,
          eid = {107021G},
        pages = {107021G},
          doi = {10.1117/12.2313403},
       adsurl = {https://ui.adsabs.harvard.edu/abs/2018SPIE10702E..1GT}
}

@INPROCEEDINGS{pfs,
       author = {{Tamura}, Naoyuki and {Takato}, Naruhisa and {Shimono}, Atsushi and {Moritani}, Yuki and {Yabe}, Kiyoto and {Ishizuka}, Yuki and {Kamata}, Yukiko and {Ueda}, Akitoshi and {Aghazarian}, Hrand and {Arnouts}, Stephan{\'e} and {Barkhouser}, Robert H. and {Balard}, Philippe and {Barette}, Rudy and {Belhadi}, Mohamed and {Burnham}, Jill A. and {Caplar}, Neven and {Carr}, Michael A. and {Chabaud}, Pierre-Yves and {Chang}, Yin-Chang and {Chen}, Hsin-Yo and {Chou}, Chueh-Yi and {Chu}, You-Hua and {Cohen}, Judith G. and {de Almeida}, Rodorigo P. and {de Oliveira}, Antonio C. and {de Oliveira}, L{\'\i}gia S. and {Dekany}, Richard G. and {Dohlen}, Kjetil and {dos Santos}, Jesulino B. and {dos Santos}, Leandro H. and {Ellis}, Richard S. and {Fabricius}, Maximilian and {Ferreira}, Decio and {Furusawa}, Hisanori and {Garcia-Carpio}, Javier and {Golebiowski}, Mirek and {Gross}, Johannes and {Gunn}, James E. and {Hammond}, Randolph and {Harding}, Albert and {Hart}, Murdock and {Heckman}, Timothy M. and {Ho}, Paul T.~P. and {Hope}, Stephen C. and {Hover}, David J. and {Hsu}, Shu-Fu and {Hu}, Yen-Shan and {Huang}, Ping-Jie and {Jamal}, Sara and {Jaquet}, Marc and {Jeschke}, Eric and {Jing}, Yipeng and {Kado-Fong}, Erin and {Karr}, Jeniffer L. and {Kimura}, Masahiko and {King}, Matthew E. and {Koike}, Michitaro and {Komatsu}, Eiichiro and {Le Brun}, Vincent and {Le F{\`e}vre}, Olivier and {Le Fur}, Arnaud and {Le Mignant}, David and {Ling}, Hung-Hsu and {Loomis}, Craig P. and {Lupton}, Robert H. and {Madec}, Fabrice and {Mao}, Peter H. and {Marchesini}, Danilo and {Marrara}, Lucas S. and {Medvedev}, Dmitry and {Mineo}, Sogo and {Minowa}, Yosuke and {Murayama}, Hitoshi and {Murray}, Graham J. and {Ohyama}, Youichi and {Onodera}, Masato and {Orndorff}, Joseph and {Pascal}, Sandrine and {Peebles}, Josh and {Pernot}, Guillaume and {Pourcelot}, Raphael and {Reiley}, Daniel J. and {Reinecke}, Martin and {Roberts}, Mitsuko and {Rosa}, Josimar A. and {Rousselle}, Julien and {Schmitt}, Alain and {Schwochert}, Mark A. and {Seiffert}, Micheal D. and {Siddiqui}, Hassan and {Smee}, Stephen A. and {Sodr{\'e}}, Laerte and {Steinkraus}, Aaron J. and {Strauss}, Michael A. and {Surace}, Christian and {Tait}, Philip J. and {Takada}, Masahiro and {Tamura}, Tomonori and {Tanaka}, Masayuki and {Tanaka}, Yoko and {Thakar}, Aniruddha R. and {Verducci}, Orlando and {Vibert}, Didier and {Wang}, Shiang-Yu and {Wang}, Zuo and {Wen}, Chih-Yi and {Werner}, Suzanne and {Yamada}, Yoshihiko and {Yan}, Chi-Hung and {Yasuda}, Naoki and {Yoshida}, Hiroshige and {Yoshida}, Michitoshi},
        title = "{Prime Focus Spectrograph (PFS) for the Subaru telescope: ongoing integration and future plans}",
    booktitle = {Ground-based and Airborne Instrumentation for Astronomy VII},
         year = 2018,
       editor = {{Evans}, Christopher J. and {Simard}, Luc and {Takami}, Hideki},
       series = {Society of Photo-Optical Instrumentation Engineers (SPIE) Conference Series},
       volume = {10702},
        month = jul,
          eid = {107021C},
        pages = {107021C},
          doi = {10.1117/12.2311871},
       adsurl = {https://ui.adsabs.harvard.edu/abs/2018SPIE10702E..1CT}
}

@INPROCEEDINGS{weave,
       author = {{Dalton}, Gavin and {Trager}, Scott C. and {Abrams}, Don Carlos and {Carter}, David and {Bonifacio}, Piercarlo and {Aguerri}, J. Alfonso L. and {MacIntosh}, Mike and {Evans}, Chris and {Lewis}, Ian and {Navarro}, Ramon and {Agocs}, Tibor and {Dee}, Kevin and {Rousset}, Sophie and {Tosh}, Ian and {Middleton}, Kevin and {Pragt}, Johannes and {Terrett}, David and {Brock}, Matthew and {Benn}, Chris and {Verheijen}, Marc and {Cano Infantes}, Diego and {Bevil}, Craige and {Steele}, Iain and {Mottram}, Chris and {Bates}, Stuart and {Gribbin}, Francis J. and {Rey}, J{\"u}rg and {Rodriguez}, Luis Fernando and {Delgado}, Jose Miguel and {Guinouard}, Isabelle and {Walton}, Nic and {Irwin}, Michael J. and {Jagourel}, Pascal and {Stuik}, Remko and {Gerlofsma}, Gerrit and {Roelfsma}, Ronald and {Skillen}, Ian and {Ridings}, Andy and {Balcells}, Marc and {Daban}, Jean-Baptiste and {Gouvret}, Carole and {Venema}, Lars and {Girard}, Paul},
        title = "{WEAVE: the next generation wide-field spectroscopy facility for the William Herschel Telescope}",
    booktitle = {Ground-based and Airborne Instrumentation for Astronomy IV},
         year = 2012,
       editor = {{McLean}, Ian S. and {Ramsay}, Suzanne K. and {Takami}, Hideki},
       series = {Society of Photo-Optical Instrumentation Engineers (SPIE) Conference Series},
       volume = {8446},
        month = sep,
          eid = {84460P},
        pages = {84460P},
          doi = {10.1117/12.925950},
       adsurl = {https://ui.adsabs.harvard.edu/abs/2012SPIE.8446E..0PD}
}

\begin{appendix} 
\section{Validation with calibration sample}\label{sec:Test}
To verify the accuracy of our results, we selected two calibration samples (CS). The first consists of 19 stars (16 RRab and 3 RRc) from the du Pont catalog (CS-D), with an average of approximately 116 measurements per star. The second includes 56 stars (39 RRab and 17 RRc) from the Gaia DR3 catalog (CS-G), with an average of about 30 measurements per star. The calibration stars were selected visually, ensuring that each had RV measurements covering the full pulsation cycle. We used these CS to validate both the $V_{\gamma}$ and the RV amplitudes (Amp(RV)) derived in this work. A good agreement between our measurements and literature values, without significant systematic offsets, would confirm the reliability of our method. 

\begin{figure}[!htbp]
\centering
\includegraphics[width=0.45\textwidth]{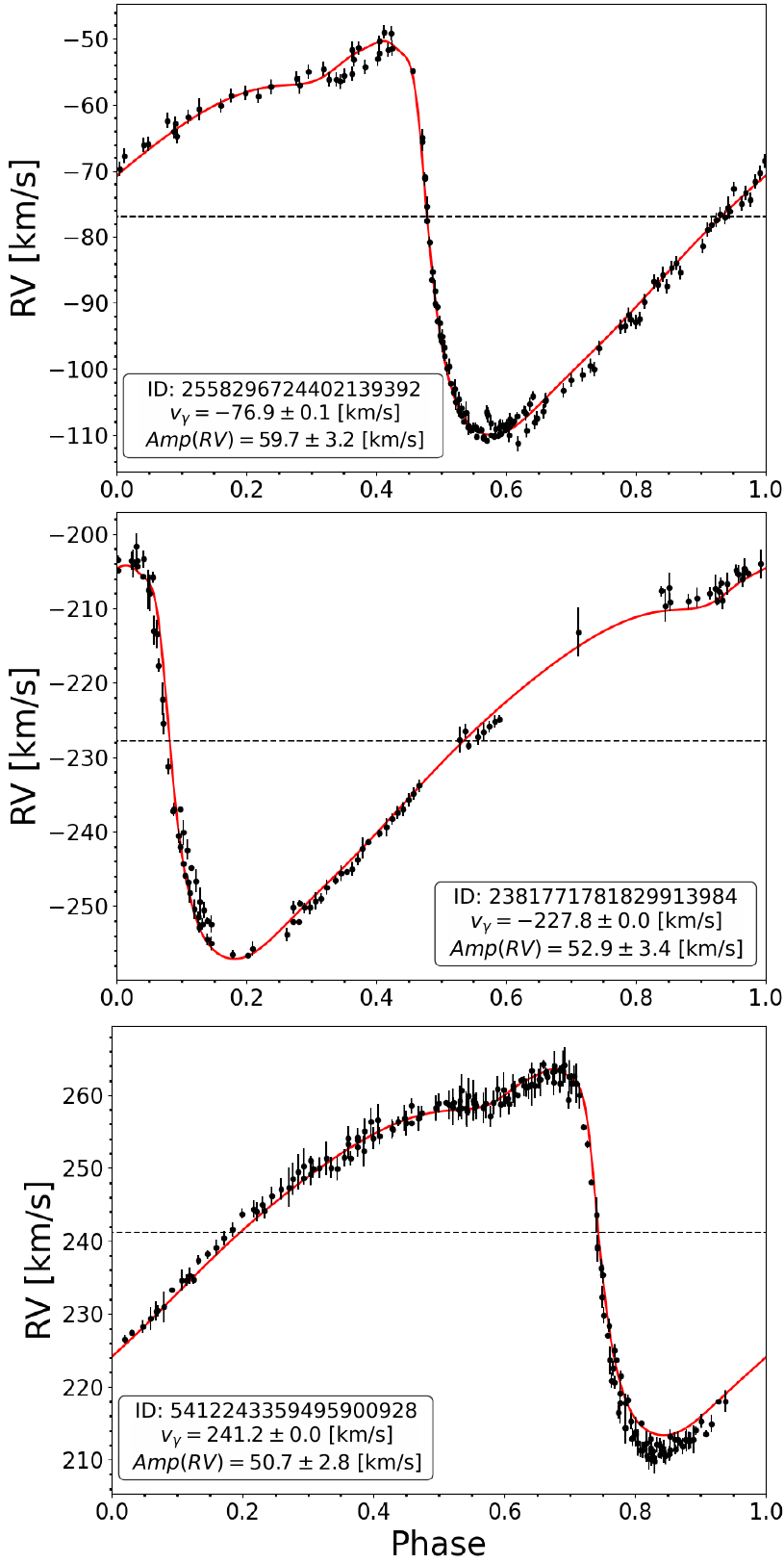}
\caption{Example of the RV curve as a function of phase for three RRab stars in the calibration sample. Individual measurements are marked with black dots, and the solid red line represents the RVC template based on Fe lines, using the free-amplitude fitting method. The dashed black line indicates the mean of $V_{\gamma}$ and the RV amplitude obtained is labeled.}\label{fig:CS_Fe}
\end{figure}

Fig.~\ref{fig:CS_Fe} shows examples of RV curves as a function of phase for three fundamental mode stars in the calibration sample, based on Fe lines using the free-amplitude fitting method.

\subsection{Statistical validation of barycentric radial velocity and RV amplitude estimates from template fitting}

To evaluate the robustness of the $V_{\gamma}$ estimates obtained through template fitting compared with the traditional approach based on averaging RV measurements, we performed a statistical validation test. For each star in our CS, we repeated the procedure 100 times by randomly selecting subsets containing 3, 4, 5, 6, 9, and 12 RV measurements. In each iteration, we estimated $V_{\gamma}$ using three different methods: (i) the simple mean of the RV measurements; (ii) template fitting with a fixed-amplitude based on Fe lines; and (iii) template fitting with a free-amplitude, also based on Fe lines, adopted only when more than three RV measurements were available.

\begin{figure*}[!htbp]
\centering
\includegraphics[width=0.95\textwidth]{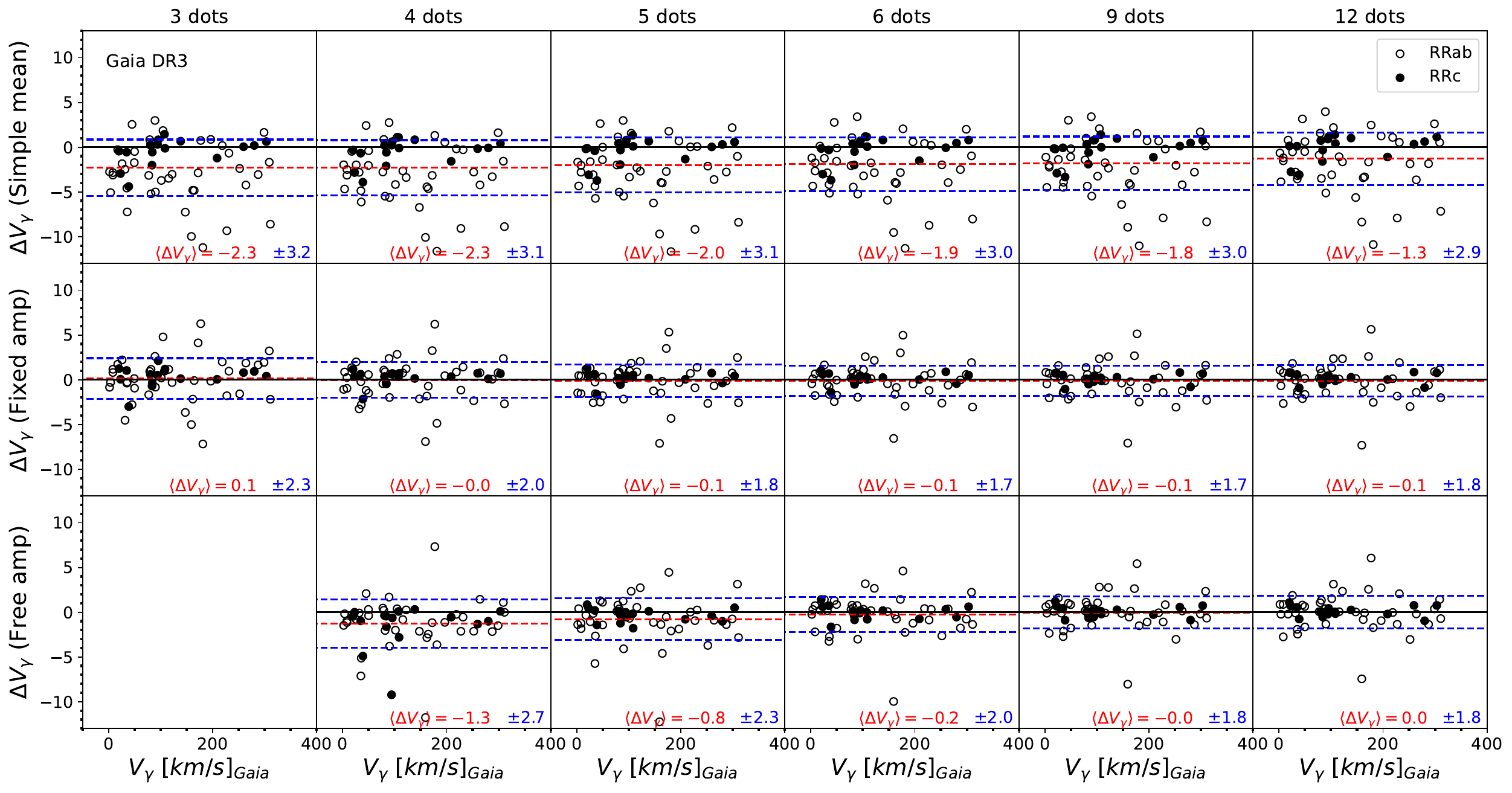}
\caption{Comparison of the $\Delta V_\gamma$ using three different methods: simple mean of RV measurements (top row), template fitting with fixed-amplitude (middle row), and template fitting with free-amplitude (bottom row). The RRab variables are shown as open circles, while RRc are shown as filled circles. Each column corresponds to a different number of RV measurements retained per iteration (3, 4, 5, 6, 9, and 12). The black line indicates $\Delta V_\gamma = 0$; the dashed red line represents the mean $\Delta V_\gamma$ across all stars, and the dashed blue lines show the mean plus and minus the standard deviation. The corresponding mean and standard deviation values are indicated in each subplot.}\label{fig:T1}
\end{figure*}

To quantify the accuracy of each method, we calculated the $ \Delta V_\gamma$ value defined as

\begin{equation*}
\Delta V_\gamma = \langle V_\gamma^{method} - V_\gamma^{literature}\rangle,
\end{equation*}

where $V_\gamma^{\mathrm{literature}}$ refers to the values reported in Gaia DR3, in which the RVC of the target star is modeled using a truncated Fourier series when they have more than 3 RV measurements. Fig.~\ref{fig:T1} shows the final results obtained using the CS-G. The results reveal that the simple mean method consistently exhibits a negative bias in $\Delta V_\gamma$, with mean offsets ranging from -2.3 km s$^{-1}$ to -1.3 km s$^{-1}$ and standard deviations between 2.9 and 3.2 km s$^{-1}$. This highlights the sensitivity of this approach to both the distribution of individual measurements along the RVC and the number of available data points.

In contrast, the fixed-amplitude method yields mean offsets much closer to zero, with significantly smaller scatter (1.7-2.3 km s$^{-1}$), demonstrating greater robustness to phase sampling. The free-amplitude template method also performs better than the simple mean, though in some cases it exhibits a residual bias (e.g., -1.2 km s$^{-1}$ for four observations), possibly due to degeneracies between the shape and amplitude of the template when phase coverage is sparse. Its dispersion remains comparable to that of the fixed-amplitude method (1.8-2.7 km s$^{-1}$).

The difference between the simple mean and the fixed-amplitude method decreases as the number of phase points increases (from 2.4 km s$^{-1}$ with 3 observations to 1.2 km s$^{-1}$ with twelve), indicating that the bias introduced by the simple mean diminishes with improved phase coverage. However, it remains less precise than template-based approaches. Differences between the fixed- and free-amplitude templates are consistently small ($<$1.3 km s$^{-1}$ in all cases) and tend to converge with increasing observational sampling, suggesting that free-amplitude fitting can be a valuable alternative when amplitude estimates are uncertain or unavailable.

The RRc variables in the three adopted methods show a better performance, meaning smaller residuals and less scatter, compared to RRab variables. Quantitatively, RRc stars exhibit dispersions that are systematically lower by $\sim$1-2 km s$^{-1}$ compared to RRab stars, depending on the method and number of observations. This behavior is expected, since RRc light curves are more sinusoidal, and in turn,  more amenable to template fitting and less sensitive to sampling effects.

These findings support the conclusion that, for reliable $V_{\gamma}$ determinations in RRLs, template fitting methods are superior to the simple mean, even with a limited number of observations. Among them, fixed-amplitude template fitting generally provides the best trade-off between precision and robustness, especially when a representative template and accurate amplitude estimate are available.

In addition, we applied the same methodology to evaluate the consistency of RV amplitude estimates for RRLs, using the peak-to-peak amplitude of the RVC as provided by Gaia DR3. To quantify the accuracy of each method, we calculated the $ \Delta Amp(RV)$ value defined as

\begin{equation*}
\Delta Amp(RV)= \langle Amp(RV)^{method} - Amp(RV)^{literature}\rangle.
\end{equation*}

\begin{figure*}[!htbp]
\centering
\includegraphics[width=0.95\textwidth]{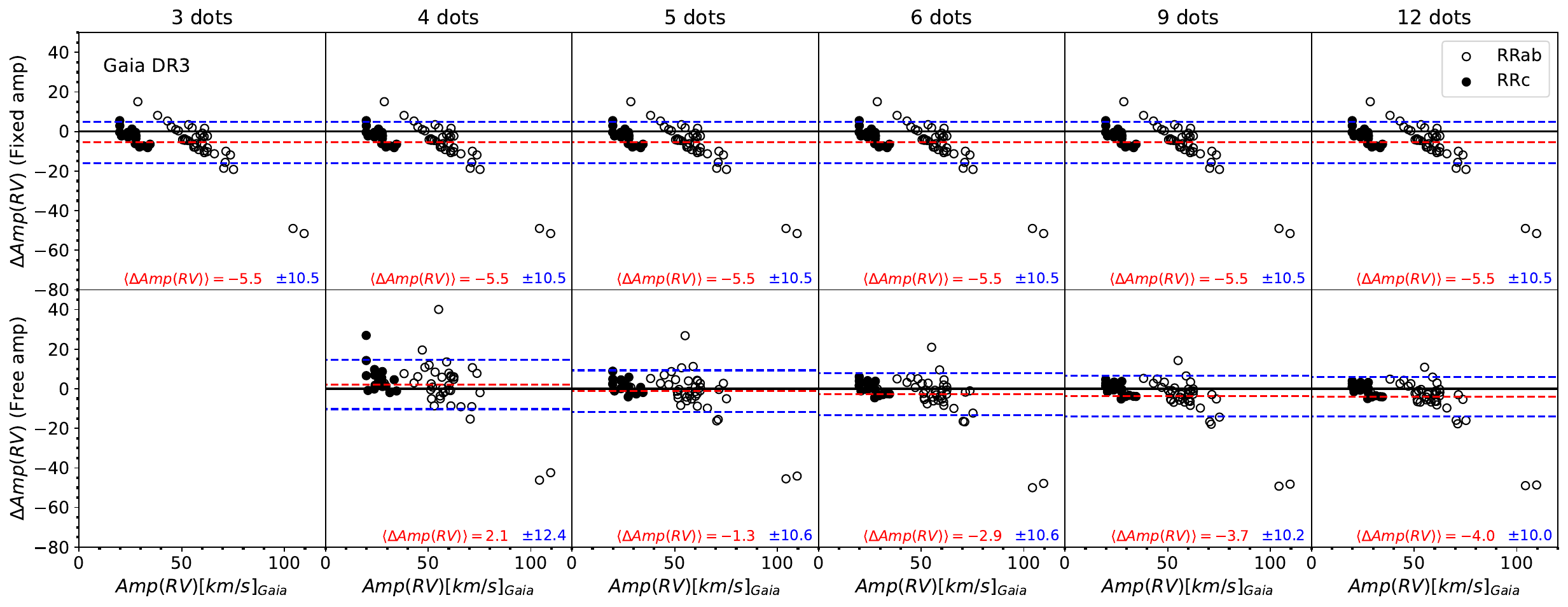}
\caption{Same as Fig.~\ref{fig:T1}, but showing the differences in RV amplitude ($\Delta$Amp(RV)) between the estimated values from both template-fitting methods and the Gaia DR3 reference amplitudes.}\label{fig:T1_amp}
\end{figure*}

\begin{figure}[!htbp]
\centering
\includegraphics[width=0.45\textwidth]{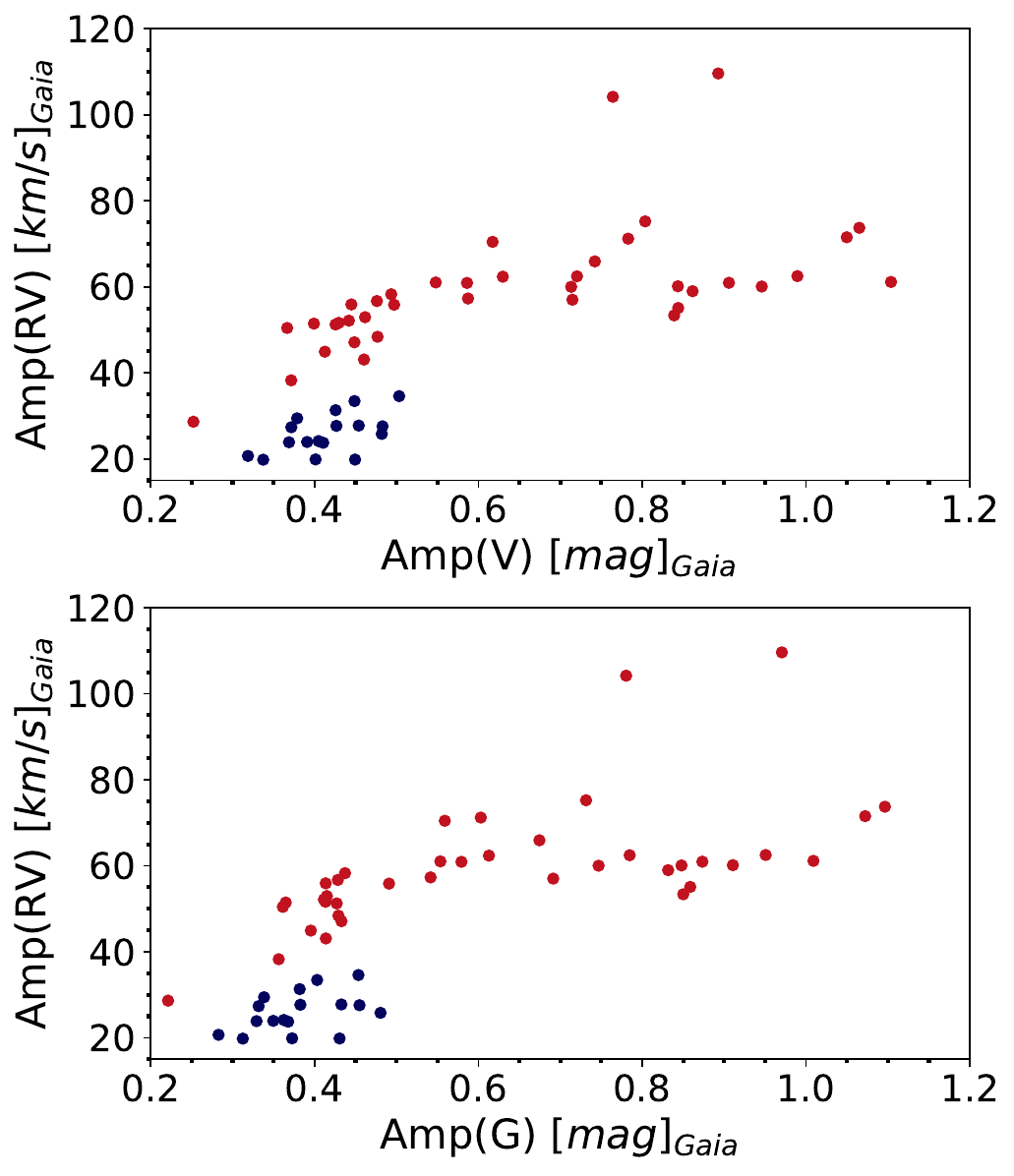}
\caption{Comparison between the RV amplitudes (Amp(RV)) provided by the literature (Gaia DR3) and the optical amplitudes in the Gaia V (top panel) and G (bottom panel) bands. Red and blue points correspond to RRab and RRc variables, respectively.}\label{fig:GaiaopticalRV}
\end{figure}

It is important to note that in the fixed-amplitude template method, the expected RV amplitude is not measured directly from the RVC but rather inferred from the photometric $V$-band amplitude (see \citet{Braga2021} for a detailed discussion). As a result, the outcome does not vary with the number of available data points.
Fig.~\ref{fig:T1_amp} shows that, for the fixed-amplitude method, the mean residual with respect to the reference value remains constant at –5.5 km s$^{-1}$, with a standard deviation of 10.5 km s$^{-1}$.
It is also important to highlight that Gaia optical and RV amplitudes are not related by a linear correlation (see Fig.~\ref{fig:GaiaopticalRV}). Therefore, when a linear transformation is adopted to convert optical amplitudes into RV amplitudes, the resulting RV values do not match those measured by Gaia, but instead follow a linear relation with them. In other words, this is most likely due to the fact that RV time series in Gaia have fewer phase points with respect Gaia \textit{G}-band light curves, and thus provide an increasing underestimation of the R\textit{V}-band amplitude with increasing \textit{G}-band amplitude.

In contrast, the free-amplitude method showed a progressive convergence toward the fixed-amplitude result as the number of data points increased. For four RV measurements, the mean offset was 2.1 km s$^{-1}$ with a standard deviation of 12.4 km s$^{-1}$, whereas for 12 measurements, the offset decreased to –4.0 km s$^{-1}$ with a standard deviation of 10.0 km s$^{-1}$. The absolute difference between the two methods also decreased, from 7.7 km s$^{-1}$ (four points) to 1.5 km s$^{-1}$ (12 points), indicating that the free-amplitude method becomes increasingly consistent with the fixed-amplitude approach as the number of RV measurements increases. 
In conclusion, the free-amplitude method offers greater flexibility, particularly when $A_V$ measurements are uncertain or unavailable. For observations with sparse phase coverage, the use of template-based methods is essential to mitigate amplitude underestimation and to obtain reliable pulsation parameters.

\subsection{Comparison of barycentric radial velocity and RV amplitudes with literature values}

\begin{figure}[!htbp]
\centering
\includegraphics[width=0.45\textwidth]{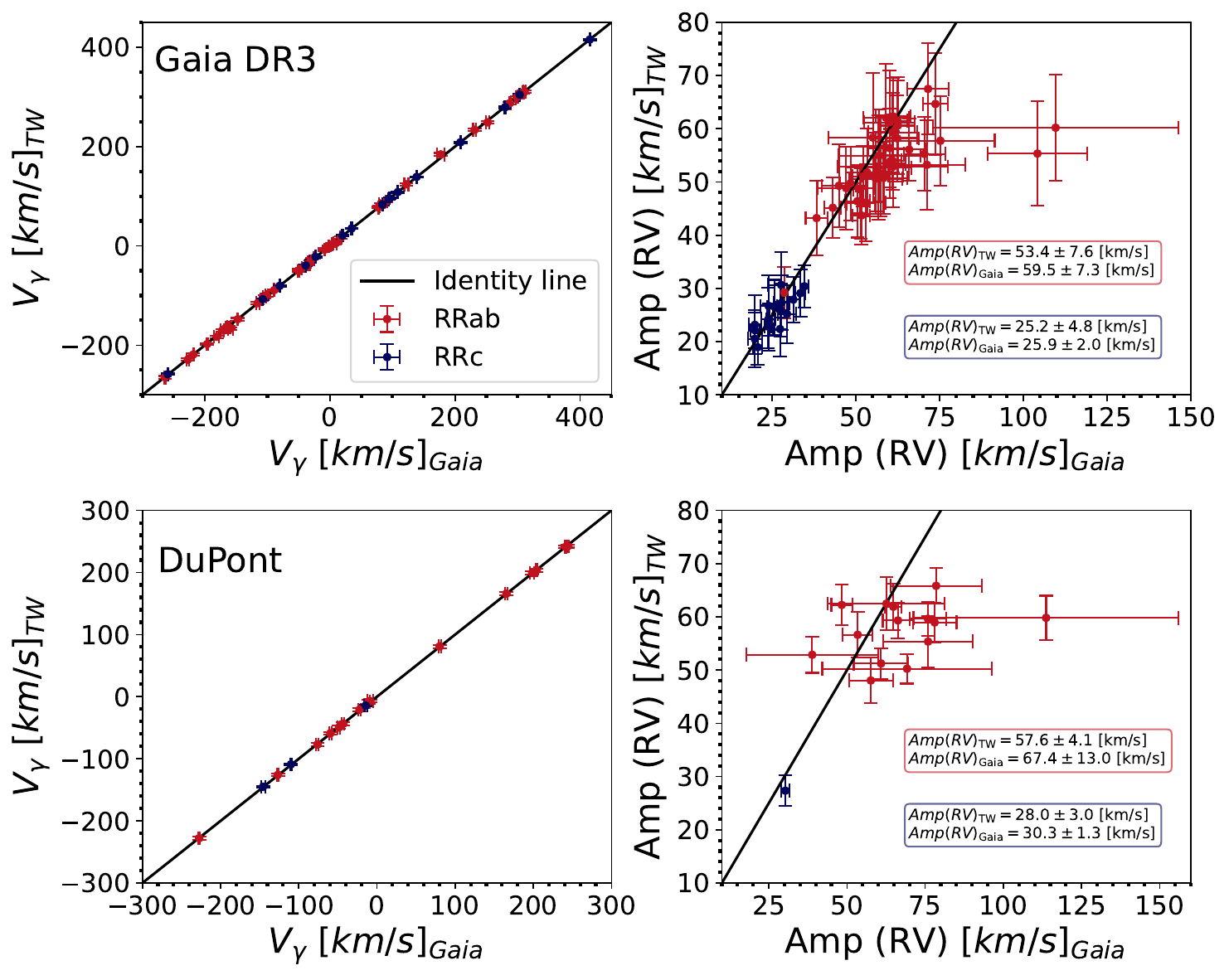}
\caption{Comparison of the accuracy in estimating the $V_{\gamma}$ and the RV amplitude (Amp(RV)), obtained in this work (TW) versus literature values (Gaia DR3). The left panels show $V_{\gamma}$, while the right panels display Amp(RV). The top row corresponds to the CS-G, and the bottom row to the CS-D. Red and blue dots represent RRab and RRc variables, respectively, while the black line denotes the identity line. The mean amplitude values for RRab and RRc stars are also indicated.}\label{fig:CS}
\end{figure}

We compare the final values of $V_{\gamma}$ and Amp(RV) estimates obtained from the free-amplitude templates in this work (TW) with those reported in the literature. The results are presented in Fig.~\ref{fig:CS}, showing a clear and consistent agreement between our $V_{\gamma}$ measurements and the published values for both CS and both RRab and RRc variable. 

Regarding the RV amplitude estimates, for the CS-G sample, we find that the mean RV amplitude for RRab stars in TW is $53.4 \pm 7.6$ km s$^{-1}$, whereas the literature reports a mean value of $59.5 \pm 7.3$ km s$^{-1}$. This corresponds to a relative difference of approximately 10$\%$. However, we note that in some cases the literature reports amplitudes exceeding 100 km s$^{-1}$, which are far outside the typical range. Therefore, part of the observed discrepancy may stem from methodological differences rather than an actual underestimation in our measurements. In particular, RV amplitudes derived from metallic lines (e.g., Fe) typically range between 30-80 km s$^{-1}$ for RRab stars, and values significantly outside this range may reflect systematic overestimation \citep[see, e.g.,][]{Bono2020,Prudil2024}. One key methodological difference might be the absence of a sigma clipping step in their fitting procedure. Without this step, outliers could significantly bias the amplitude estimates in some cases. We reviewed the analysis presented by \citet{Katz2023}, where several filters are applied to remove noisy data. However, there is no explicit mention of a star-by-star sigma clipping procedure as implemented in our analysis, which may contribute to the observed differences. 
In contrast, for RRc stars, the agreement is significantly better, with a mean amplitude of $25.2 \pm 4.8$ km s$^{-1}$ from TW compared to $25.9 \pm 2.0$ km s$^{-1}$ from the literature. This suggests that our methodology performs particularly well for RRc variables in the CS-G sample, with a relative difference of less than 3 $\%$.

In the CS-D sample, a similar pattern is observed. For RRab stars, our mean amplitude of $57.6 \pm 4.1$ km s$^{-1}$ is below the literature value of $67.4 \pm 13.0$ km s$^{-1}$, indicating a relative difference of $\sim15\%$. However, the difference remains within $1\sigma$ when uncertainties are taken into account. This offset may once again reflect systematic differences in how the amplitudes were derived. For RRc stars, our estimated amplitude of 28.0 $\pm$ 3.0 km s$^{-1}$ is in good agreement with the literature value of 30.3 $\pm$ 1.3 km s$^{-1}$, with a difference well within 1$\sigma$ when uncertainties are taken into account.

In summary, our method provides robust estimates of $V_\gamma$ across both calibration sets and pulsation types. While our RV amplitudes for RRab stars tend to be lower than literature values, this is consistent with expectations from using templates based on Fe lines. These results validate the use of our templates for deriving the $V_{\gamma}$ and RV amplitude parameters in RRLs.

\section{Additional figures}\label{sec:figures}

\begin{figure}[!ht]
\centering
\includegraphics[width=0.49\textwidth]{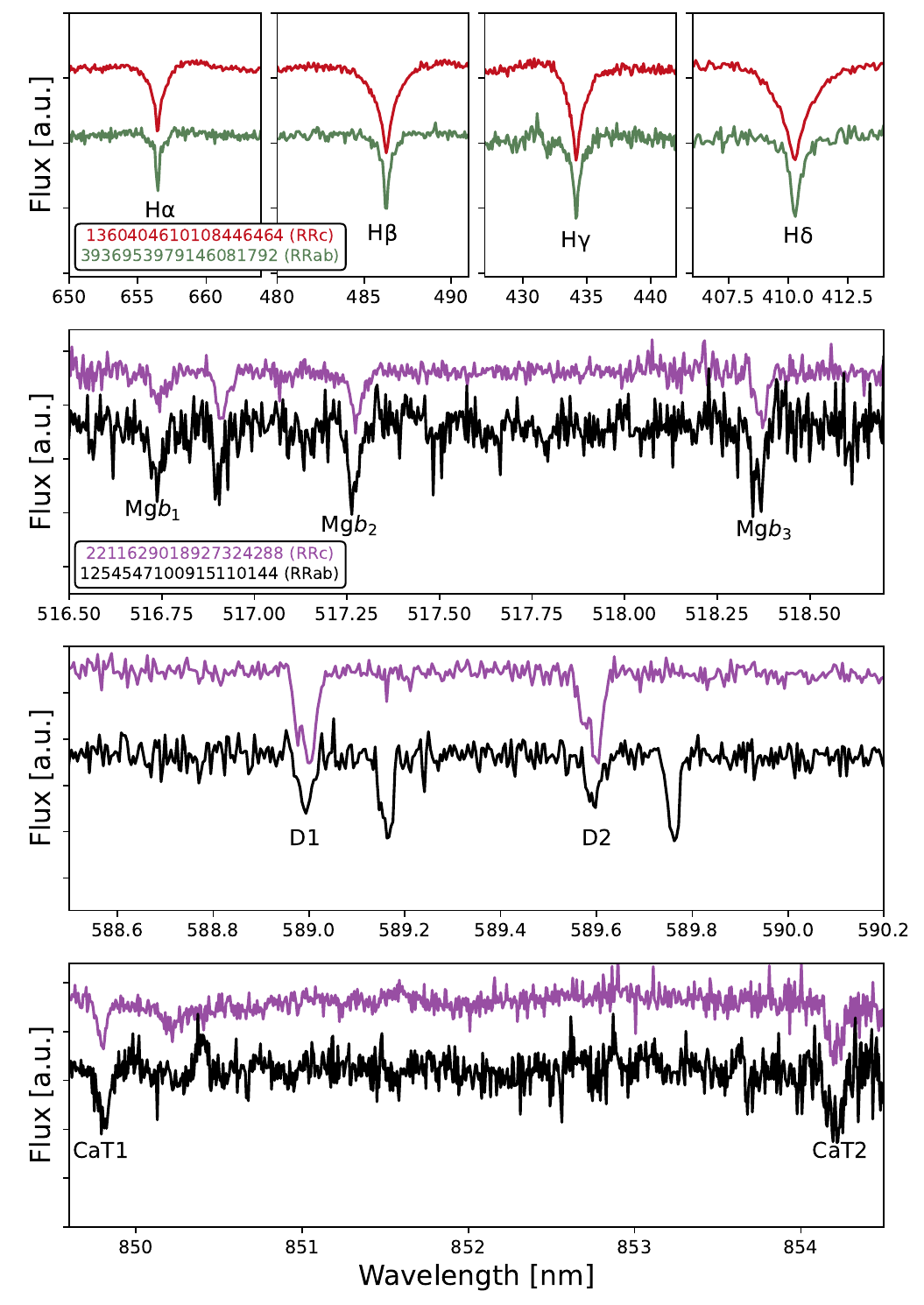}
\caption{From top to bottom, the panels show selected spectra of four RRL stars. Red and violet spectra correspond to RRc variables, while green and black spectra correspond to RRab variables. The four subplots display, from top to bottom, the Balmer lines (from the DESI EDR catalog), the Mg I b triplet, the Na I doublet, and two lines of the Ca II triplet (from the STELLA catalog), all used to derive the RV measurements. Flux units are arbitrary. The Gaia DR3 IDs are labeled.}\label{fig:lines}
\end{figure}

\clearpage

\section{Additional tables}\label{sec:tables}

\begin{table}[!htbp]

\footnotesize
\caption{Key properties of the spectroscopic datasets.}
\label{tab:datasets}
\begin{center}
\begin{tabular}{lccc}
\hline\hline
Datasets & $N_{\text{RRL}}$ & $R$ & $\lambda$ range (\AA) \\
\hline
\multicolumn{3}{c}{Low Resolution}\\
DESI EDR & 594  &2000-5500 & 3600-9800 \\
LAMOST-LR    & 7284 & 2000 &  3700-9000  \\
SDSS DR18    & 5984 & 2000 & 3600-10400   \\
\multicolumn{3}{c}{Medium Resolution}\\
LAMOST-MR     &  639  & 7500 &  4950-6800 \\
Gaia DR3     &  1091  & 11500 &  8450-8720 \\
\multicolumn{3}{c}{High Resolution}\\
STELLA    &   14 & 55000 & 3900-8700 \\
du Pont    &  34& 35000  &  3600-9820\\
\hline\hline
\end{tabular}
\end{center}
\tablefoot{The DESI spectrograph uses three arms-blue (B), red (R), and infrared (Z)-to cover a resolution range of 2000 (at 3600 \AA) to 5500 (at 9800 \AA).}
\end{table}

\begin{table}[!htbp]
\footnotesize
\centering
\caption{Wavelengths ($\lambda$) of the lines adopted for RV measurements.}
\label{tab:wave}

\begin{tabular}{lcc}
\hline\hline
Species & Line ID & $\lambda(\text{\AA})$ \\
\hline

\multicolumn{3}{l}{\hspace{0.1cm}Balmer lines} \\
\hline
$\mathrm{H}_\alpha$ & $\mathrm{H}_\alpha$ & 6562.80 \\
$\mathrm{H}_\beta$  & $\mathrm{H}_\beta$  & 4861.36 \\
$\mathrm{H}_\gamma$ & $\mathrm{H}_\gamma$ & 4340.46 \\
$\mathrm{H}_\delta$ & $\mathrm{H}_\delta$ & 4101.74 \\

\hline
\multicolumn{3}{l}{\hspace{0.1cm}Fe group} \\
\hline
Fe I  & Fe1 & 4045.81 \\
Fe I  & Fe2 & 4063.59 \\
Fe I  & Fe3 & 4071.74 \\
Sr II & Sr  & 4077.71 \\

\hline
\multicolumn{3}{l}{\hspace{0.1cm}Mg group}\\
\hline
Mg I & $\mathrm{Mg} \mathrm{b}_1$ & 5167.32 \\
Mg I & $\mathrm{Mg} \mathrm{b}_2$ & 5172.68 \\
Mg I & $\mathrm{Mg} \mathrm{b}_3$ & 5183.60 \\

\hline
\multicolumn{3}{l}{\hspace{0.1cm}Ca II group}\\
\hline
Ca II & \CaTone & 8498.02 \\
Ca II & \CaTtwo & 8542.09 \\

\hline
\multicolumn{3}{l}{\hspace{0.1cm}Na group}\\
\hline
Na I & D1 & 5889.95 \\
Na I & D2 & 5895.92 \\
\hline\hline
\end{tabular}
\end{table}

\begin{table}[!htbp]
\centering
\footnotesize
\caption{Number of analyzed stars and RV measurements.}
\label{tab:npoints}
\begin{tabular}{lccc}
\hline
\hline
& \multicolumn{3}{c}{Free-method} \\
\cline{2-4}
Line 
& $N_{\rm stars}$ 
&Mean($N_{RV}$) 
& Std($N_{RV}$) \\
\hline
\Ha     & 2851& 8.0& 15.0 \\
\Hb     & 2990&8.0 &  15.4 \\
\Hg     & 2887&7.9 & 15.6 \\
\Hd     &  2527& 8.0& 16.4 \\
\CaTone &256 & 6.6& 3.3  \\
\CaTtwo &  257& 7.0 &  3.6  \\
Mg      &170 & 30.6& 56.4 \\
Na      & 40  &94.7 &  81.4 \\
Fe      &1112 &20.7& 22.1 \\
\hline
& \multicolumn{3}{c}{Fixed-method} \\
\cline{2-4}
\Ha     &4839& 6.0&  11.7 \\
\Hb     & 5057&6.1 & 12.1 \\
\Hg     &4962&5.9 &12.1 \\
\Hd     & 4488&5.8 &  12.6 \\
\CaTone & 578 &4.6 & 2.9  \\
\CaTtwo & 583 &4.8 &  3.1  \\
Mg      & 199 & 26.6&53.0 \\
Na      & 40  &94.7 & 81.4 \\
Fe      & 1113&20.6 & 22.1 \\
\hline
& \multicolumn{3}{c}{Simple mean} \\
\cline{2-4}
Balmer   &  2119& 5.3&  5.2 \\
Metallic & 148  &2.0& 4.8 \\
Calcium  & 1861 &2.4&  1.7 \\
\hline
\hline
\end{tabular}
\tablefoot{The table reports the number of analyzed stars and the mean number of RV measurements, $\langle N_{\rm RV}\rangle$, together with the corresponding standard deviations.}
\end{table}

\begin{table*}[!htbp]
\footnotesize
\caption{Number of stars in common between the different datasets.} 
\label{tab:starscommon}
\begin{center}
\begin{tabular}{lccccccc}
\hline\hline
&DESI EDR & LAMOST-LR & SDSS DR18 & LAMOST-MR & Gaia DR3 & STELLA & du Pont \\
\hline
DESI EDR & 596 & 119 &   186     &   2     & 0       &    0&0    \\
LAMOST-LR &  $\dotsc$&   7284     &  1567      &   393     & 225       &  3  &   3 \\
SDSS DR18 & $\dotsc$&  $\dotsc$  & 5984   &  24      &  1      &     0   &   0     \\
LAMOST-MR& $\dotsc$ & $\dotsc$ &  $\dotsc$      &   639     &   131     &  0   &  1 \\
Gaia DR3 & $\dotsc$ & $\dotsc$  &$\dotsc$        &  $\dotsc$      &  1095      &   11   & 28 \\
STELLA&  $\dotsc$&   $\dotsc$ &$\dotsc$        &  $\dotsc$      & $\dotsc$       &   14    & 0\\
du Pont &$\dotsc$ &  $\dotsc$ &$\dotsc$        &   $\dotsc$     &  $\dotsc$      & $\dotsc$    &  34 \\
\hline\hline
\end{tabular}
\end{center}
\end{table*}

\begin{table*}[!htbp]
\caption{Properties of the RRL stars employed in establishing the amplitude scaling relation.}     
\label{table:properties}
\centering
\footnotesize
\begin{tabular}{r l c l l c r}   
\hline\hline
ID\tablefootmark{a} & Type & Period\tablefootmark{b} & Line &
Amp(RV) & eAmp(RV)$_{flag}$\tablefootmark{c} & N\tablefootmark{d} \\ 
& & (day) & & (km s$^{-1}$) \\ 
\hline          
1793460115244988800 & RRab & 0.390382 & \Ha & 85.9 $\pm$ 4.9 & 0 & 166 \\
1436609218405063936 & RRc & 0.319095 & \Hb & 18.2 $\pm$ 4.4 & 0 & 5 \\
5510293236607430656 & RRab & 0.390745 & \Hg & 77.9 $\pm$ 6.8 & 0 & 178 \\
66719297584339968 & RRc & 0.411452 & \Hd & 26.5 $\pm$ 4.8 & 0 & 12 \\
6768845304331810560 & RRab & 0.442621 & Fe & 75.2 $\pm$ 16.3 & 0 & 10 \\
5022411786734718208 & RRc & 0.377356 & Mg & 26.9 $\pm$ 2.8 & 0 & 54 \\
3479598373678136832 & RRab & 0.567974 & Na & 49.4 $\pm$ 7.0 & 0 & 102 \\
3946316423735761536 & RRc & 0.448578 & \CaTone & 25.9 $\pm$ 7.4 & 0 & 16 \\
919801507393822464 & RRab & 0.616423 & \CaTtwo & 55.1 $\pm$ 10.4 & 0 & 10 \\
\hline\hline
\end{tabular}
\tablefoot{
This table is available entirely at the CDS. A portion is provided here as an example to illustrate its structure and content.\\
\tablefoottext{a}{Gaia DR3 identification number.}
\tablefoottext{b}{Period values adopted from PR3C.}
\tablefoottext{c}{Method used to obtain the error associated with the amplitude.}
\tablefoottext{d}{Number of RV measurements.}
}
\end{table*}

\end{appendix}
\end{document}